\documentclass{aa} 

\usepackage{natbib}
\usepackage{graphicx}
\usepackage{epsfig}
\usepackage{txfonts}
\usepackage{hyperref}
\usepackage{url}
\usepackage{xcolor}
\hypersetup{
colorlinks,
linkcolor={red!80!black},
citecolor={blue!80!black},
urlcolor={blue!80!black}
}

\def\cm3{cm$^{-3}$}
\def\kms{km~s$^{-1}$}
\def\Mms{Mm~s$^{-1}$}
\newcommand{\Teff}{\hbox{$T_{\hbox{\small eff}}$}}

\def\msunyr{M$_{\odot}$\,yr$^{-1}$}

\def\rsun{R$_{\odot}$}

\def\msun{M$_{\odot}$}

\def\one{\ts {\,\sc i}}
\def\two{\ts {\,\sc ii}}
\def\three{\ts {\,\sc iii}}
\def\four{\ts {\,\sc iv}}
\def\five{\ts {\sc v}}

\def\beq{\begin{equation}}
\def\eeq{\end{equation}}

\def\lesssim{\mathrel{\hbox{\rlap{\hbox{\lower4pt\hbox{$\sim$}}}\hbox{$<$}}}}
\def\gtrsim{\mathrel{\hbox{\rlap{\hbox{\lower4pt\hbox{$\sim$}}}\hbox{$>$}}}}

\def\isoni{$^{56}{\rm Ni}$}\def\isoco{$^{56}{\rm Co}$}

\def\one{{\,\sc i}}
\def\two{{\,\sc ii}}
\def\three{{\,\sc iii}}
\def\four{{\,\sc iv}}
\def\five{{\sc v}}

\def\v1d{{\sc v1d}}

\def\mesa{{\sc mesa}}
\def\cmfgen{{\sc cmfgen}}

\newcommand{\iso}[2]{\ensuremath{^{#1}\rm{#2}}}

\def\apjs{ApJS}
\def\apjl{ApJL}
\def\aap{A\&A}

\def\nifs{\iso{56}Ni}

\begin{document}

   \title{On the photometric and spectroscopic diversity of Type II supernovae}
   \titlerunning{Type II SN diversity}

\author{D. John Hillier\inst{\ref{inst1}}
  \and
   Luc Dessart\inst{\ref{inst2}}
  }

\institute{
    Department of Physics and Astronomy \& Pittsburgh Particle Physics,
    Astrophysics, and Cosmology Center (PITT PACC),  University of Pittsburgh,
    3941 O'Hara Street, Pittsburgh, PA 15260, USA.\label{inst1}
\and
Unidad Mixta Internacional Franco-Chilena de Astronom\'ia (CNRS, UMI 3386),
    Departamento de Astronom\'ia, Universidad de Chile,
    Camino El Observatorio 1515, Las Condes, Santiago, Chile.\label{inst2}
  }

   \date{Received; accepted}

  \abstract{
Hydrogen-rich (type II) supernovae (SNe) exhibit considerable photometric and spectroscopic diversity. Extending previous work that focused exclusively on photometry, we simultaneously model the multi-band light curves and optical spectra of Type II SNe using RSG progenitors that are characterized by their H-rich envelope masses or the mass and extent of an enshrouding cocoon at the star's surface. Reducing the H-rich envelope mass yields faster declining light curves, a shorter duration of the photospheric phase, broader line profiles at early times, but only a modest boost in early-time optical brightness. Increasing the mass of the circumstellar material (CSM) is more effective at boosting the early-time brightness and producing a fast-declining light curve while leaving the duration of  the photospheric phase intact. It also makes the optical color bluer, delays the onset of recombination, and can severely reduce the speed of the fastest ejecta material. The early ejecta interaction with CSM is conducive to producing featureless spectra at $10-20$\,d and a weak or absent H$\alpha$ absorption during the recombination phase. The slow decliners SNe 1999em, 2012aw, and 2004et can be explained with a $1.2 \times 10^{51}$\,erg explosion in a compact ($\sim$\,600\,\rsun) RSG star from a 15\,\msun\ stellar evolution model. A small amount of CSM ($<0.2$\,\msun) improves the match to the SN photometry at times $<$ 10\,d. With more extended RSG progenitors, one predicts lower ejecta kinetic energies, but the SN color stays blue for too long and the spectral line widths are too narrow. The fast decliners SNe 2013ej and 2014G may require $0.5-1.0$\,\msun\ of CSM, although this depends on the CSM structure. A larger boost to the luminosity (as for fast decliners SNe 1979C or 1998S) requires interaction with a more spatially extended CSM, which might also be detached from the star.
  }

   \maketitle

\section{Introduction}

Type II supernovae (SNe) exhibit diverse photometric and spectroscopic properties. They cover a range of brightness or luminosity  (at peak or time-averaged over the high-brightness phase), visual decline rates, photospheric phase duration, nebular phase brightness \citep{hamuy_03,anderson_2pl,sanders_sn2_15}. They exhibit a range of spectral line widths,  absorption to emission line equivalent-width ratios (e.g., for H$\alpha$), and metal line strengths \citep{zpap,gutierrez_pap1_17,gutierrez_pap2_17,gutierrez_z_18}. The H$\alpha$ emission and absorption strengths are correlated with the SN decline rate \citep{gutierrez_pap2_17,chen_2p_2l_ha_18}. At early times, narrow emission lines are sometimes seen (e.g., \citealt{fassia_98S_01,smith_ptf11iqb_15,yaron_13fs_17})  whereas one expects the spectrum to form in the fastest expanding material. There is much diversity in line profile morphology \citep{gutierrez_ha_14}. Polarization is also seen, sometimes large, especially at the  transition to the nebular phase (see e.g., \citealt{leonard_04dj_06,leonard_12aw_12}).

All these features have been studied and connected to a physical model of the progenitor star and explosion. Explosion energies are expected to vary by a factor of about 10 as a result of the different iron core structures in  lower and higher mass red-supergiant (RSG) stars (see, e.g., \citealt{ugliano_ccsn_12,sukhbold_ccsn_16}). This can explain the range in plateau brightness and spectral line width, as well as the nebular-phase brightness (from variations of the \nifs\ yield). Progenitors with greater mass-loss rates (through stellar wind or mass exchange with a companion) will die with a smaller H-rich envelope mass, which favors a shorter photospheric-phase duration \citep{litvinova_sn2p_85,bartunov_blinnikov_92,popov_93}. The exact outcome depends on the uncertain RSG mass loss rates \citep{meynet_rsg_15} as well as the role of binarity (physics of mass exchange, frequency of binaries etc; \citealt{yoon_ibc_10,yoon_ib_17,eldridge_sn2_19}). The presence of circumstellar material (CSM) can produce excess brightness at early times \citep{morozova_2l_2p_17,moriya_13fs_17,d18_13fs} as well as narrow lines and blue optical spectra \citep{groh_13cu,grafener_vink_13cu_16,D16_2n,yaron_13fs_17,d18_13fs}. If this interaction is sustained, or if the dense shell  formed from swept-up CSM is massive, it can quench the absorption and boost the emission in lines like H$\alpha$ (as seen in SN\,1998S; \citealt{D16_2n}). Line profile diversity or polarization should also arise from asymmetry \citep{shapiro_sutherland_82}, associated with a distortion of the continuum photosphere \citep{hoeflich_87A_91,jeffery_87_pol_91,leonard_98S_00}, the presence of \nifs\ ``blobs" \citep{chugai_04dj_06}, or a combination of an asymmetric distribution of scatterers and of the flux \citep{DH11_pol}.

There is, however, much degeneracy in the outcomes from RSG star explosions.  The high brightness phase is mostly sensitive to the energy (kinetic and radiative) stored in the shocked H-rich envelope, while the He core material has only a small influence \citep{DH19}. Because stars with a very different ZAMS mass can have the same H-rich envelope mass at explosion, and thus similar LCs, photometric information is of little use to constrain the ejecta or the progenitor mass. The explosion energy is also  hard to constrain since a significant fraction of the explosion energy is used to unbind the progenitor He core. A modest explosion in a 12\,\msun\ star can produce the same ejecta kinetic energy at infinity as a powerful explosion in a 25\,\msun\ star. There are also multiple combinations of energy, mass, and progenitor radius that can deliver the same Type II SN brightness \citep{litvinova_sn2p_85}. The remnant mass and envelope fallback are also poorly constrained so that the inferred ejected \nifs\ mass is not a clean and direct measure of the explosive nucleosynthesis (see, e.g., \citealt{zhang_fallback_08}).

To limit the impact of all these shortcomings and degeneracies, one should use all observational constraints. However, the focus is often only on light curve modeling. For example, the study of Type II SNe by \citet{morozova_2l_2p_17}  is based exclusively on photometric data -- none of their models is tested for dynamical adequacy. In general, however, the modeling is performed using a combination of photometric data and spectroscopic data. The latter is typically limited to a single line (e.g., Fe\two\,5169\,\AA) which is used as a proxy to constrain the ejecta expansion rate.  However, Fe\two\,5169\,\AA\ is not present prior to the recombination phase and thus provides no information on the outer, and hence fastest, ejecta material. It can also underestimate or overestimate the photospheric velocity \citep{D05_epm}. The line width eventually stops decreasing during the recombination phase, no longer reflecting the recession of the photosphere toward the inner ejecta (see, e.g., \citealt{lisakov_08bk_17}). Not all lines behave the same way as Fe\two\,5169\,\AA, so that a given model may reproduce well the width of some spectral lines, and overestimate or underestimate the width of others (this is particularly striking in Type II-pec SNe, which show much heterogeneity; \citealt{DH19_2pec}). The disparity in the strength and width of the absorption and emission parts of various lines (e.g., H$\alpha$) is also suggestive that the conditions for line formation vary significantly amongst SNe (e.g., between fast and slow decliners; \citealt{gutierrez_pap2_17}).

In this study, we present a controlled experiment to explore the origin of the photometric and spectroscopic diversity of Type II SNe. We use two sets of progenitor models, all based on a star with solar metallicity and an initial mass of 15\,\msun. In order to generate pre-SN models with a range of envelope masses, the first set of models is produced by varying the efficiency of mass loss during the RSG phase. This variation is thought to be one mechanism for producing faster-declining light curves \citep{bartunov_blinnikov_92,blinnikov_bartunov_2l_93,snec,moriya_2l_16}. This is the {\it mdot} model set. The second set of models, the {\it ext} set, is generated from one model in the {\it mdot} set by adding an increasing amount of CSM directly above the stellar surface. With these two model sets, we can explore the key observables produced by a reduced envelope mass or by interaction of the ejecta with a confined CSM enshrouding the progenitor star. Unlike all previous studies we assess the impact on spectra, as well as correlate the spectral and light curve properties.

In the next section, we briefly discuss the diversity of photometric and spectroscopic properties of Type II SNe, and in particular how these differ between fast (II-L) and slow decliners (II-P).  Section~\ref{sect_model} then presents the models used in this study, including the pre-SN evolution, the treatment of the explosion, and the radiative-transfer modeling. Results from the radiation-hydrodynamics and the radiative-transfer simulations are presented in Sections~\ref{sect_v1d_res} and \ref{sect_cmfgen_res}. The comparison to well observed slow and fast declining Type II SNe is discussed in Section~\ref{sect_comp_obs}, with conclusion in Section~\ref{sect_conc}.

 \begin{figure}
\epsfig{file=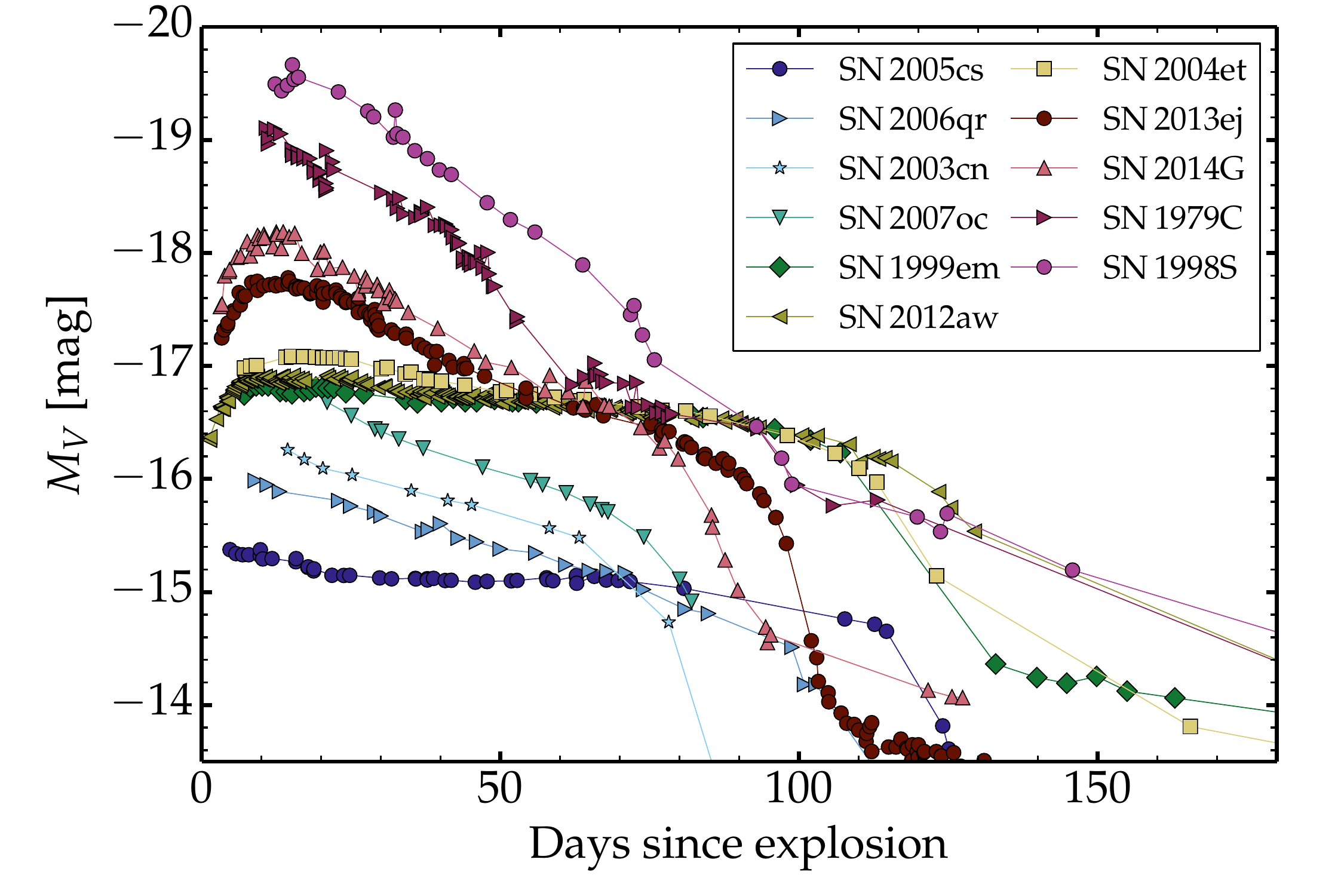, width=9cm}
\caption{Sample of observed $V$-band light curves, corrected for extinction and reddening, illustrating the
well known diversity of Type II SNe \citep[e.g.,][]{1994A&A...282..731P,pasto_low_lum_04}. This diversity is representative of that shown in \citet{anderson_2pl}, revealing Type II SNe with a range of brightness, decline rate, and duration in their high-brightness phase. The plotting order progresses from faint to bright events at maximum. [See Section~\ref{sect_phot_obs} for discussion.]
\label{fig_obs_mv}
}
\end{figure}

\section{Observational diversity}
\label{sect_obs}

\subsection{Dataset}

   A few well observed Type II SNe are used for comparisons and illustrations in this study. The photometric and spectroscopic data are taken from the SN catalog \citep{sn_catalog} and from WISEREP \citep{wiserep}. For each object, the SN characteristics (distance, reddening, redshift, explosion epoch) are adopted from the literature (Table~\ref{tab_obs}). The sample includes SN\,2005cs \citep{pastorello_etal_05cs2}, SNe 2006qr, 2003cn, 2007oc \citep{anderson_2pl}, SN\,1999em \citep{leonard_99em}, SN\,2012aw \citep{dallora_12aw_14}, SN\,2004et \citep{sahu_04et_06}, SN\,2013ej \citep{yuan_13ej_16}, SN\,2014G \citep{terreran_14G_16}, SN\,1979C \citep{panagia_79c_80}, and SN\,1998S \citep{fassia_98S_00}.  One deviation from the literature is that we use a reddening $E(B-V)=0.3$\,mag rather than 0.41\,mag for SN\,2004et (which makes this SN more similar to SNe 1999em and 2012aw; see Section~\ref{sect_comp_2p}).

   \begin{table}
\caption{Characteristics of the observed Type II SNe used in this paper, including the inferred time of explosion, the redshift, the distance, the reddening, and the reference from where these quantities and observational data were taken.
\label{tab_obs}
}
\begin{center}
\begin{tabular}{l@{\hspace{3mm}}c@{\hspace{3mm}}c@{\hspace{3mm}}
c@{\hspace{3mm}}c@{\hspace{3mm}}c@{\hspace{3mm}}
}
\hline
                             &   $t_{\rm expl}$    &   $z$     &     $d$       &       $E(B-V)$       &     Ref.   \\
                             &         [MJD]                 &              &    [Mpc]    &           [mag]        &                \\
\hline
SN\,2005cs    &  53547.6  &  0.0016 & 8.9   & 0.04 & a  \\
SN\,2003cn    & 52719.5  &  0.0181 & 78.7  & 0.059 & b  \\
SN\,2006qr    &  54062.8  & 0.0145 &  63.68& 0.124 &  b \\
SN\,2007oc    &   54388.5 &  0.0048 & 18.11 & 0.014 & b  \\
SN\,1999em   &   51474.3 & 0.0024 & 11.5 & 0.1 &  c \\
SN\,2012aw    &  56002.6  & 0.0026 & 9.9 & 0.074 &  d \\
SN\,2004et   &  53270.5  &  0.0009 & 5.5 & 0.3 &  e \\
SN\,2013ej    &  56497.5  & 0.0022 & 10.2  & 0.06 &  f \\
SN\,2014G    &  56669.6  &  0.0045 & 24.5 & 0.21 &  g \\
SN\,1979C    &  43975.0  &  0.0046 & 16.86 & 0.023 &  h \\
SN\,1998S    &  50875.2 &  0.003 & 17.0 & 0.22 &  i \\
\hline
\end{tabular}
\end{center}
\flushleft
{\bf Notes:} The references used are a:
\citet{pastorello_etal_05cs2} and \citet{dessart_05cs_06bp};
b: \citet{anderson_2pl};
c: \citet{leonard_99em} and \citet{DH06};
d: \citet{dallora_12aw_14};
e: \citet{sahu_04et_06} -- we use a lower reddening $E(B-V)$ of 0.3\,mag;
f: \citet{yuan_13ej_16};
g: \citet{terreran_14G_16};
h: \citet{panagia_79c_80};
i: \citet{fassia_98S_00}.
\end{table}

   \subsection{The $V$-band light curves of Type II SNe}
\label{sect_phot_obs}

Figure~\ref{fig_obs_mv} shows a sample of $V$-band light curves covering a range of rise times, brightnesses, declines rates, and photospheric phase durations. The brightest events in the set are type IIn SN\,1998S and Type II-L SN\,1979C. Slightly fainter are SNe 2014G and 2013ej, which exhibit a brightness above $-17$\,mag for about 50 days. In SN\,2014G, this is followed by a short plateau and a transition to the nebular phase at about 90\,d. In SN\,2013ej, the light curve has a similar fading rate from the maximum, until the sudden drop at about 100\,d. One step down in brightness is the group are the slow decliners or genuine plateau-like SNe II-P such as SNe 2004et, 2012aw, and 1999em. These SNe have a longer photospheric phase. Below are less luminous events, first with fast-decliners (SNe 2007oc, 2003cn, and 2006qr), and finally the low-energy low-brightness Type II-P SN\,2005cs. Most low-luminosity Type II SNe are slow decliners \citep{spiro_low_lum_14,lisakov_ll2p_18}.

This small set of events captures the photometric diversity of Type II SNe presented in \citet{anderson_2pl}, although Fig.~\ref{fig_obs_mv} extends to higher maximum brightness (about $-19.6$\,mag) while the larger sample of \citet{anderson_2pl} peaks at $-18.3$\,mag, with most of the events being fainter than the standard Type II-P (slow decliner) SN\,1999em. In fact, SN\,1999em is amongst the brightest Type II SN at 100\,d in the sample of \citet[only four objects are brighter, out of a sample of 116 objects]{anderson_2pl}, which seems paradoxical given that its properties can be well reproduced with a standard RSG explosion model and a standard \nifs\ mass \citep{d13_sn2p} -- polytropic and non-evolutionary RSGs progenitors have also been proposed \citep{utrobin_07_99em,bersten_11_2p} but this seems unnecessary. The duration of SN\,1999em's optically thick phase is also somewhat larger than average -- according to \citet{anderson_2pl} it is 96.0\,d where as the mean is  83.7 d.\footnote{The definition of the optically-thick phase in \citet{anderson_2pl} is distinct from the true  optically-thick phase, which lasts until the ejecta optical depth drops to 1, or 2/3. At that time, the bolometric luminosity follows the decay power, and the change of slope from the fall-off from the plateau to the nebular tail is obvious. However, even during the nebular stage optical depth effects must be taken into account when modeling spectra \citep[see, for example,][]{DH11_2p,jerkstrand_04et_12}.}

In some of these $V$-band light curves (e.g., SNe 2012aw, 2004et, 2013ej, 2014G), the rise time to a broad maximum is captured. The time of maximum is hard to measure in SN\,2012aw since the curve only bends and flattens, at about 10\,d. In SN\,2004et, the time of maximum is around 20\,d. For the brighter events SNe 2013ej and 2014G, the time of maximum is slightly greater than 10\,d. Overall, these values are larger than those of \citet{gonzalez_gaitan_2p_15}, who report a median rise time of 7.5\,d in the rest-frame g$'$-band ($\lambda 4722$) from their sample of 223 events
(but there is considerable dispersion, and some rise times approach 20 days). Their study was based on the Sloan Digital Sky Survey (SDSS) -- II Supernova Survey \citep{2018yCat.2333....0S} and the Supernova Legacy Survey \citep{2010A&A...523A...7G}. The distributions of rise times from the two surveys were somewhat different, and this was attributed to the better cadence of the SDSS data. \citet{gonzalez_gaitan_2p_15} also concluded that the radii of the SN progenitors are, on average, smaller than those of known RSGs.

The rise times for a  sample of 20 core-collapse SNe with both well-constrained explosion times and light curves have been by provided by \cite{2015A&A...582A...3G} who find mean rise times of $7.0 \pm 0.3$\,d for II-P SNe  and $13.3 \pm 0.6\,$d for Type II-L SNe. However, they note that the rise time of the Type II-L SNe might be biased upward by the most luminous events which tend to have the most well-defined explosion times.  The study also found that larger progenitor radii and higher explosion energies lead to a larger peak brightness at optical wavelengths.

These $V$-band rise times are however shorter than expected for the explosion of a RSG star. \citet{DH11_2p} presented the first non-LTE time-dependent radiative-transfer modeling of RSG star explosions, allowing for the detailed influence of lines and in particular line blanketing. They reported that for a standard RSG star progenitor of 15\,\msun\ initially, the resulting Type II SN light curve exhibits a rough plateau morphology but with a long rise time of $\sim$\,50\,d in the $V$-band. This rise time is associated with the delay until the onset of hydrogen recombination since this signals the time when the photospheric temperature is around $5000-7000$\,K and the spectral energy distribution peaks in the $V$-band. The only way for such a model to produce a shorter rise time is to produce a more rapid recombination, in analogy to what is seen in events like SN\,1987A (whose photosphere recombines after just a few days). In other words, this rise time is controlled by a color shift from the UV to the optical. Using a more compact RSG progenitor (reduced from 810 to 500\,\rsun), the type II-P SN light curve of model m15mlt3 peaks earlier to a flat maximum, at around $20-30$\,d \citep{d13_sn2p}. Importantly, there is no longer a color offset between the observations of SN\,1999em and the model m15mlt3.

The Type II SN rise time of 7.5\,d reported by \citet{gonzalez_gaitan_2p_15} is incompatible with the modeling results discussed above. Such a short rise time cannot result from the spectral energy distribution shift to optical bands as the ejecta cools and recombines because RSG stars are too big to allow for this. An alternative is that the short rise-time is associated with a bolometric boost resulting from interaction with confined CSM at the surface of the RSG progenitor. The early-time observations of SN\,2013fs provide empirical evidence for this \citep{yaron_13fs_17}. It now seems that such a CSM is a fundamental feature of RSG stars and may impact, at various levels, all Type II SNe \citep{yaron_13fs_17,morozova_2l_2p_17,d18_13fs,moriya_13fs_17,forster_csm_18}. However, since the majority of events in \citet{anderson_2pl} are fainter than SN\,1999em, the interaction with CSM (which produces a luminosity boost) cannot be the only driver for Type II SN diversity. Neither can it explain the scatter in photospheric-phase duration.

\begin{figure*}[ht]
\begin{center}
\epsfig{file=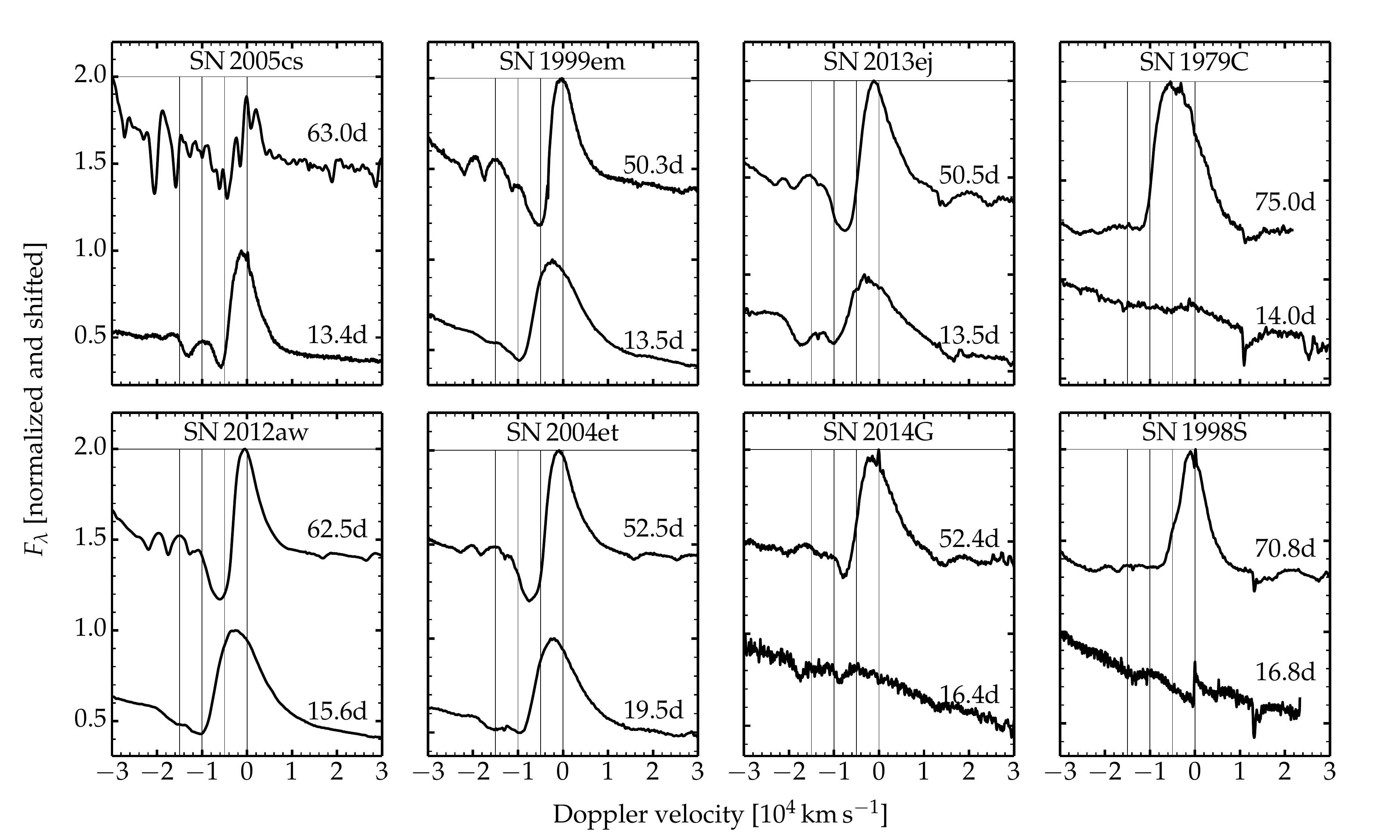, width=15cm}
\end{center}
\caption{Comparison of spectra in the H$\alpha$ region at about 15 and 60\,d after the inferred time of explosion for a set of observed Type II SNe with a range of $V$-band decline rates during the photospheric phase. The spectra have been normalized so that the peak value (in the spectral window shown) is unity, with an additional offset of unity for the upper spectrum. The left-most two columns correspond to standard Type II SNe with a slow decline rate. The third column shows SNe with a much larger brightness at early times and followed by a fast decline. The rightmost column correspond to SNe with a huge early-time brightness and a fast decline rate at all times. The photometric differences between these families of events have a clear spectroscopic counterpart [See Section~\ref{sect_spec_obs} for discussion.]
\label{fig_obs_2epochs}
}
\end{figure*}

   \subsection{The observed properties of H$\alpha$ profiles in Type II SNe}
\label{sect_spec_obs}

   Figure~\ref{fig_obs_2epochs} shows how the luminous fast decliners SNe 2013ej, 2014G, 1979C and 1998S (in order of increasing early-time brightness) show drastically different spectral properties in the H$\alpha$ region both at 15 and 60\,d after explosion compared to the more standard Type II-P (i.e., slow decliners) SNe 2005cs, 2012aw, 1999em, and 2004et.

At both epochs,  standard Type II-P SNe show a well-developed P-Cygni profile. At 15\,d (hence close to the time of $V$-band maximum), the H$\alpha$ profile has a broad absorption and a strong emission (the two leftmost columns). Both absorption and emission are Doppler broadened. The absorption is weaker and broader so it is difficult to assess the maximum velocity in the line. Si\two\,6355\,\AA\ contributes to the absorption and extends it. The maximum velocity attributed to H$\alpha$ seems however quite large, and probably in the range $10000-15000$\,\kms\ for SNe 1999em, 2012aw, and 2004et (it is less than 10000\,\kms\ in SN\,2005cs, whose slower expansion rate makes the contribution of the Si\two\ line clearly visible).  At 50\,d (i.e., during the recombination phase), the H$\alpha$ line profile is stronger and still very broad, but the absorption does not extend beyond about 10000\,\kms. This implies that these SNe II-P outer ejecta are sufficiently dense to maintain H$\alpha$ optically thick at 10000\,\kms; it also implies that there is material accelerated to such large velocities (something that a massive CSM may prevent).

   In SN\,2013ej, the H$\alpha$ profile at 13.5\,d shows a stronger absorption from Si\two\ and weaker from H$\alpha$, while the H$\alpha$ line emission strength is reduced. However, at 50.5\,d, it looks similar to the standard SNe II-P previously discussed. In contrast, in SN\,2014G, the H$\alpha$ region at 16.4\,d is essentially featureless. Two blue-shifted absorptions (probably associated with H$\alpha$ and Si\two\,6355\,\AA) without emission are now seen, while the H$\alpha$ profile at 52.4\,d is clearly present, but with a weak absorption relative to standard SNe II-P. This absorption is also less extended in velocity space.

   The right-most column shows the properties for SNe\,1979C and 1998S, which are analogous to those discussed for SN\,2014G but more extreme. The H$\alpha$ region at about 15\,d is almost featureless while at 70\,d the H$\alpha$ profile shows no absorption component, even though the emission line strength at this epoch is similar to the other SN .

\begin{table*}
\begin{center}
\caption{Summary of progenitor and ejecta properties. All models start with the same mass of 15\,\msun\ on the zero age main sequence. All ejecta have a kinetic energy of $1.2 \times$\,10$^{51}$\,erg. $V_m$ is equal to $\sqrt{2 E_{\rm kin} / M_{\rm ej}}$.  $V_{\rm e}$(H) corresponds to the innermost ejecta velocity above which the H mass fraction is greater than 0.3 (this depends both on the progenitor structure and the adopted mixing). (See Section~\ref{sect_model} for discussion.)
\label{tab_presn}}
\begin{tabular}{lccccccccc}
\hline
Model     &  $M_{\rm f}$  &    $R_{\star}$     & $M_{\rm He,c}$  & $M_{\rm H,e}$  & $M_{\rm ej}$ & $V_m$ & $V_{\rm e}$(H) & $M_{^{56}{\rm Ni}_0}$ & $M_{\rm csm}$ \\
          &    [\msun]   &     [\rsun]       &    [\msun]    &   [\msun] & [\msun]  & [\Mms] & [\Mms] & [\msun]       & [\msun]      \\
\hline
 x1p5     &     13.75    &      586.7        &  4.27         & 9.48 &  12.12 &   3.16  &  1.52    &  0.056 &    0.0  \\
 x2p0     &     13.48    &      582.2        &  4.24         & 9.24 &  11.87 &   3.19  &  1.52    &  0.036 &    0.0  \\
 x3p0     &     12.66    &      582.2        &  4.18         & 8.48 &  11.12 &   3.29  &  1.58    &  0.018 &    0.0  \\
 x4p0     &     10.89    &      610.5        &  4.07         & 6.82 &   9.35 &   3.59  &  1.99    &  0.024 &    0.0  \\
 x5p0     &     10.07    &      598.5        &  4.05         & 6.02 &   8.57 &   3.75  &  2.12    &  0.031 &    0.0  \\
 x6p0     &      9.09    &      633.0        &  4.05         & 5.04 &   7.57 &   3.99  &  2.47    &  0.020 &    0.0  \\
 x7p0     &      8.16    &      651.8        &  4.04         & 4.12 &   6.67 &   4.25  &  2.78    &  0.036 &    0.0  \\
 x8p0     &      7.22    &      658.5        &  3.99         & 3.23 &   5.80 &   4.56  &  3.36    &  0.007 &    0.0  \\
 x9p0     &      6.05    &      656.7        &  4.05         & 2.00 &   4.53 &   5.16  &  4.20    &  0.036 &    0.0  \\
 x1e1     &      4.96    &      680.8        &  4.05         & 0.91 &   3.36 &   5.99  &  5.83    &  0.022 &    0.0  \\
\hline
 x1p5     &     13.75    &      586.7        &  4.27         & 9.48 &  12.12 &   3.16  &   1.52    &  0.060 &    0.0     \\
 x1p5ext3 &     13.75    &      586.7        &  4.27         & 9.48 &  12.12 &   3.16  &  1.52     &  0.053 &    0.246   \\
\hline
 x3p0     &     12.66    &      582.2        &  4.18         & 8.48 &  11.12 &   3.29  &   1.58  &  0.018 &    0.0     \\
 x3p0ext1 &     12.66    &      582.2        &  4.18         & 8.48 &  11.12 &   3.29  & 1.74    &  0.018 &    0.022   \\
 x3p0ext2 &     12.66    &      582.2        &  4.18         & 8.48 &  11.12 &   3.29  &  1.77   &  0.018 &    0.049   \\
 x3p0ext3 &     12.66    &      582.2        &  4.18         & 8.48 &  11.12 &   3.29  &  1.75   &  0.018 &    0.213   \\
 x3p0ext4 &     12.66    &      582.2        &  4.18         & 8.48 &  11.12 &   3.29  &  1.84   &  0.018 &    0.496   \\
 x3p0ext5 &     12.66    &      582.2        &  4.18         & 8.48 &  11.12 &   3.29  &  1.83   &  0.018 &    0.937   \\
 x3p0ext6 &     12.66    &      582.2        &  4.18         & 8.48 &  11.12 &   3.29  &  1.76   &  0.018 &    1.973   \\
\hline
\end{tabular}
\end{center}
\end{table*}


  SN\,1998S has been modeled by \citet{D16_2n}, who find that the interaction of a standard RSG explosion with 0.4\,\msun\ of CSM at 10$^{15}$\,cm reproduces both the light curve and the spectral evolution. This evolution includes the presence of narrow lines early on (at that time, the spectrum forms in unshocked ionized slow-moving CSM), followed by a blue featureless spectrum with blueshifted  absorptions (the spectrum forms in the dense shell formed by the swept-up CSM), and finally a more typical SN II spectrum but with weak signs of blanketing and a strong H$\alpha$ emission with no associated absorption.

It thus appears that the spectral properties shown in Fig.~\ref{fig_obs_2epochs} correspond to a continuum of events in which the mass of CSM grows from small to significant (from left to right). This interpretation agrees with the photometric properties discussed in Section~\ref{sect_phot_obs}. While the photometry can help constrain the amount of CSM, the spectral information can help constrain the impact of the CSM on the ejecta dynamics, as we discuss below.

\section{Numerical setup}
\label{sect_model}

   The numerical approach used for this study is similar to our previous works on Type II SNe  with and without CSM  (see, e.g., \citealt{d13_sn2p,d18_13fs}). It involves stellar evolution calculations with \mesa\ \citep{mesa1,mesa2}, radiation hydrodynamics simulations of the explosion with \v1d\ \citep{livne_93,dlw10a,dlw10b}, and non-LTE time-dependent radiative transfer simulations with \cmfgen\ \citep{HD12}.

\begin{figure*}[t]
\epsfig{file=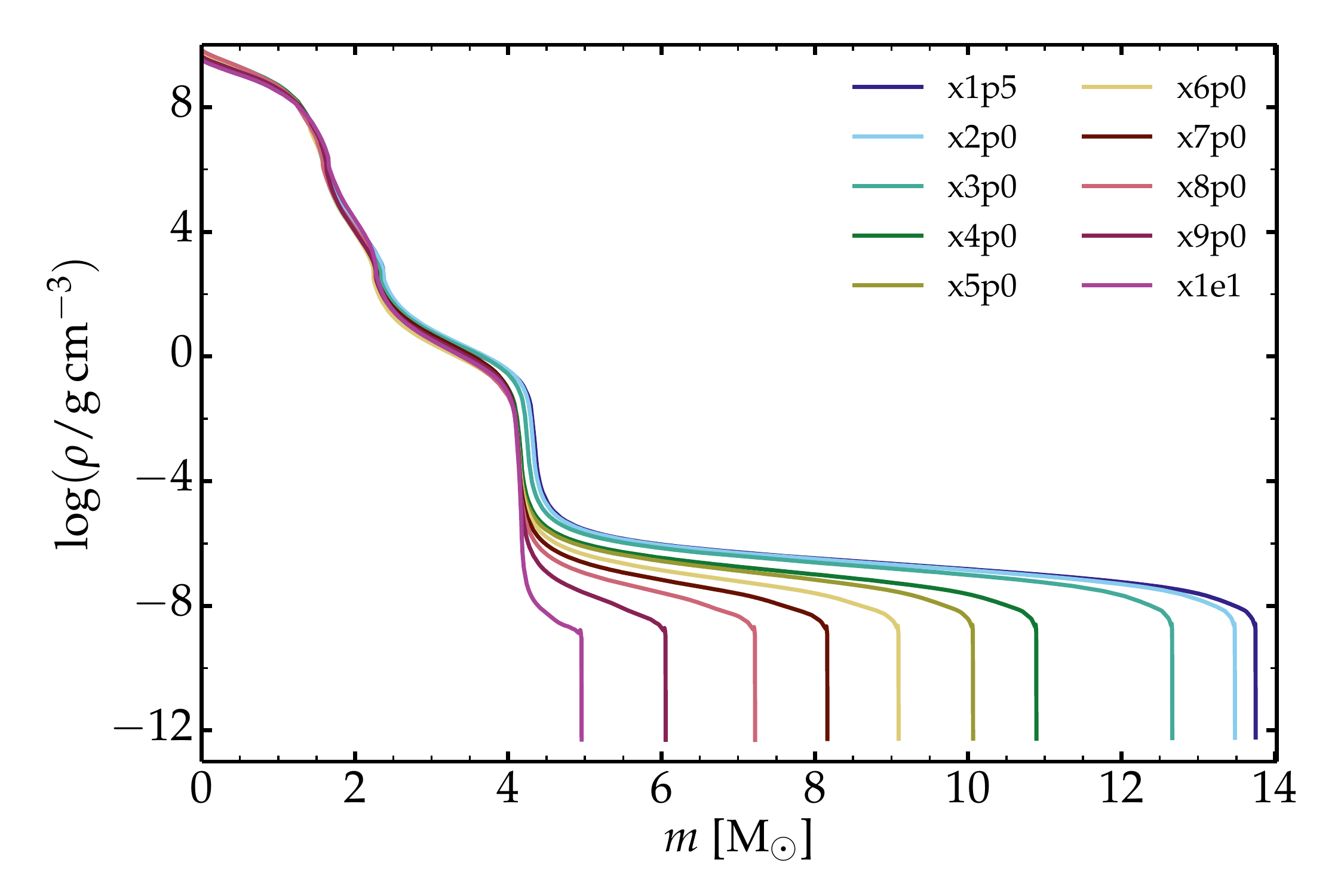,width=9.2cm}
\epsfig{file=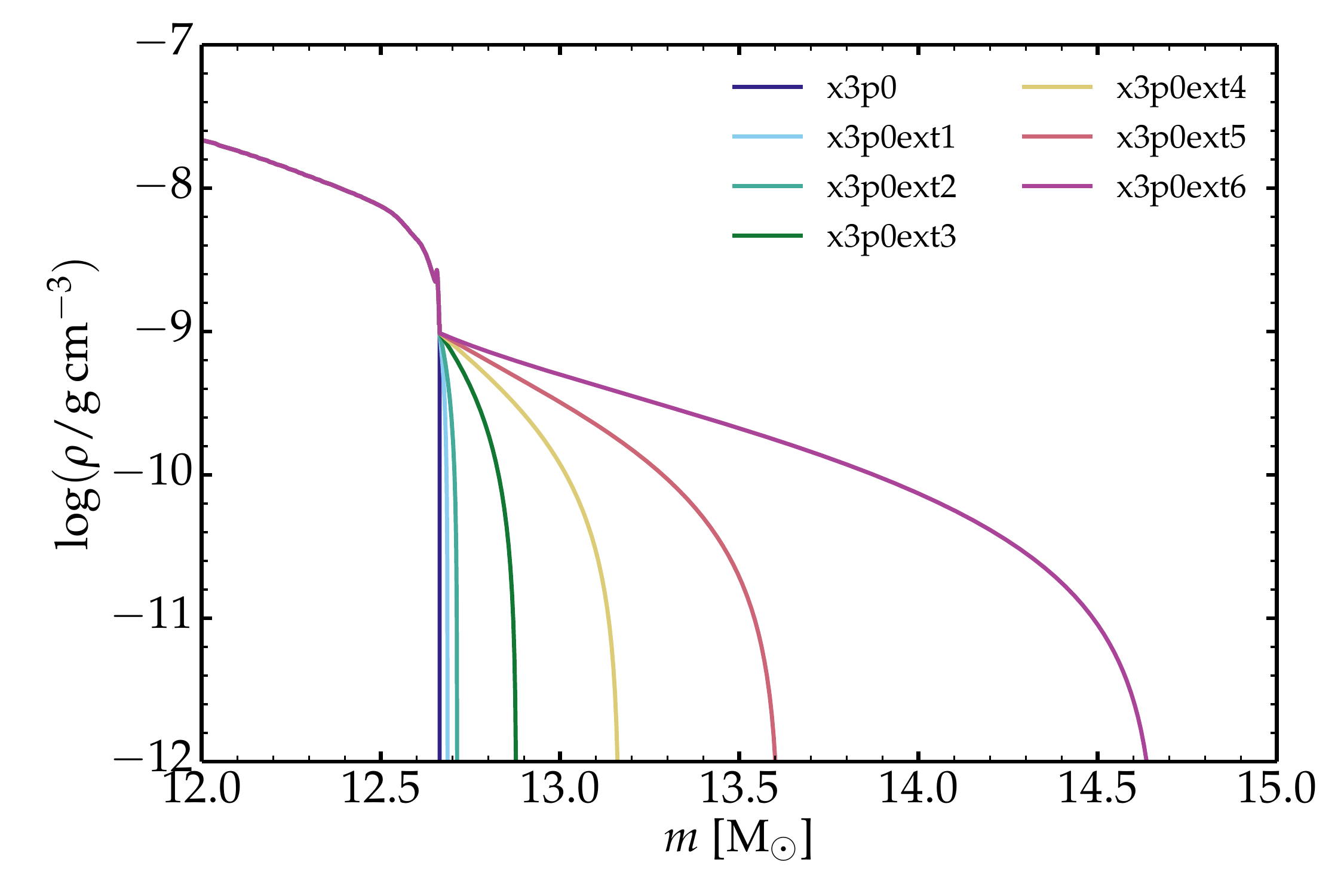,width=9.2cm}
\epsfig{file=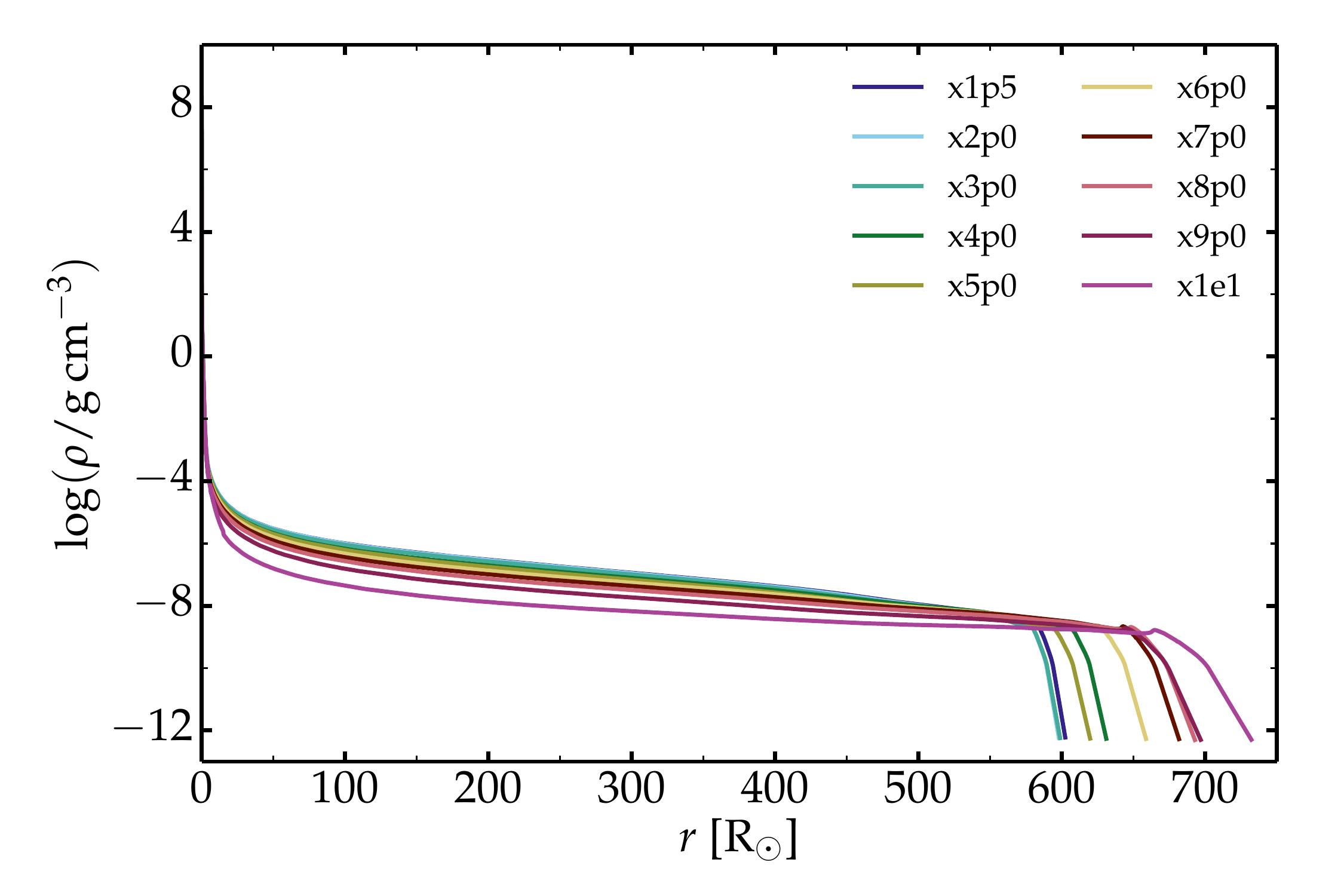,width=9.2cm}
\epsfig{file=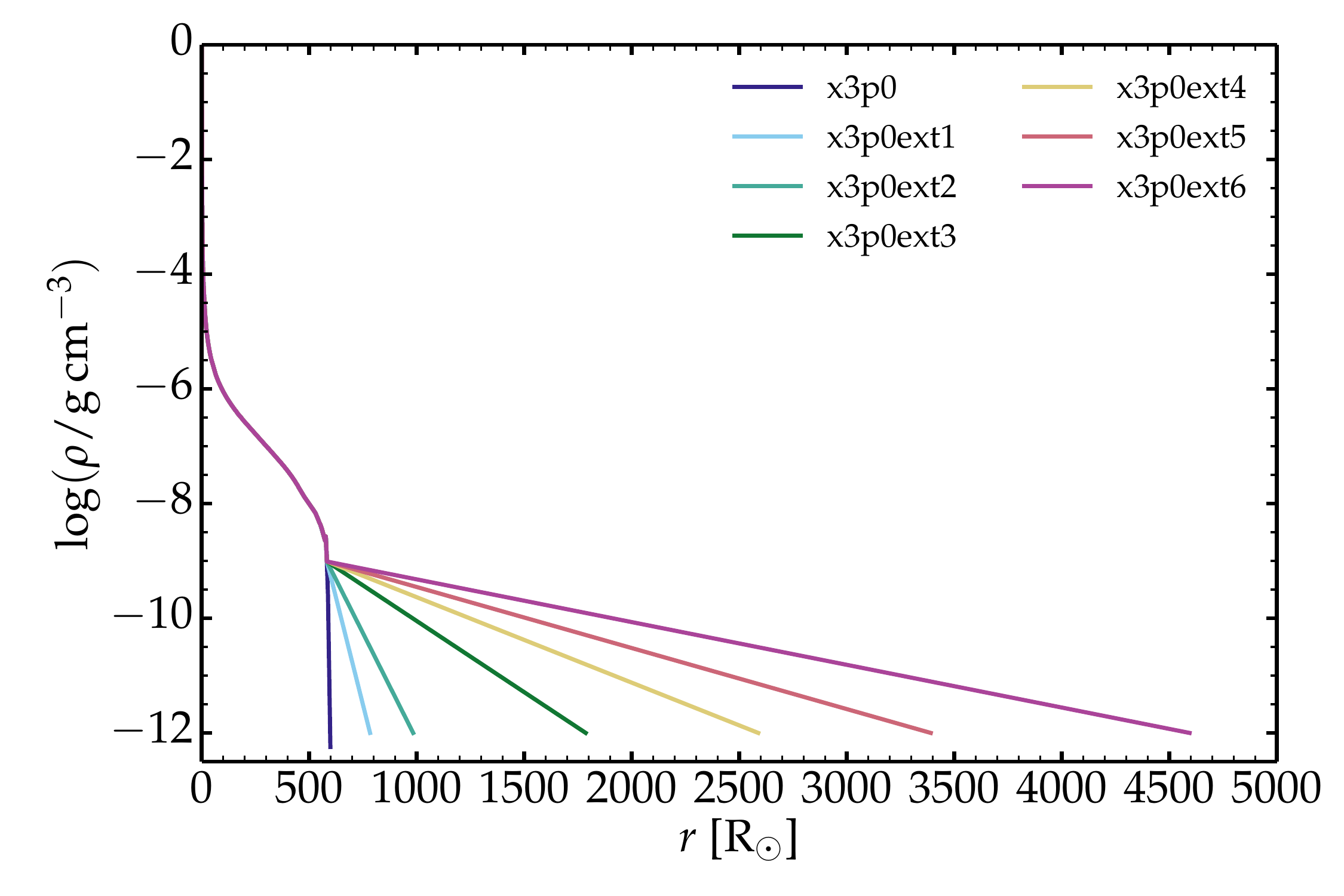,width=9.2cm}
\caption{Left: Mass density versus lagrangian mass (top) and radius (bottom) at the onset of core collapse for the set of 15\,\msun\ simulations produced with \mesa\ using a mixing-length  parameter of 3 and a variety of mass-loss rate scalings during the RSG phase. Right: Same as left, but now for variants of model x3p0 in which some CSM has been added. The corresponding model properties are given in Table~\ref{tab_presn}.
\label{fig_presn_mlt3}
}
\end{figure*}

All simulations presented here are based on a star with solar metallicity and an initial mass of 15\,\msun. The evolution, until the onset of core collapse is performed,  with \mesa\ version 4670 using the default parameters and the modifications specified in \citet{d13_sn2p}. The reason why such an old \mesa\ version is used is because all \mesa\ simulations were performed in Feb. 2013, when this project was started. The mixing length parameter $\alpha_{\rm MLT}$ is changed from 1.6 to 3. This is necessary to produce more compact RSG stars at the time of explosion, since very extended RSG stars produce SNe II-P with a delayed recombination, in conflict with observations \citep{d13_sn2p}.

For the {\it mdot} model set, the mass loss rate is scaled by a factor 1.5 (model x1p5), 2.0 (model x2p0 etc), 3.0, 4.0, 5.0, 6.0, 7.0, 8.0, 9.0, and 10.0 when the model effective temperature drops below 4000\,K (i.e., the scaling therefore applies during the RSG phase only; the higher effective temperature of these RSG star models implies that by default their mass loss rates would be lower than for standard RSG star models computed with a lower mixing length parameter). This yields pre-SN progenitors with similar surface radii (from about 580 to 680\,\rsun), similar He core masses (from 3.99 to 4.27\,\msun), but different H-rich envelope masses (from 0.91 to 9.48\,\msun).

The {\it ext} model set is generated from the pre-SN model x3p0 by adding an atmosphere with a density scale height that varies from 0.05  (model x3p0ext1) to 1.0\,$R_\star$ (model x3p0ext6) and extending out to a radius where the density drops to 10$^{-12}$\,g\,cm$^{-3}$. One additional model (x1p5ext3) was done based on model x1p5 because it was found to be well suited for SN\,2012aw. The corresponding mass of this CSM increases from 0.022 (model x3p0ext1) to 1.973\,\msun\ (model x3p0ext6). Figure~\ref{fig_presn_mlt3} illustrates the density structure for these two models sets, both versus Lagrangian mass and radius. Table~\ref{tab_presn} summarizes the main model properties.  In that table $M_{\rm H,e}$ is the envelope mass defined as the mass above a hydrogen mass fraction of $\sim$0.3, while $M_{\rm He,c}$ is the core mass, and is simply $M_f- M_{\rm H,e}$ where $M_f$ is the progenitor mass at core collapse.

The surface radius of RSG progenitors is a fundamental parameter that impacts both the brightness (see, e.g., \citealt{litvinova_sn2p_85}) and color evolution of Type II SNe \citep{d13_sn2p}. Recently, \citet{mesa4} studied Type II SNe and used a mixing-length parameter of 3 for the progenitor evolution, as here and in \citet{d13_sn2p}. In contrast, \citet{morozova_2l_2p_17} took the massive star models of \citet{WH07}, which are computed with a lower mixing-length parameter. As consequence, their model set of $12-30$\,\msun\ progenitors reach collapse as RSG stars with a surface radius between 640 (12\,\msun) and 1550\,\rsun\ (30\,\msun). Their 15\,\msun\ model has a surface radius about 30\% larger than those in our {\it mdot} set. Such large radii are in tension with the color evolution of SNe II-P but may be possible in events where the SN radiation stems predominantly from interaction (and thus do not look like SNe II-P).

In addition to issues with the radii of SN progenitors, there are significant issues with the structure of RSG photospheres, mass-loss rates, and the immediate CS environment. Betelgeuse is the nearest RSG and has been the subject of numerous studies. Its atmosphere is extended over a few stellar radii, shows a non-monotonic temperature structure, inhomogeneities, large convection cells, and strong asymmetries \citep[e.g.,][]{2017A&A...602L..10O}.  
Betelgeuse also has a shell of atomic hydrogen at a distance of 0.24\,pc from the star \citep{2012MNRAS.422.3433L}. From the properties of this shell it is inferred that Betelgeuse has been losing mass at a rate of $1.2 \times 10^{-6}$\,\msunyr\ for about $8 \times 10^4$\,yr \citep{2012MNRAS.422.3433L}. \cite{mackey_rsg_shell_14} argue that
the shell is maintained by pressure from the photoionized wind which is ionized by external sources. A recent review of high spatial resolution observations of RSGs is by \cite{2017IAUS..329...97O}. 

It is also unclear whether the CSM when the RSG explodes is similar to that found earlier in its evolution.
Changes in global properties of the star (over the final 10,000 years) will potentially lead to large changes in mass loss and pulsation characteristics, thus affecting the photosphere and CSM \citep{heger_rsg_97}. \cite{heger_rsg_97} also postulate that a superwind may occur prior to collapse.  However archival studies of several SN progenitor sites  by \cite{2018MNRAS.480.1696J} indicate that the magnitudes for the RSG progenitors of 4 SNe were stable to within 10\%.

Each model in the {\it mdot} and {\it ext} sets are exploded by means of a piston, placed at a Lagrangian mass of 1.6\,\msun, to deliver an asymptotic ejecta kinetic energy of $1.2 \times 10^{51}$\,erg. Because the models have a slightly different core structure, the explosion produces different \nifs\ masses, from 0.007 up to 0.056\,\msun. When these models were computed in 2013, no attempt was made to correct for this. Because the focus of the study in on the diversity of Type II SN light curves during the high brightness phase, the \nifs\ mass was left as it was and the resulting ejecta models have, consequently, a different brightness at the end of the photospheric phase and beyond.\footnote{Scaling the \nifs\ abundance to match a desired value would be inconsistent since all abundances should be scaled. This means that these models should be recomputed with a different piston location or piston trajectory until they produce the same explosive yields. This is beyond the scope of this study.}

 At $10-15$\,d after the piston trigger, the \v1d\ simulations are remapped into \cmfgen. Homology is enforced, the SN age is set to $R/V$, which causes a small adjustment to the \nifs\ mass. The models in the {\it mdot} ({\it ext}) set were exploded with \v1d\ in 2013 (2018) and a different approach for mixing was used in each set. For the {\it mdot} set, a boxcar algorithm is used to mix the composition. The resulting mixing is weak. In the {\it ext} set, we make the material within the 5\,\msun\ both homogenous and of constant density, and then operate a gaussian smoothing on both the density and the composition.  As a consequence, the H mass fraction in the innermost ejecta varies from $\sim$0.1 in model x1p5 to almost 0 in model x1e1, but it is 0.3 in all the {\it  ext} models. One reason for doing this is that physically, the reverse shock should lead to a smearing of the H/He interface in the ejecta and a strong mixing of the He core material with the base of the H-rich envelope \citep{mesa4,utrobin_sn2p_17}. A density profile with strong gradients can lead to problems with mass conservation across a time sequence computed with \cmfgen; as noted above, such strong gradients are an artifact of 1-D).

 The \cmfgen\ models are evolved until 200 to 300\,d after explosion. The model atoms are similar to those of \citet{d13_sn2p}. The $\gamma$-ray energy deposition is computed using a gray pure-absorption radiative transfer solver and a depth-dependent opacity of $0.06 Y_{\rm e}$\,cm$^2$\,g$^{-1}$ (where $Y_{\rm e}$ is the electron fraction). Non-thermal processes are treated as in \citet{li_etal_12_nonte}. In \cmfgen, the ejecta is assumed to expand freely in a vacuum. There is thus no ongoing interaction considered, for example, with the progenitor RSG wind material. A comparison of the \cmfgen\ light curves with those of \v1d\ (Fig.~\ref{fig_lbol_v1d}) for the x3p0 model set, together with a discussion on the causes of the (relatively small) differences is provided in Appendix~\ref{sect_cmf_v1d_lc}. Figure~\ref{fig_lbol_v1d}
 also illustrates the influence of the different CSM structures on the light curve for the x3p0 model set.
We illustrate the temporal evolution  of the temperature for two of the models (x3p0 and x3poext4) in Figure~\ref{fig_temp_evol} in Appendix~\ref{sect_temp_evol}.

By starting at $10-15$\,d after explosion, the \cmfgen\ simulations miss the earlier evolution. This is not a major problem. Studies dedicated to the earlier, dynamical, evolution can be done using a different technique \citep{d18_13fs}. As such, the present \cmfgen\ simulations are complementary. Observations at the earliest times are also rare. Early interaction with CSM, for example, leaves an imprint on the ejecta that is visible for weeks after explosion. So, the current modeling with \cmfgen\ is not strongly affected by this shortcoming. We also note that numerous studies focus exclusively on light curve modeling, and thus are unable to constrain the complexity of the CSM interaction, whose information is contained in the spectra that these studies ignore.

These \cmfgen\ simulations use a slightly improved numerical procedure (in place since 2015) to improve energy conservation in time. In particular, the transfer equations to be solved are now written in the form
 \begin{equation}
  {1 \over cr^4}  {D(r^4 J{_\nu})  \over Dt} + {1 \over r^2} {\partial (r^2 H_\nu)  \over \partial r}
  - {V \over rc} { \partial \nu J_\nu \over \partial \nu } = \eta_\nu - \chi_\nu J_\nu
 \label{eq_zero_mom_alt}
 \end{equation}
rather than
\begin{equation}
  {1 \over cr^3}  {D(r^3 J{_\nu})  \over Dt} + {1 \over r^2} {\partial (r^2 H_\nu)  \over \partial r}
  - {V \nu \over rc} { \partial J_\nu \over \partial \nu } = \eta_\nu - \chi_\nu J_\nu
 \label{eq_zero_mom}
 \end{equation}
 with a similar modification for the flux equation. This modification leads to improved energy conservation
during the photospheric phase. The meaning of the symbols is the same as in \cite{HD12}, with the exception of the velocity, for which we use the symbol $V$ rather than $v$ to avoid confusion with the frequency $\nu$. The procedure used to examine the energy conservation in time is described in Appendix \ref{App_en_con}.

 \section{Results from radiation hydrodynamics simulations}
\label{sect_v1d_res}

\begin{figure*}[ht]
\epsfig{file=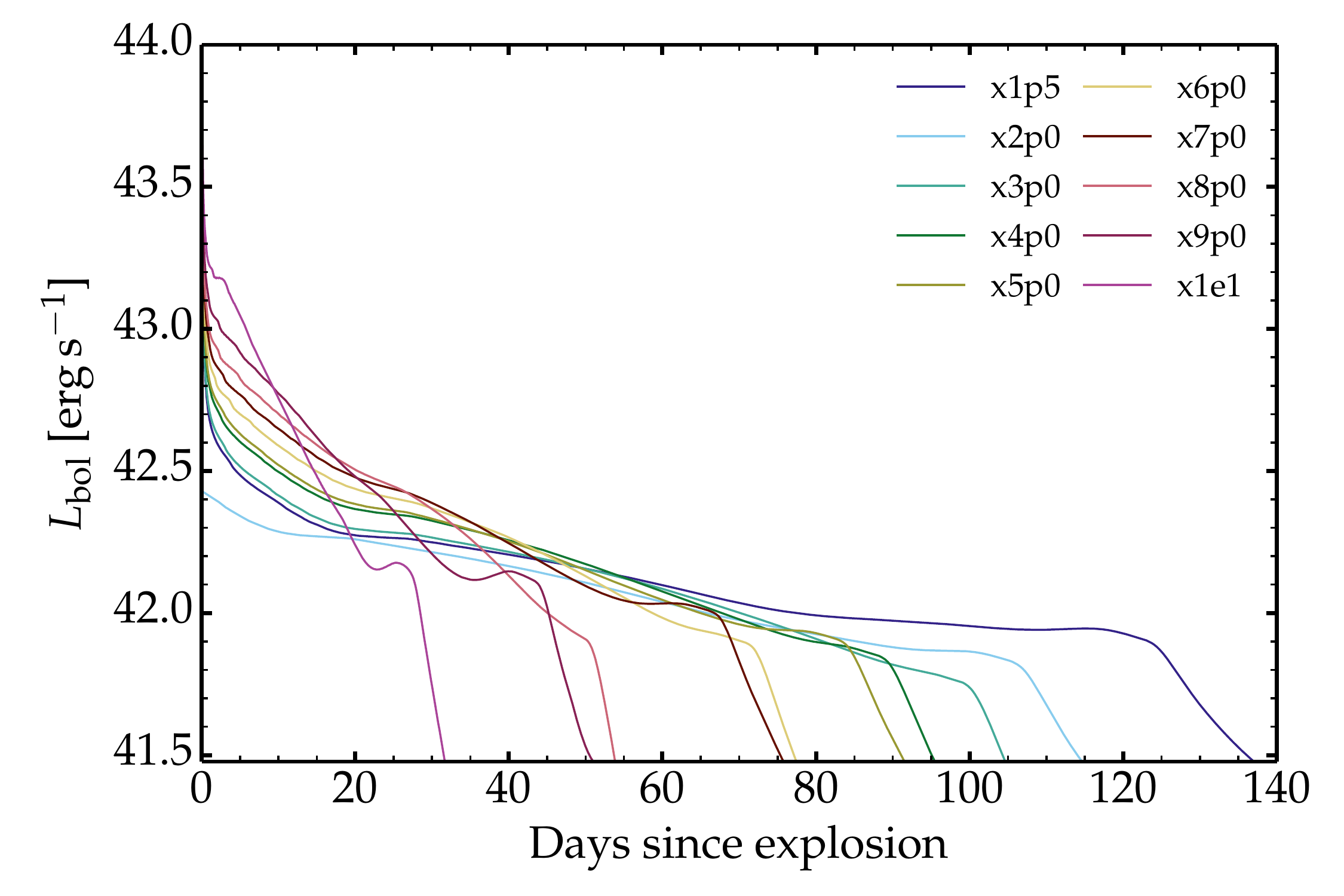  , width=9.2cm}
\epsfig{file=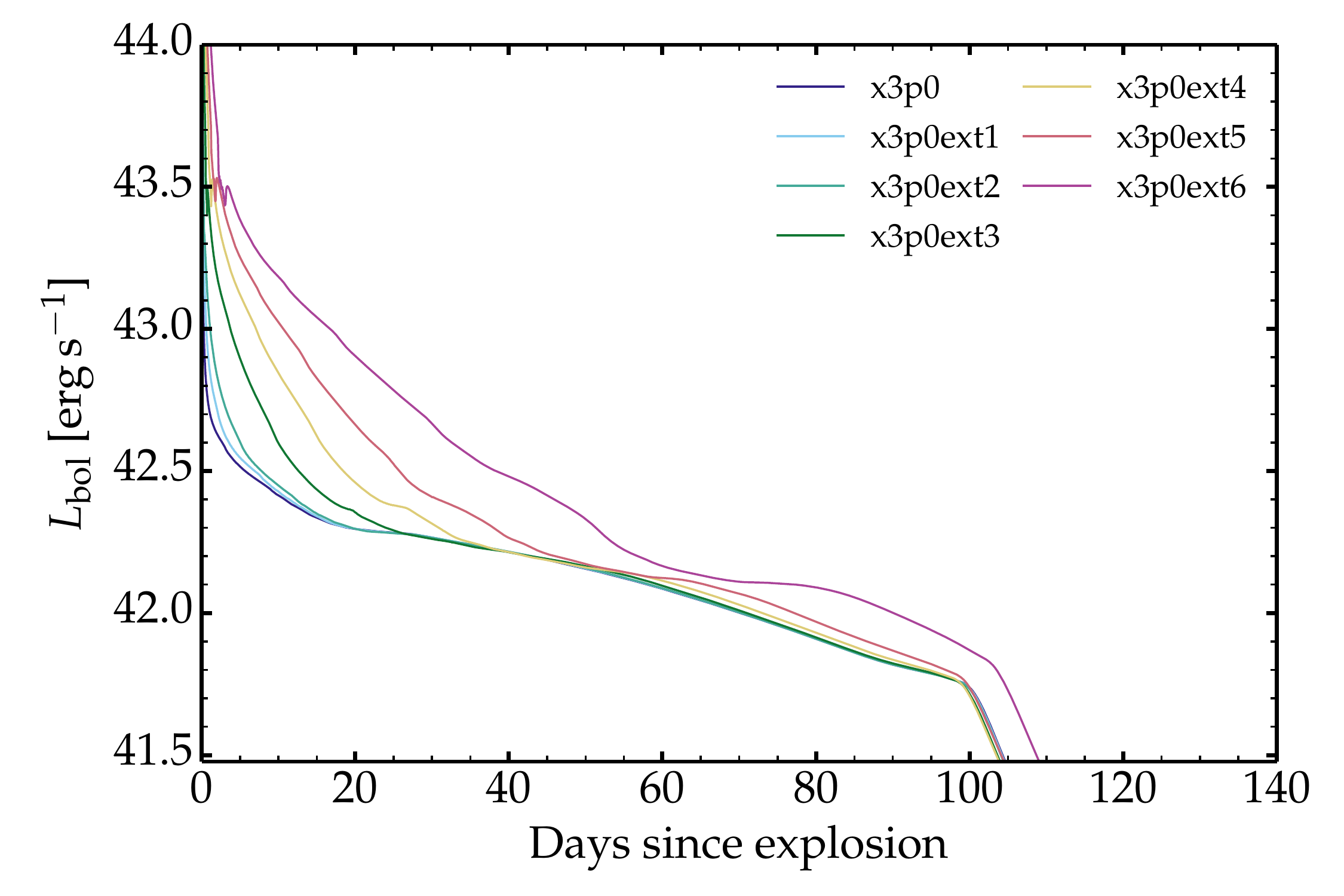, width=9.2cm}
\epsfig{file=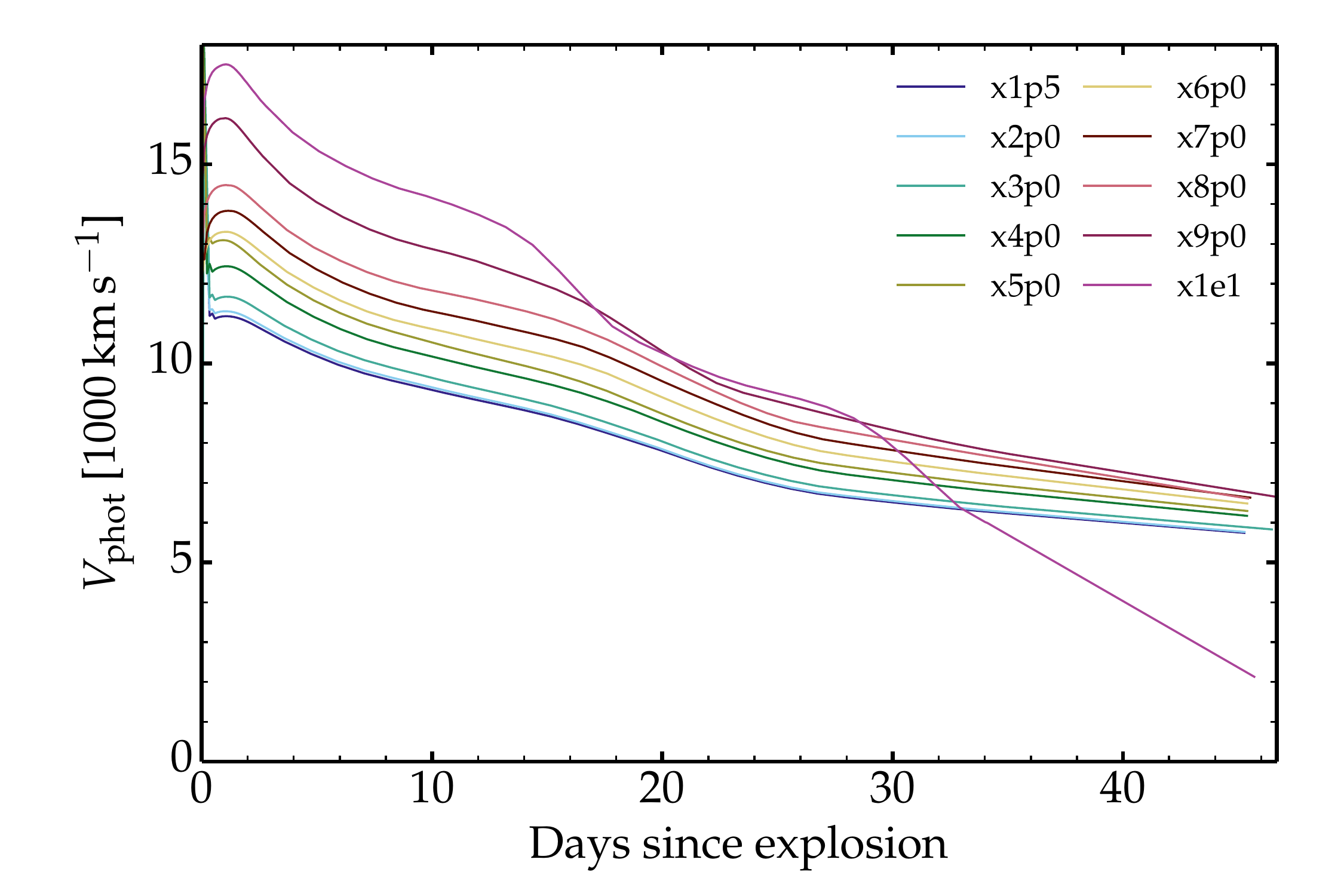, width=9.2cm}
\epsfig{file=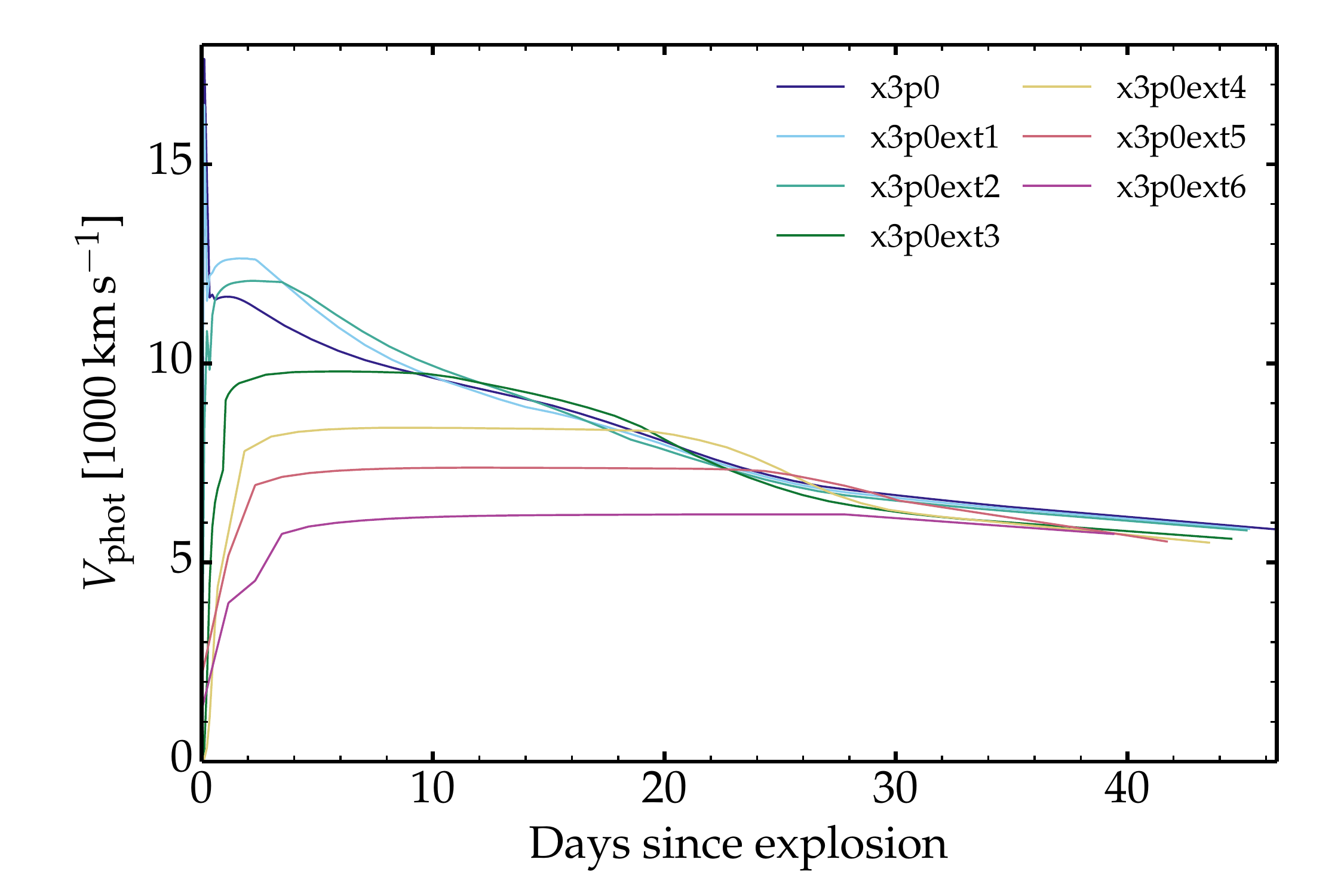, width=9.2cm}
\epsfig{file=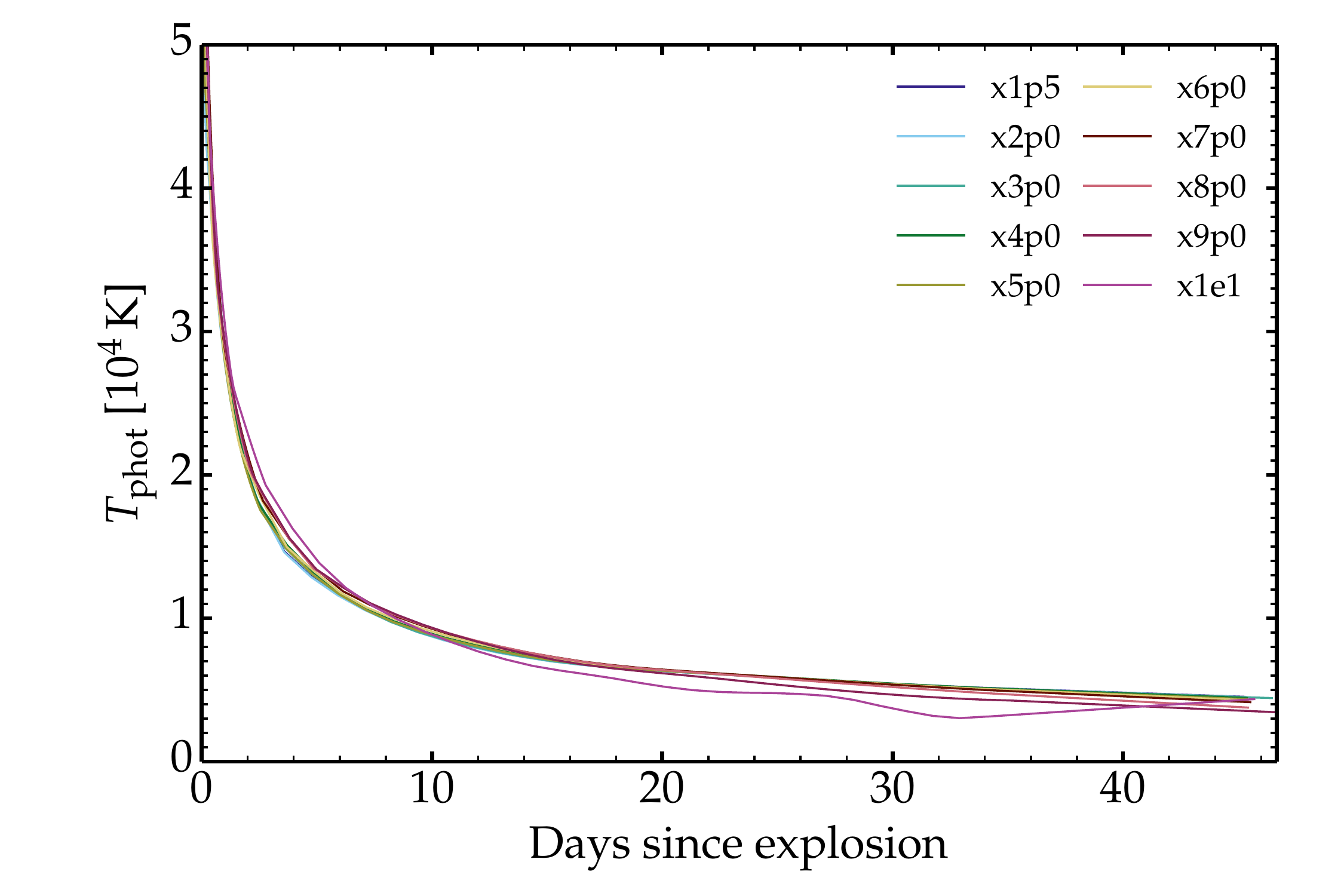 , width=9.2cm}
\epsfig{file=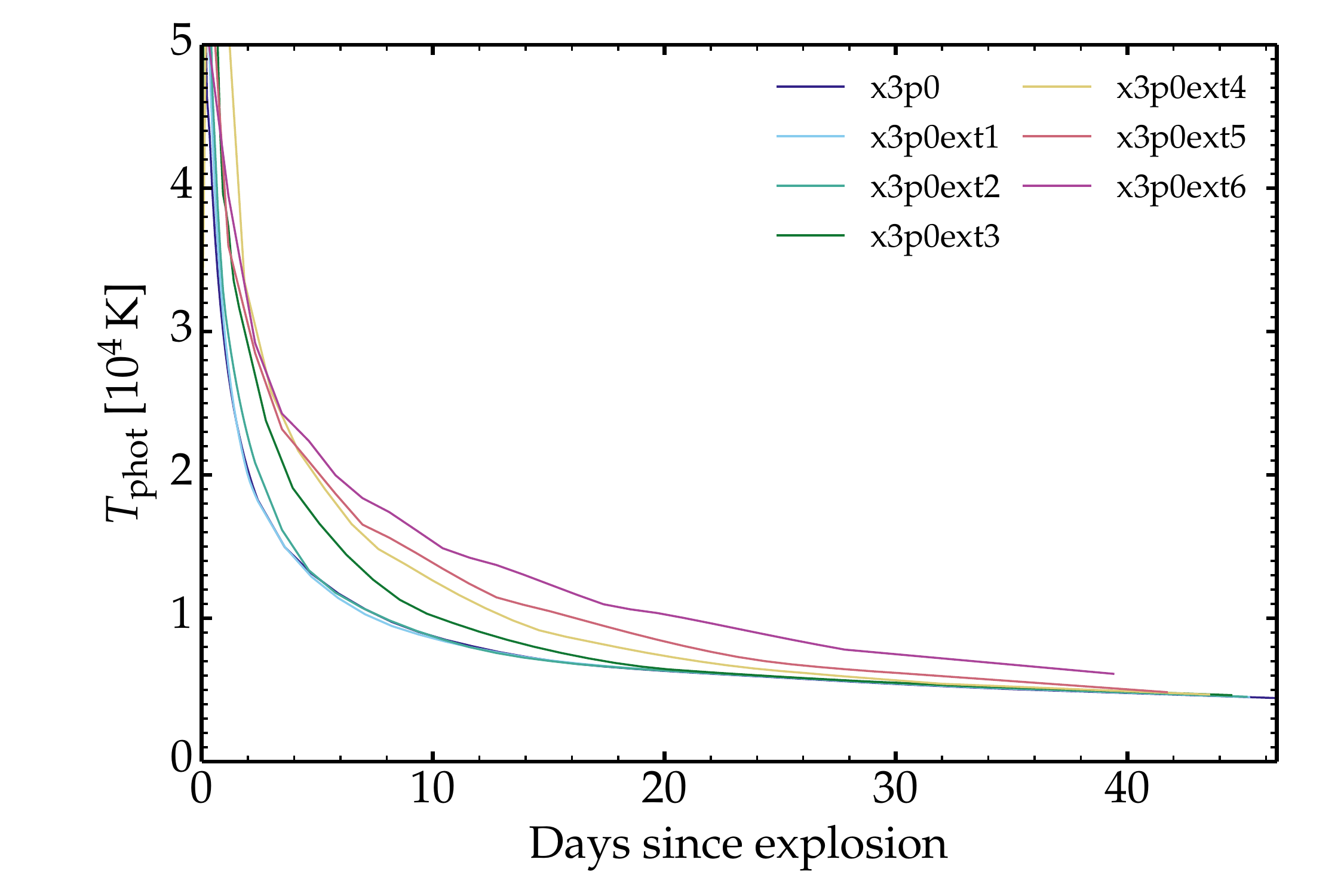  , width=9.2cm}
\caption{Left: Evolution of the bolometric luminosity (top), the photospheric velocity (middle), and the photospheric temperature (bottom) for the explosion models computed by \v1d\ and based on the {\it mdot} model set. Right: Same as left, but now for the {\it ext} model set. [See Section~\ref{sect_v1d_res} for discussion.]
\label{fig_v1d_res}
}
\end{figure*}

We first discuss the results from the radiation hydrodynamics simulations. Figure~\ref{fig_v1d_res} shows the bolometric luminosity (top), the photospheric velocity (middle), and the photospheric temperature (bottom) for both model sets.

If the ejecta mass is reduced for a fixed ejecta kinetic energy ({\it mdot} set; left column), the bolometric light curves progressively shift to a faster decline and an earlier transition to the nebular phase, as expected \citep{blinnikov_bartunov_2l_93,snec,moriya_2l_16}. The larger $E_{\rm kin}/M_{\rm ej}$ implies larger photospheric velocities early on, but since the ejecta optical depth is lower, the photosphere reaches the slower inner ejecta earlier. This causes a faster decline of the photospheric velocity. However, all models show the same evolution for the photospheric temperature evolution. The temperature for H recombination (i.e, around 7000\,K) is reached after about 20\,d.  The principal cause for the variation in the light curves is  the variation in $M_{\rm H,e}$ -- there is a direct effect due to the lower envelope mass and indirect effect arising from more rapid expansion of the envelope.

For a fixed ejecta mass and kinetic energy (corresponding to those of model x3p0), an increasing amount of CSM ({\it ext} set; right column) causes a stronger and longer-lived boost to the luminosity. For a CSM of about 0.2\,\msun\ (model x3p0ext3), the boost is limited to times prior to 20\,d, but for the highest CSM mass of about 2\,\msun\ (model x3p0ext6), nearly the entire photospheric phase is affected. The photospheric phase duration is in general not affected by the presence of CSM, except in model x3p0ext6 in which the reverse shock (caused by the CSM) reached down to the inner ejecta layers, slowing it down and depositing additional internal energy. The boost in luminosity arises here from interaction, converting kinetic energy to radiation energy. The luminosity boost is thus associated with a reduction of the outer ejecta kinetic energy and maximum velocities reached. The evolution of the photospheric velocity is a quasi monotonic decrease from 12000 to 5000\,\kms\ in model x3p0 (no CSM) but is instead a plateau at 5000\,\kms\ for the first 40\,d in model x3p0ext6.

In the {\it ext} models, the reduced maximum velocities is not exactly a braking due to interaction with the CSM. This certainly occurs, but unlike in SNe ejecta interacting at large distances, the interaction takes place here when the shock reaches the stellar surface. At that time, about half the total energy is radiation, and the other half is kinetic. By interacting with CSM at moderate optical depth, a sizable fraction of the shock energy is lost to escaping radiation. If the configuration was adiabatic (i.e., no radiative losses), the kinetic energy would first drop and be converted into radiation energy. Then, this radiation energy would be tapped to accelerate the ejecta again. This second step does not happen fully here because of radiation leakage, which causes the boost to the emergent luminosity.

These results have been discussed in \citet{d18_13fs}, both for the hydrodynamics of the SN shock interacting with a variety of CSM structures, as well as their effect on the SN photometric and spectroscopic properties during the first 15\,d after shock breakout. Hence, additional details that can be found there are not repeated here.

\begin{figure*}
\epsfig{file=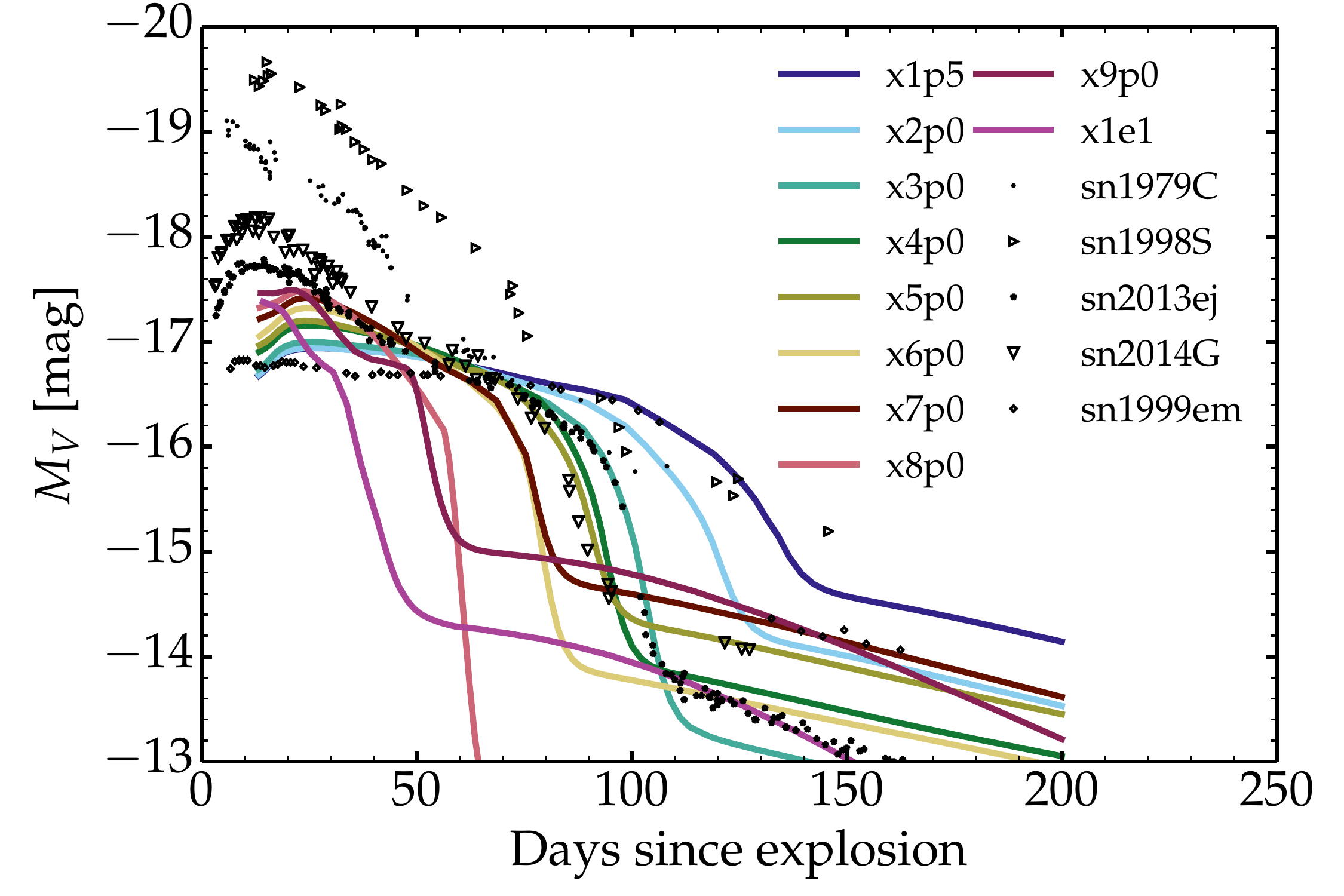, width=9.2cm}
\epsfig{file=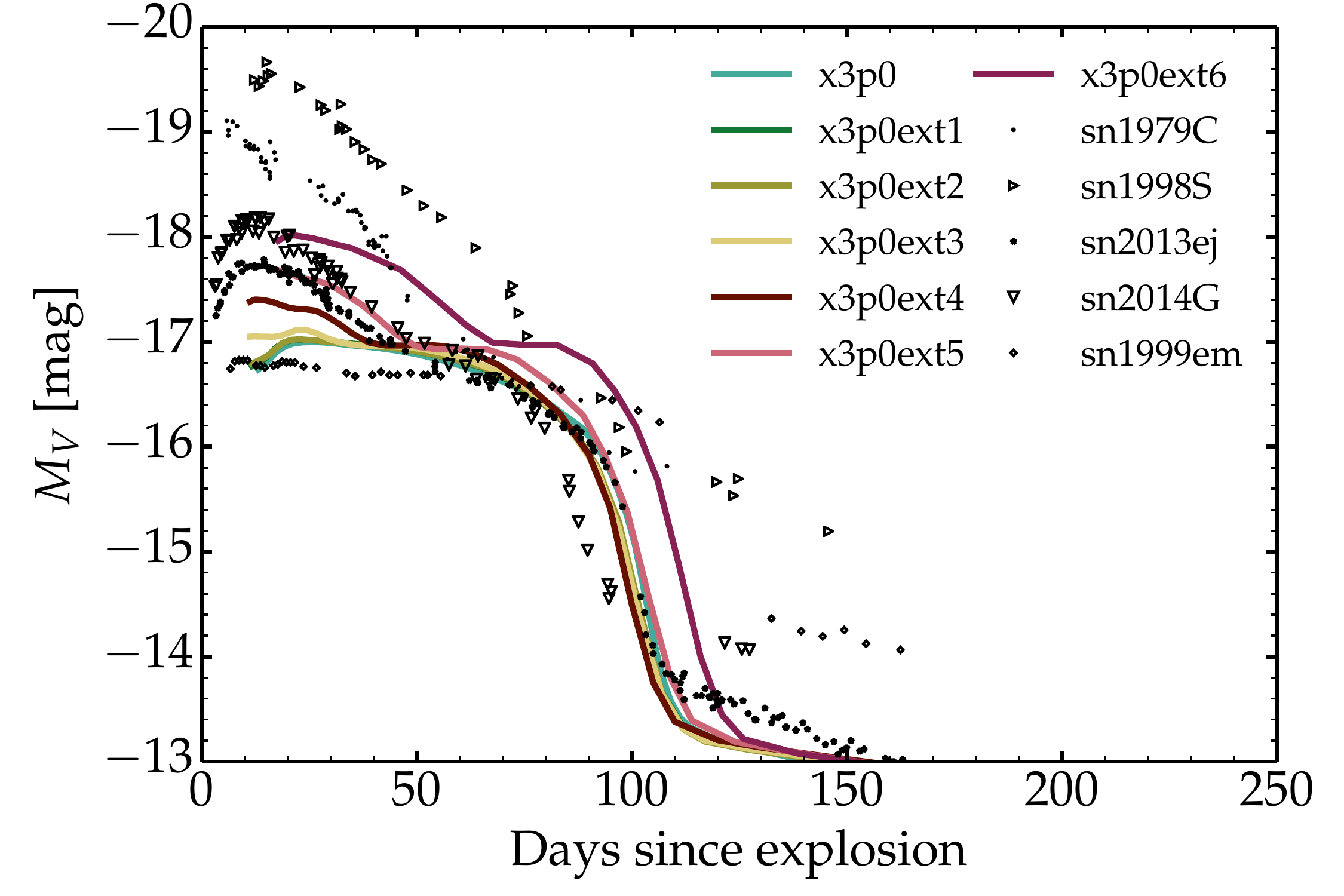, width=9.2cm}
\epsfig{file=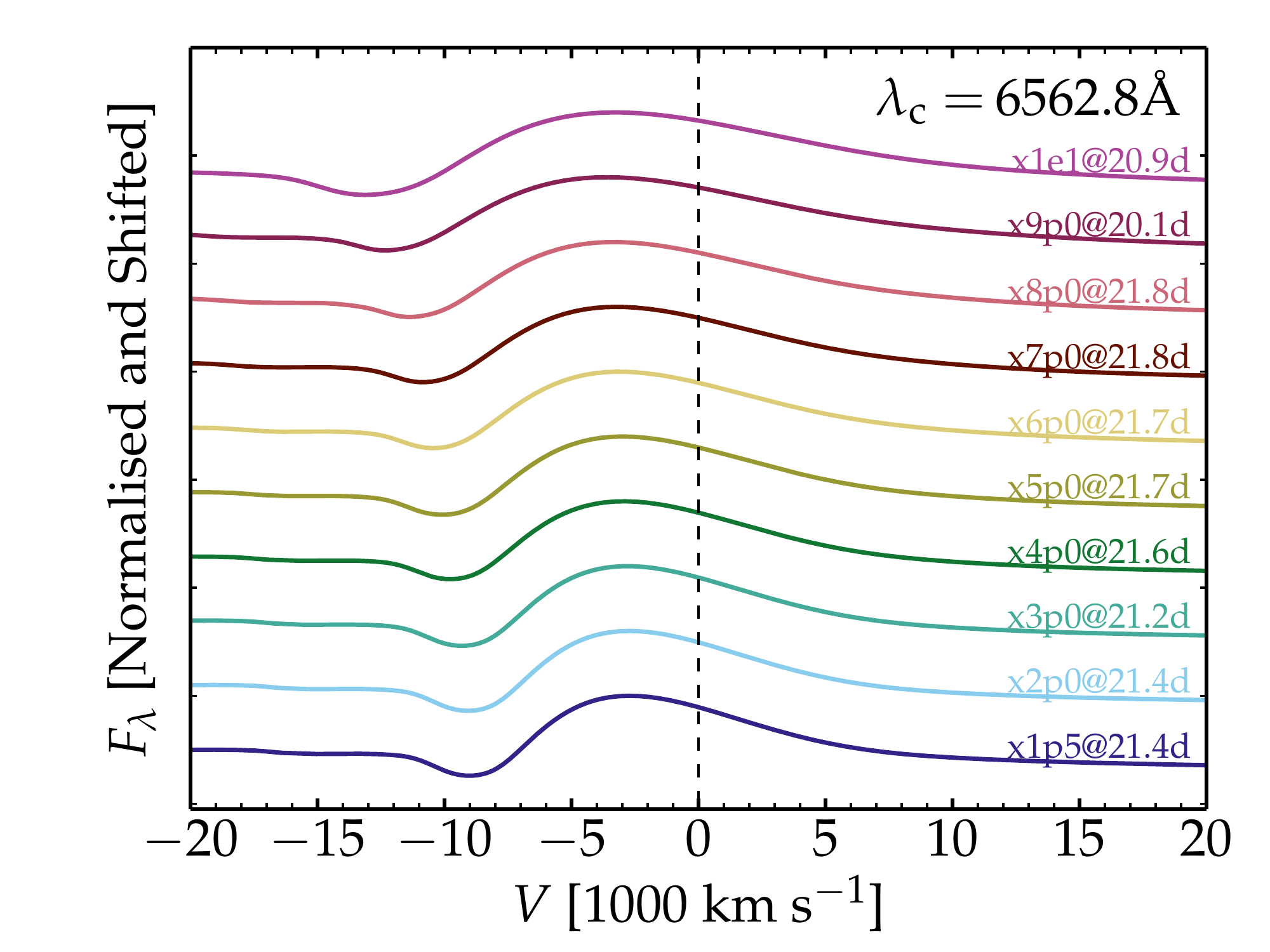, width=9.2cm}
\epsfig{file=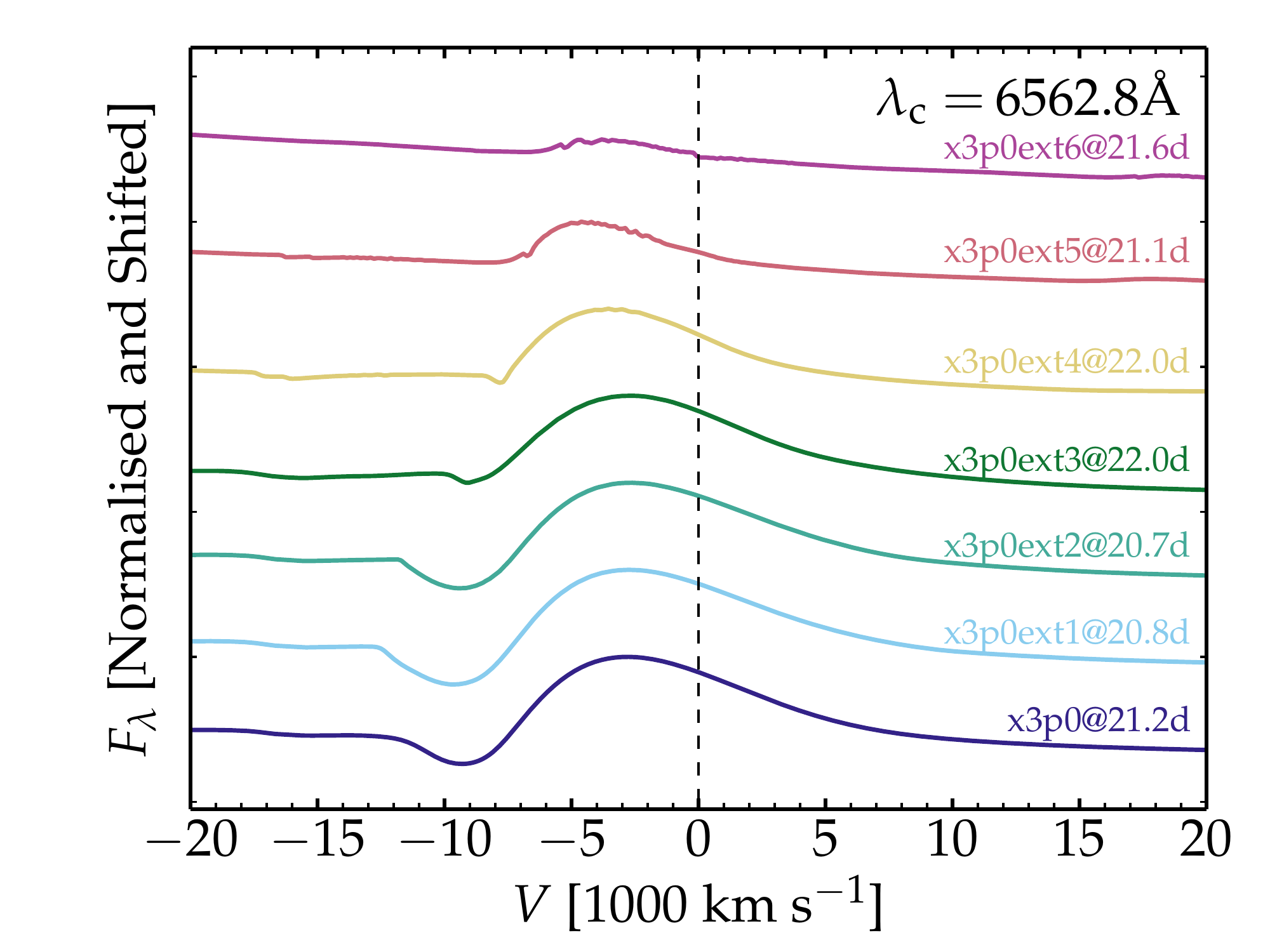, width=9.2cm}
\epsfig{file= 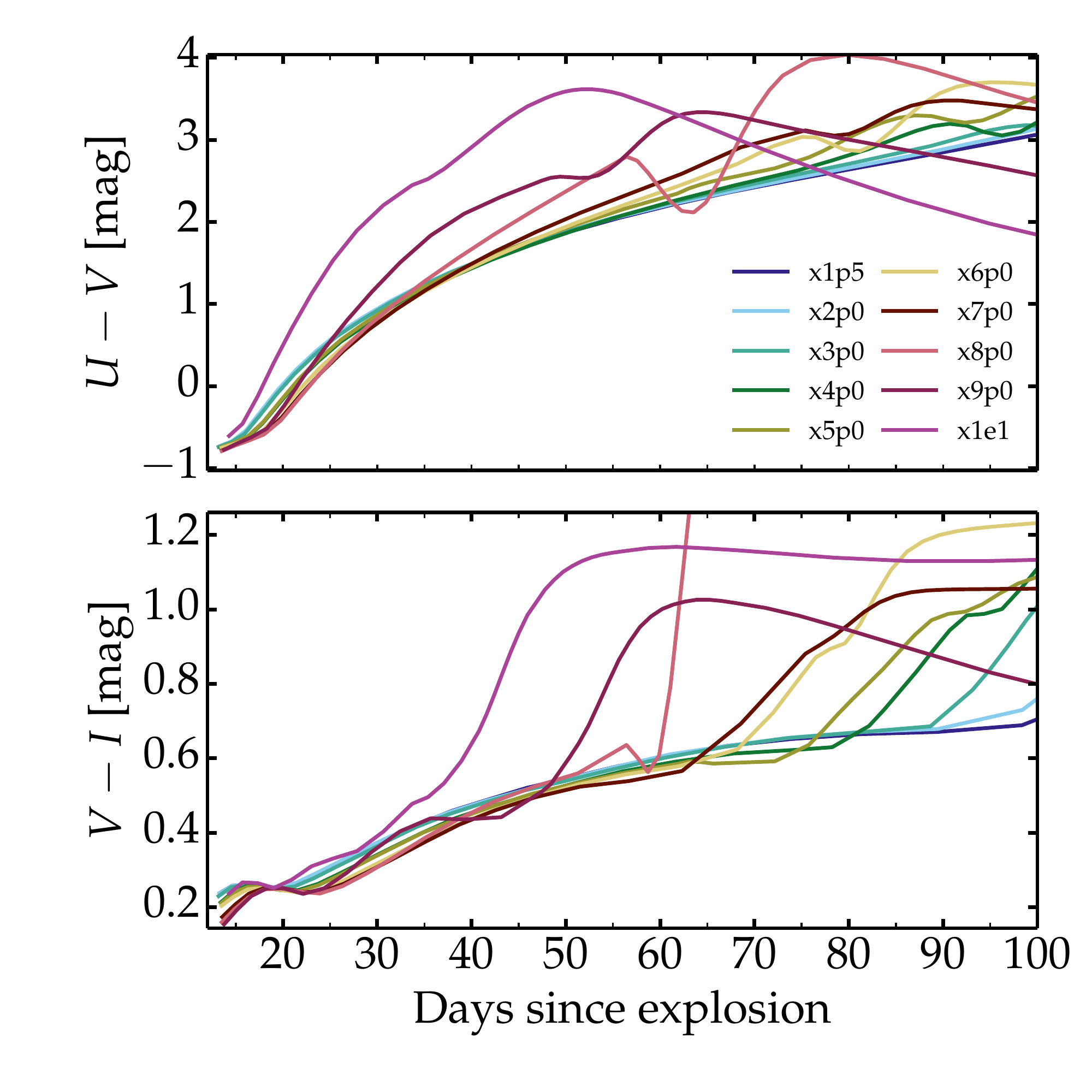, width=9.2cm}
\epsfig{file= 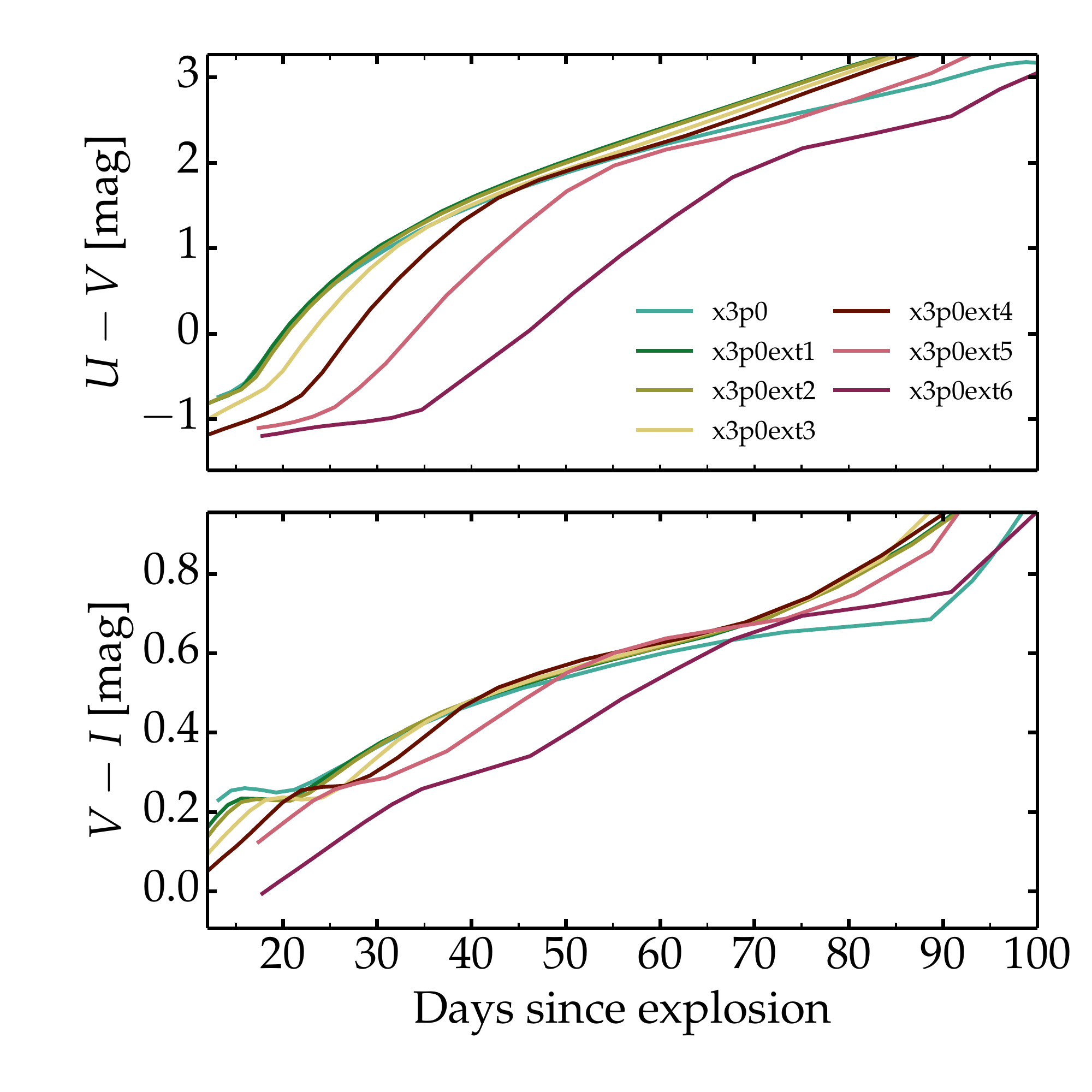 , width=9.2cm}
\caption{Left: Illustration of the absolute $V$-band light curves (top), the spectral region centered on H$\alpha$ at about 20\,d after explosion (middle), and the $U-V$ and $V-I$ color curves (bottom) for the {\it mdot} simulations performed with \cmfgen. Right: Same as left, but now for the {\it ext} model set.  [See Section~\ref{sect_cmfgen_res} for discussion.]
\label{fig_cmfgen_res}
}
\end{figure*}

In the {\it ext} models, the evolution of the photospheric temperature is qualitatively similar between models, but quantitatively, the greater the CSM mass, the greater the photospheric temperature. This arises from the excess energy dissipated in the outer progenitor layers during the interaction. Because this takes place at the largest possible radii in the star, this energy is not strongly degraded by expansion. In contrast to the {\it mdot} models, we expect a color shift of the SN radiation in the {\it ext} models, associated with a delayed recombination.

Overall, both the {\it mdot} and {\it ext} model sets tend to produce more linearly declining bolometric light curves as the H-rich envelope mass is reduced or the CSM mass is enhanced. However, the boost in luminosity is large and sustained only for the {\it ext} case. This is not surprising since interaction gives an efficient means to extract energy from where it is the most abundant (i.e., the ejecta kinetic energy).

In nature, fast decliners may stem from these two scenarios (and perhaps others, not yet identified). But from Fig.~\ref{fig_v1d_res}, the two above scenarios for the production of fast declining Type II SNe can be easily distinguished. If fast decliners primarily arise from a reduction of the progenitor H-rich envelope mass, the maximum ejecta velocities in II-L should be larger, the photospheric phase shorter, and the color similar to Type II-P. If fast decliners instead arise from interaction with CSM, the maximum ejecta velocities of II-L should be smaller, the photospheric velocity evolution should be flatter, the photospheric phase should have (statistically) the same duration as Type II-P, and the color should be bluer for longer compared to Type II-P SNe. Furthermore, this configuration would produce SNe that may appear as Type IIn at early times. The two scenarios may occur simultaneously, so that fast decliners may come from progenitors with a reduced H-rich envelope mass and enshrouded within a massive and confined CSM. For example, the greater mass loss rates in higher mass progenitors may lead to both a greater CSM mass and a reduced H-rich envelope mass. Such a correlation could be expected from stellar evolution theory.

\begin{figure*}
\begin{center}
\epsfig{file=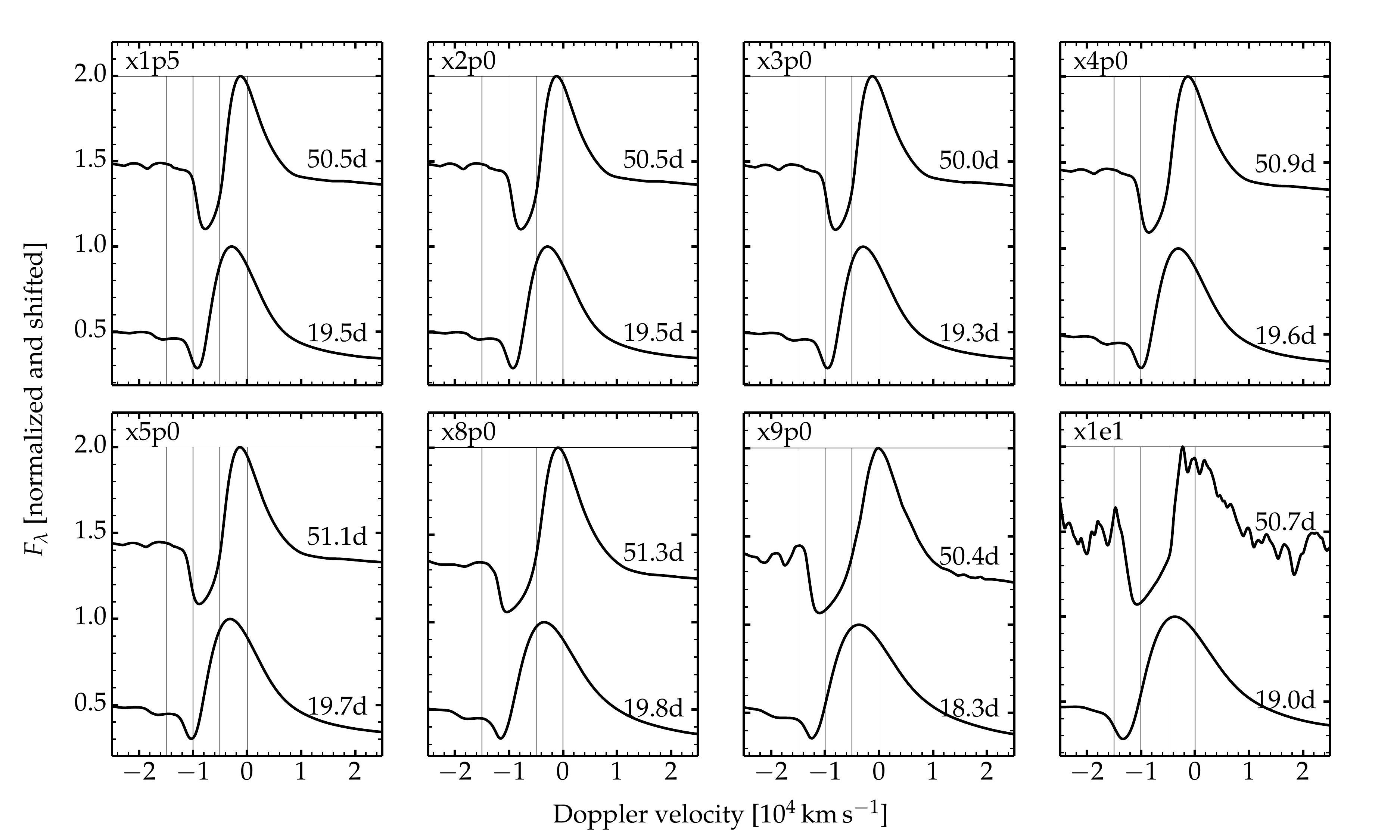, width=14cm}
\end{center}
\caption{Comparison of spectra in the H$\alpha$ region at about 15 and 50\,d after explosion for the model set with decreasing H-rich envelope mass (from x1p5 to x1e1).
\label{fig_mdot_2epochs}
}
\end{figure*}

\begin{figure*}
\begin{center}
\epsfig{file=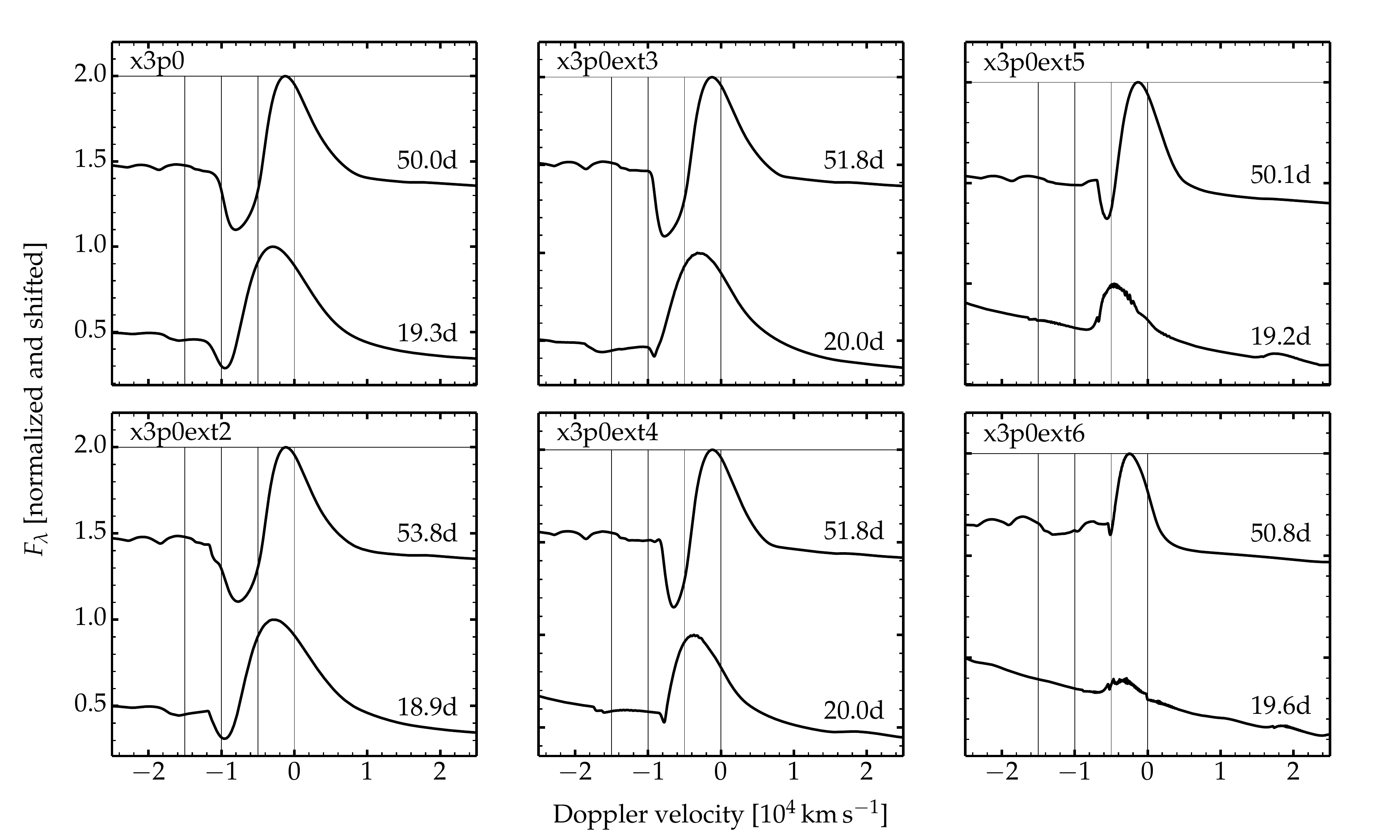, width=14cm}
\end{center}
\caption{Same as Fig.~\ref{fig_mdot_2epochs}, but now for the model set with increasing CSM mass (from x3p0 to x3p0ext6).
\label{fig_csm_2epochs}
}
\end{figure*}

\section{Results from the non-LTE time-dependent radiative transfer simulations}
\label{sect_cmfgen_res}

Figure~\ref{fig_cmfgen_res} shows some of the results from the {\sc cmfgen} simulations based on the {\it mdot} and the {\it ext} simulations undertaken with \v1d\ and \mesa. This figure is a counterpart of Fig.~\ref{fig_v1d_res}, with {\it mdot} models on the left and {\it ext} models on the right, but now showing the absolute $V$-band light curves (which reflect in part the bolometric light curve, modulo the change in $V$-band bolometric correction), the H$\alpha$ profile in velocity space (which reflects the behavior in photospheric velocity), and the color evolution (which reflects the evolution of the photospheric temperature). We also add photometric observations for comparison, including slow (SN\,1999em) and fast decliners (same SNe as shown in Fig~\ref{fig_obs_mv}).

In the {\it mdot} model set, reducing the H-rich envelope mass leads to a $<1$\,mag increase in $V$-band brightness at 15\,d, while it causes a progressive shortening of the high brightness phase. The shorter the photospheric phase, the faster the decline, so that the brightness becomes lower than standard earlier (taking SN\,1999em as representative). Increasing the kinetic energy would raise the luminosity at early times but the decline would be even faster. As seen in the properties of the photospheric velocity (middle left panel of Fig.~\ref{fig_v1d_res}), the models with lower H-rich envelope mass exhibit a broader H$\alpha$ profile at 20\,d after explosion. Reflecting the similar photospheric temperature evolution, the color evolution for the {\it mdot} set is similar during the photospheric phase. The optical colors redden and diverge between models when the ejecta turns optically thin.

In the {\it ext} set, the presence of CSM reduces the rise time in the $V$-band so that in model x3p0ext3, the $V$-band light curve is essentially flat at and beyond 10\,d. Increasing the CSM mass yields an increase in $V$-band brightness that spans between the values for SN\,1999em and SN\,2014G. SNe 1979C and 1998S require either more CSM mass or a different configuration. This could be an interaction at larger distances \citep{D16_2n}, or a sustained interaction with a dense pre-SN wind over very large distances \citep{blinnikov_bartunov_2l_93}. Around 1\,\msun\ of CSM seems necessary to explain the early-time $V$-band brightness of SNe 2013ej and 2014G. This depends on the both the mass of CSM and its spatial distribution. It also depends on the $V$-band bolometric correction, which can be hard to estimate accurately since a significant fraction of the flux falls in the UV at early times.

As expected from  Fig.~\ref{fig_v1d_res}, the lower velocities in the outer ejecta in the {\it ext} models strongly impacts the H$\alpha$ profile at 20\,d (the interaction is over by then). As the CSM mass increases, the H$\alpha$ profile becomes weaker in both absorption and emission. The absorption component vanishes in model x3p0ext5 and x3p0ext6, while only a residual emission subsists in model x3p0ext6. The enhanced CSM mass also causes the optical color to be bluer (the spectrum formation region is hotter for longer).

The primary reason for the weakness of the H$\alpha$ profile at early times is that in these models the photosphere is located in a dense shell formed from the interaction of the ejecta with the CSM. As this shell is moving at near constant velocity and because the density profile is very steep, the P~Cgyni profile is weak or absent. At later times the P~Cgyni absorption is weak or absent primarily because of the absence of high velocity material. For SN\,2006bp we reproduced the weakness of H$\alpha$ using a steep density profile ($\rho \propto r^{-50}$; \citealt{dessart_05cs_06bp}). The same effect is inferred for SN\,1998S, although in this case the interaction of the ejecta with a massive CSM is thought to have occurred at large distances \citep{D16_2n}. Finally, the influence of the CSM on the SN radiation is extensively discussed and explained in \citet{d18_13fs}, to which the reader is referred.

To summarize, a fast decline with the {\it mdot} model set inevitably leads to a short photospheric phase duration. Reducing the H-rich envelope mass is not conducive to producing a strong boost to the early-time brightness. This modest boost in brightness is correlated with the H$\alpha$ line width (i.e., brighter broader) while the color during the photospheric phase is independent of the brightness boost.

With the {\it ext} model set, the boost in $V$-band brightness is larger, compatible with the observations of SN\,2013ej and 2014G, but too small to reach that seen for SN\,1979C and SN\,1998S. The boost in luminosity is anti-correlated with the width of H$\alpha$, which may also appear as pure emission early on (Type IIn features may be seen at earlier times; \citealt{d18_13fs}). The color also correlates with the brightness boost in that the greater the boost the bluer the color and the more delayed is the onset to recombination.

Figures~\ref{fig_mdot_2epochs} and \ref{fig_csm_2epochs} illustrate the differences in the H$\alpha$ line region between $\sim$\,15 and $\sim$\,50\,d after explosion, as shown earlier for a sample of Type II SNe with a range of $V$-band brightnesses and decline rates (Fig~\ref{fig_obs_2epochs}). The {\it mdot} models show the same evolution as that seen in standard slow decliners: the H$\alpha$ profile is a well developed P-Cygni profile at both epochs and its width narrows in time (the extent of the absorption part is also affected by Si\two\,6355\,\AA\ at 15\,d). In the {\it ext} model set, the contrast between models is strong. As the CSM mass is enhanced from model x3p0 to x3pext6, we see that H$\alpha$ progressively shows a weaker absorption and emission at 15\,d. At 50\,d, the line shows a weaker absorption, but the emission remains strong and broad.

The trend seen from left to right in Fig.~\ref{fig_csm_2epochs} is analogous to that displayed in Fig.~\ref{fig_obs_2epochs} (the CSM mass increases as we progress from slow to (luminous) fast decliners). Another feature of interest is the presence of a high-velocity notch in the H$\alpha$ absorption in model x3p0ext2 at 50\,d. Such a notch is observed in numerous Type II SNe \citep{gutierrez_pap1_17} and has been associated with the dense shell that forms out of the swept-up RSG wind material \citep{chugai_hv_07}. Here, the process is similar except that the dense swept-up shell is associated with the CSM originally around $R_\star$.  In models with less CSM, the dense shell is not dense enough. In models with more CSM, the dense shell moves more slowly. In model x3p0ext3, its associated absorption merges with the absorption from lower velocities so no notch is visible. The same applies to model x3p0ext5 but the absorption is now filled in by emission from the dense shell. In model x3p0ext6, the shell speed is even lower, its density higher, and the absorption vanished. It may be that in Nature, the high-velocity notch observed in H$\alpha$ comes primarily from swept-up CSM during the first few days after shock breakout.

\begin{figure*}
\epsfig{file=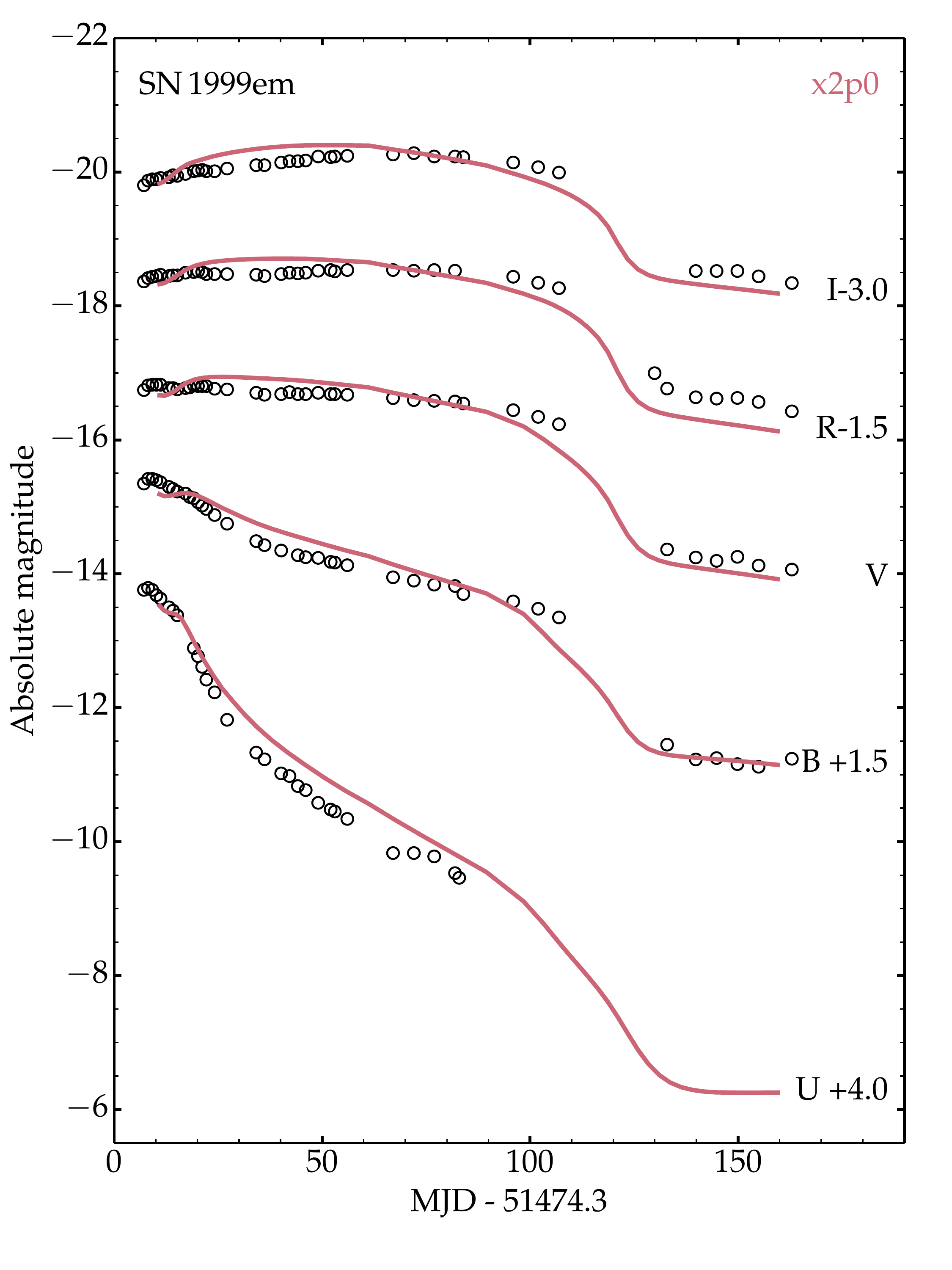, width=9.2cm}
\epsfig{file=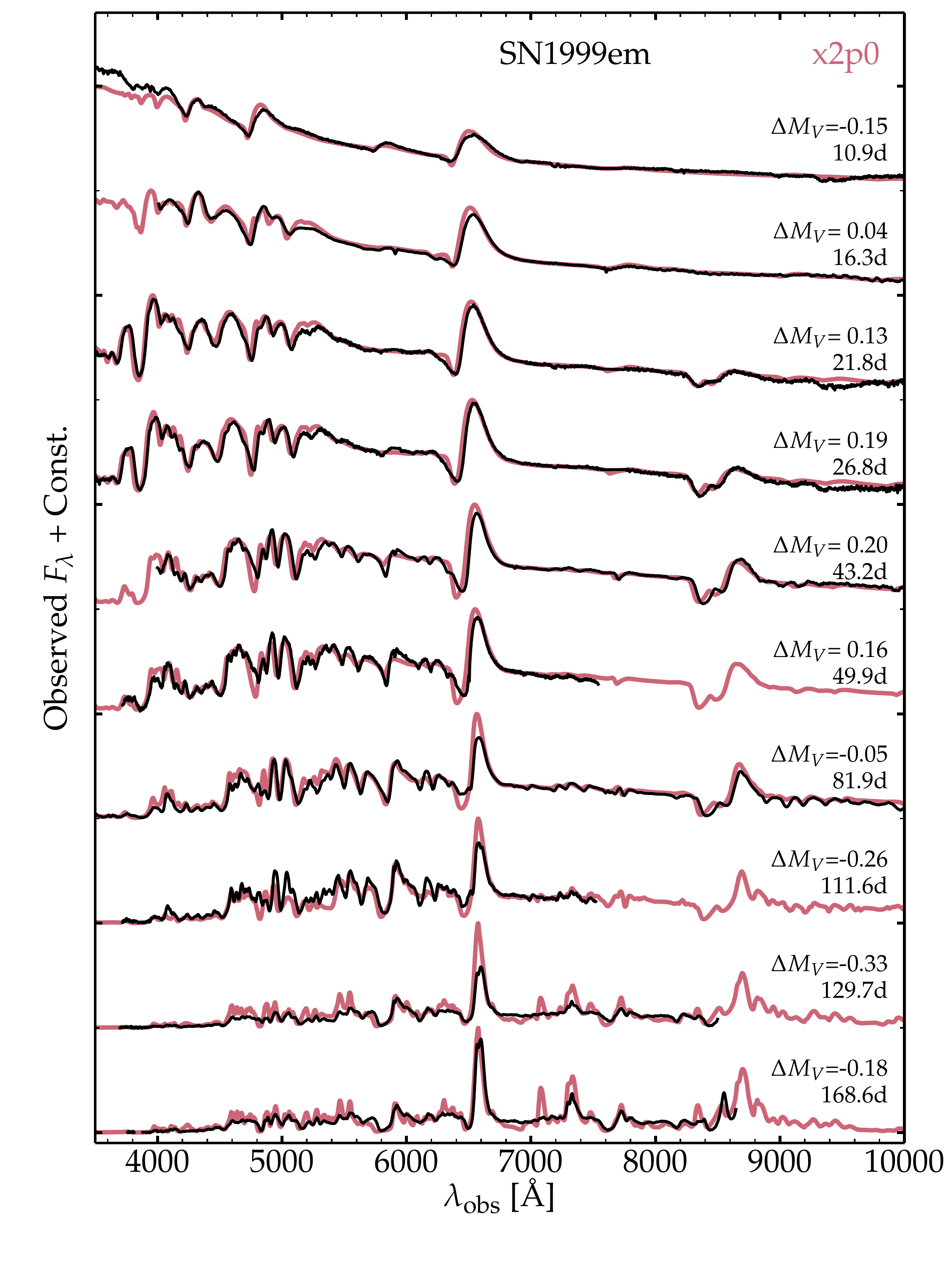, width=9.2cm}
\caption{Comparison of multi-band light curves (top) and multi-epoch spectra (bottom) for SN\,1999em and model x2p0. The time origin is the inferred time of explosion. For the spectral comparison, the model is redshifted and reddened, and the label $\Delta M_V$ gives the $V$-band magnitude offset at each epoch. The spectra are normalized to each other and shifted for better visibility.
\label{fig_comp_99em}
}
\end{figure*}

\begin{figure*}
\epsfig{file=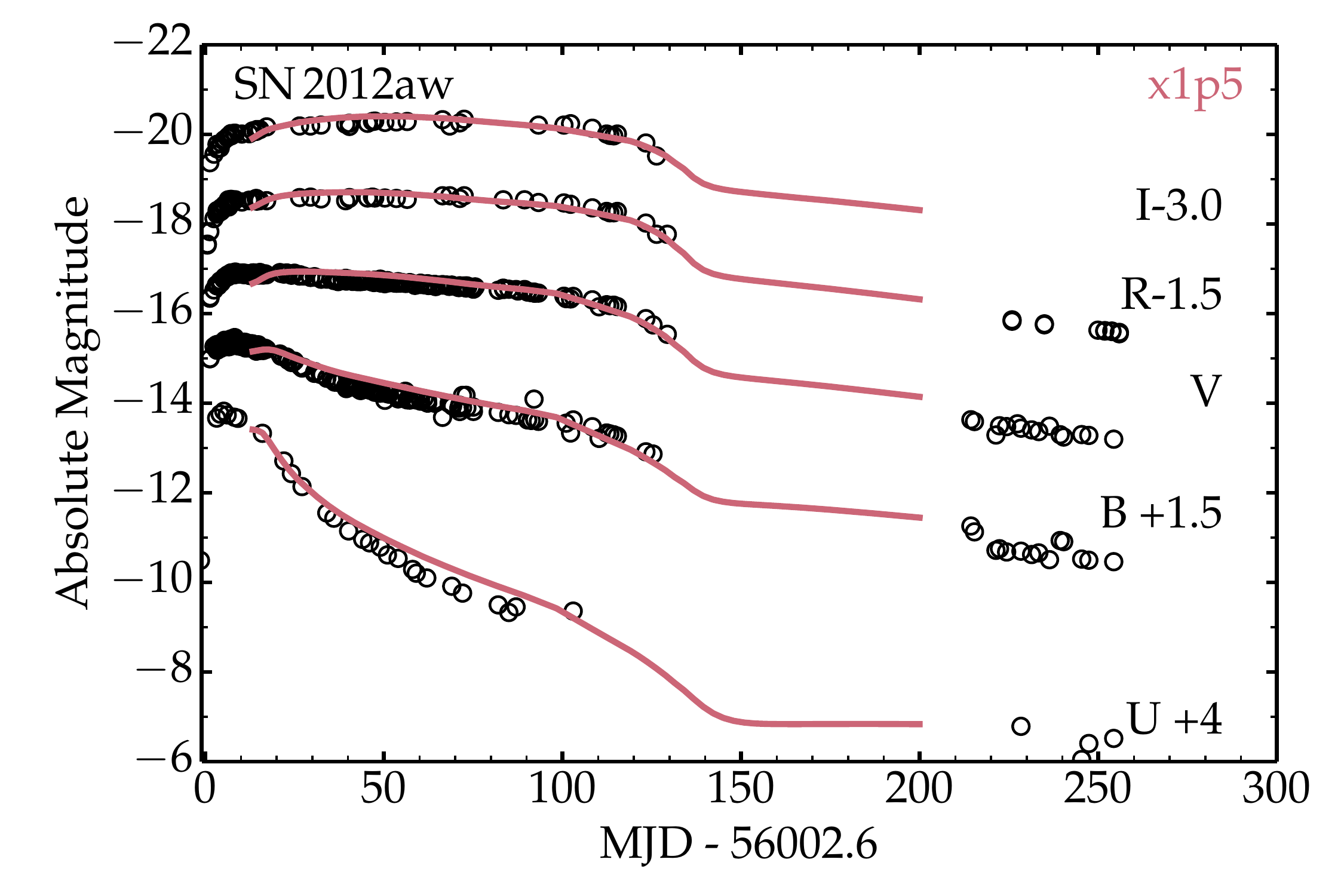, width=9.2cm}
\epsfig{file=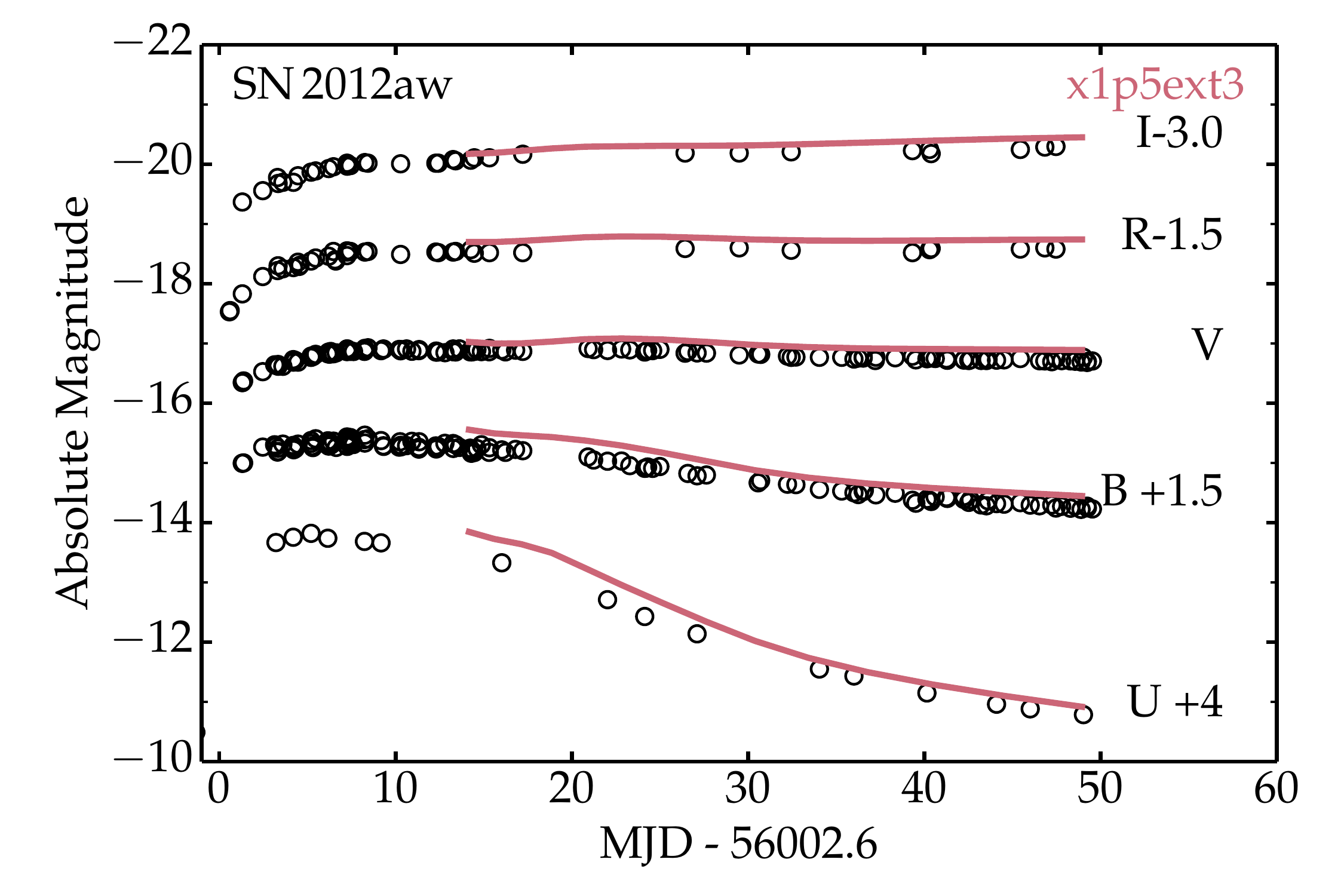, width=9.2cm}
\epsfig{file=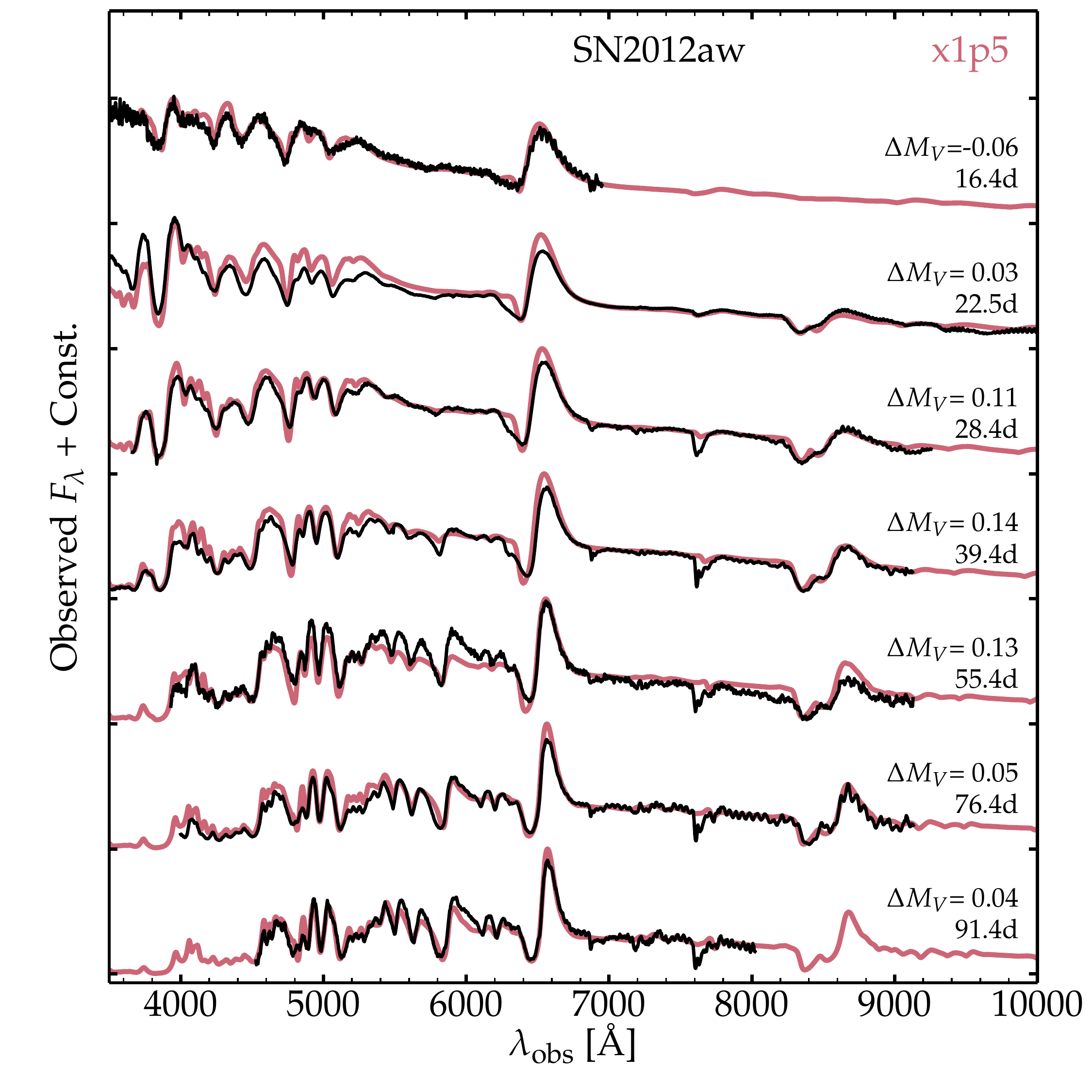, width=9.2cm}
\epsfig{file=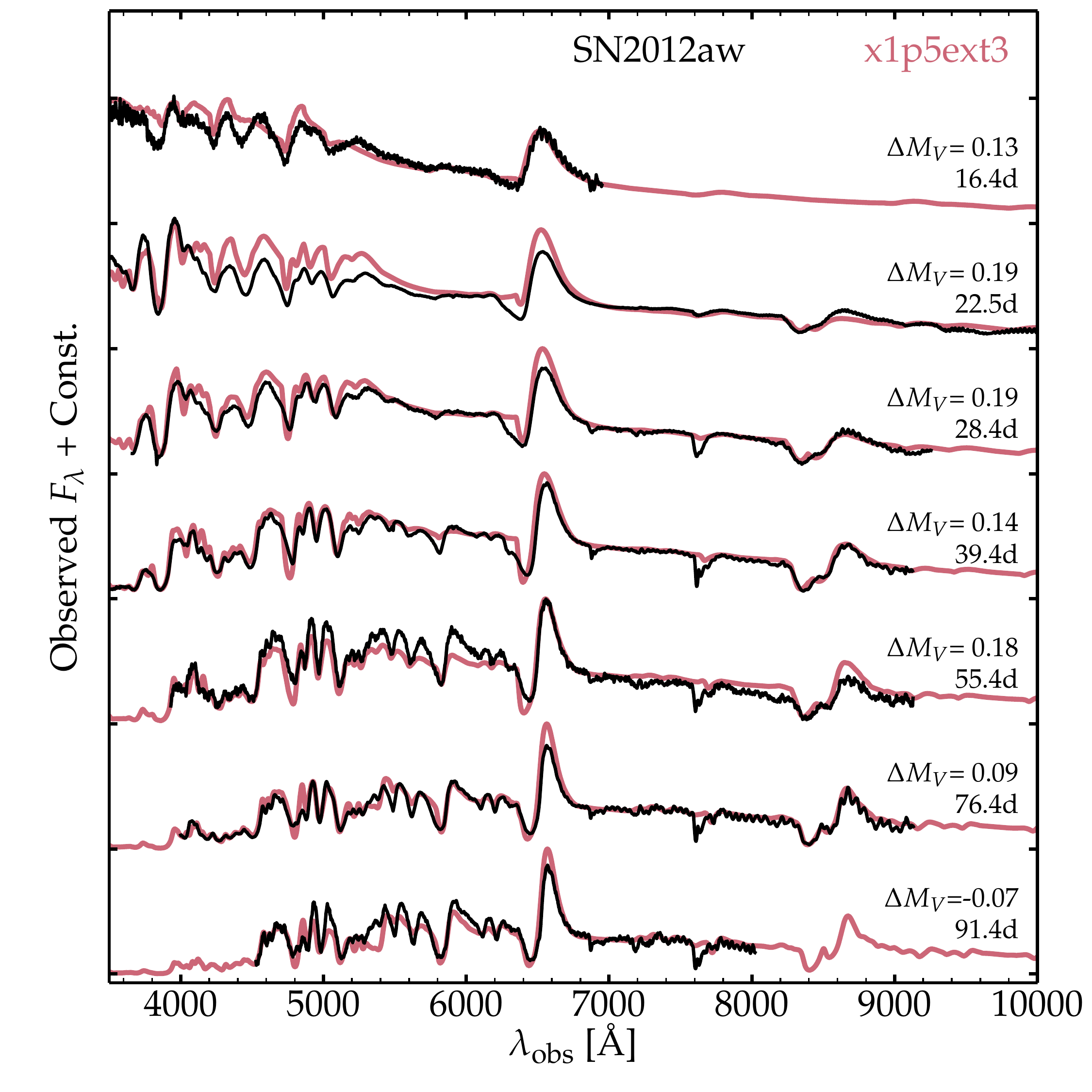, width=9.2cm}
\caption{Same as Fig.~\ref{fig_comp_99em}, but now showing a comparison of SN\,2012aw with the
model x1p5 (no CSM; left) and model x1p5ext3 (with CSM; right). The main effect of the CSM is to produce a slightly more luminous SN which peaks earlier, and which is in better agreement with observations. However the model is slightly too blue for the first month. This could probably be remedied by changes in the structure of the CSM.
The spectroscopic comparisons after 30\,d are very similar. 
\label{fig_comp_12aw}
}
\end{figure*}

\section{Comparison to observations}
\label{sect_comp_obs}

  It is beyond the scope of this study to make a comparison to a large sample of Type II SNe. Instead, this section presents a comparison of a few models from the {\it mdot} and {\it ext} sets to a few representative SNe exhibiting different $V$-band decline rates. Unlike all previous studies, we compare both multi-band light curves and multi-epoch optical spectra.

To avoid confusion, let us stress again that this section presents comparisons and not fits to observations. When one aims to produce a fit to observations, one performs a large number of simulations and then select the model that produces the best $\chi^2$.  In this study, we have produced a handful of models with sizable differences between them so a good ``fit'' to data would be largely incidental. Our models are therefore presented as comparisons in this section. When evaluating the offset between the model and the data, the reader is asked to evaluate whether a simple change in parameters could resolve the offset. The documented dependences between ejecta and progenitor properties on the one hand, and the observables on the other can be used for this. For example, a change of 1\,\msun\ in ejecta mass lengthens the plateau duration by 10\,d in the low-energy model X for SN\,2008bk \citep{lisakov_08bk_17}.

\subsection{Comparison to slow decliners}
\label{sect_comp_2p}

\subsubsection{SN\,1999\lowercase{em}}

SN\,1999em has been extensively studied. It was the first Type II-P SN detected at both radio and X-ray wavelengths \citep{PLF02_SN1998S}. \cite{PLF02_SN1998S} argue that the X-ray observations indicate a pre-SN wind with a mass-loss rate of approximately $2\times 10^{-6}$\,\msunyr\ and a speed of $10$\,\kms. Extensive photometric and spectroscopic observations have been discussed by \cite{EDC03_SN1999em} who indicate that dust formed after day 465. Several studies of SN\,1999em have used the ``expanding photosphere method" (EPM)  \citep{HPM01_1999em,leonard_99em}, or a variant, to determine its distance \citep{BNB04_1999em,DH06}.
The later give distance estimates more consistent with that obtained using cepheids.

Figure~\ref{fig_comp_99em} presents a comparison of model x2p0 ({\it mdot} set); no CSM) with the observations of SN\,1999em, including the $UBVRI$ light curves (top) and the multi-epoch spectra from 10.9\,d until 168.6\,d after the inferred time of explosion. The model qualitatively reproduces  the photometric and spectroscopic data, from the early photospheric phase until well into the nebular phase. The multi-band light curves are well matched in all bands. The slight underestimate of the optical brightness at 10\,d suggests that a small amount of CSM would help. Matching the transition to the nebular phase better merely requires a slight adjustment to the H-rich envelope mass, the explosion energy, and the \nifs\ mass. Here, the model x2p0 reproduces roughly this transition (it occurs about 10\,d too early). The nebular-phase brightness is underestimated by about 20\% (it depends on the filter considered), so an increase of the \nifs\ mass from 0.036 to about 0.043\,\msun\ would reduce the offset at nebular times and also lengthen the plateau (and thus reducing the offset mentioned earlier).

Taken individually, some lines show slight offsets. For example, Na\one\,D strengthens more slowly than observed;  the H$\alpha$ absorption is a little too broad (but the width of the emission is well matched); H$\beta$ disappears in the observations during the second part of the plateau but is always present in the model. Some discrepancies emerge at late times, like the overestimate of the He\one\,7065\,\AA. All these offsets are worth further investigation but they are small and do not alter the conclusions that can be drawn from the comparison. Importantly, we have shown that the model qualitatively, and to a lesser extent, quantitatively, matches simultaneously the multi-band light curves and spectra of SNe 1999em. In other words, a 15\,\msun\ progenitor model exploding with an H-rich envelope mass of about 9\,\msun, and producing an ejecta with about $1.2 \times 10^{51}$\,erg kinetic energy and 0.036\,\msun\ of \nifs\   is broadly compatible with observations. A small contribution from CSM would improve the match at $<15$\,d (see next section on SN\,2012aw),and a 20\% greater \nifs\ mass would improve the agreement at $>$100\,d.

The model of \citet{utrobin_07_99em} for SN\,1999em is similar to model x2p0 except that the ejecta mass is 19\,\msun\ and the progenitor density structure is crafted. Using a polytropic density structure for the progenitor, \citet{bersten_11_2p} propose a similar ejecta mass as \citet{utrobin_07_99em} but a larger progenitor radius (800\,\rsun\ instead of 500\,\rsun) and a \nifs\ mass (0.056\,\msun\ rather than 0.036\,\msun). Our results, obtained using a model  evolved with \mesa\ from main sequence to core collapse, show that it is not necessary to invoke a non-evolutionary model to reproduce the observations.  More recently \cite{utrobin_sn2p_17} considered neutrino-driven 3D explosion models of a 15\,\msun\ progenitor. While their best model matches reasonably well the light curve of SN\,1999em, the predicted photospheric velocities (prior to 20\,d, and after 50\,d) are significantly lower than observed.

Using the progenitor models of \citet{WH07}, \citet{morozova_sn2p_18} propose an ejecta kinetic energy of about $0.5 \times 10^{51}$\,erg, an ejecta mass of 14.5\,\msun, 0.0536\,\msun\ of \nifs, a progenitor radius of about 1100\,\rsun, a CSM mass of 0.31\,\msun. As argued above, some CSM would improve our fit to the light curves during the first 10 days after the explosion. Our discrepancy with \citet{morozova_sn2p_18} is the progenitor radius (twice larger than for x2p0) and the kinetic energy (0.4 times that of x2p0). As discussed in \citet{d13_sn2p}, such large progenitor radii are in tension with the color evolution of Type II-P SNe. This problem may be reduced in events with a large CSM mass and in which interaction-power is sustained for a long time. However, if the influence of the CSM ebbs after $10-20$\,d, the issue of the progenitor radius remains since the delay to recombination will still be too long. In practice, interaction with CSM delays rather than hastens recombination (see Figs.~\ref{fig_v1d_res} and \ref{fig_cmfgen_res}). The colors computed by {\sc snec} are based on LTE and therefore cannot address this point convincingly. Using a very large progenitor radius boosts the plateau luminosity, which can then be tuned by dropping the kinetic energy. But a kinetic energy of $0.5 \times 10^{51}$\,erg seems incompatible with the width of Doppler-broadened lines in the spectra of SN\,1999em (our model x2p0 is a little too energetic with $1.2\times 10^{51}$\,erg, but probably not by a factor of 2.4). This problem may be exacerbated because of the CSM of 0.3\,\msun\ in the model of \citet{morozova_sn2p_18}, which acts as a damper for the outer ejecta kinetic energy. \citet{morozova_sn2p_18} only use photometric constraints and thus do not address these dynamical aspects and associated constraints.

\begin{figure}
\epsfig{file=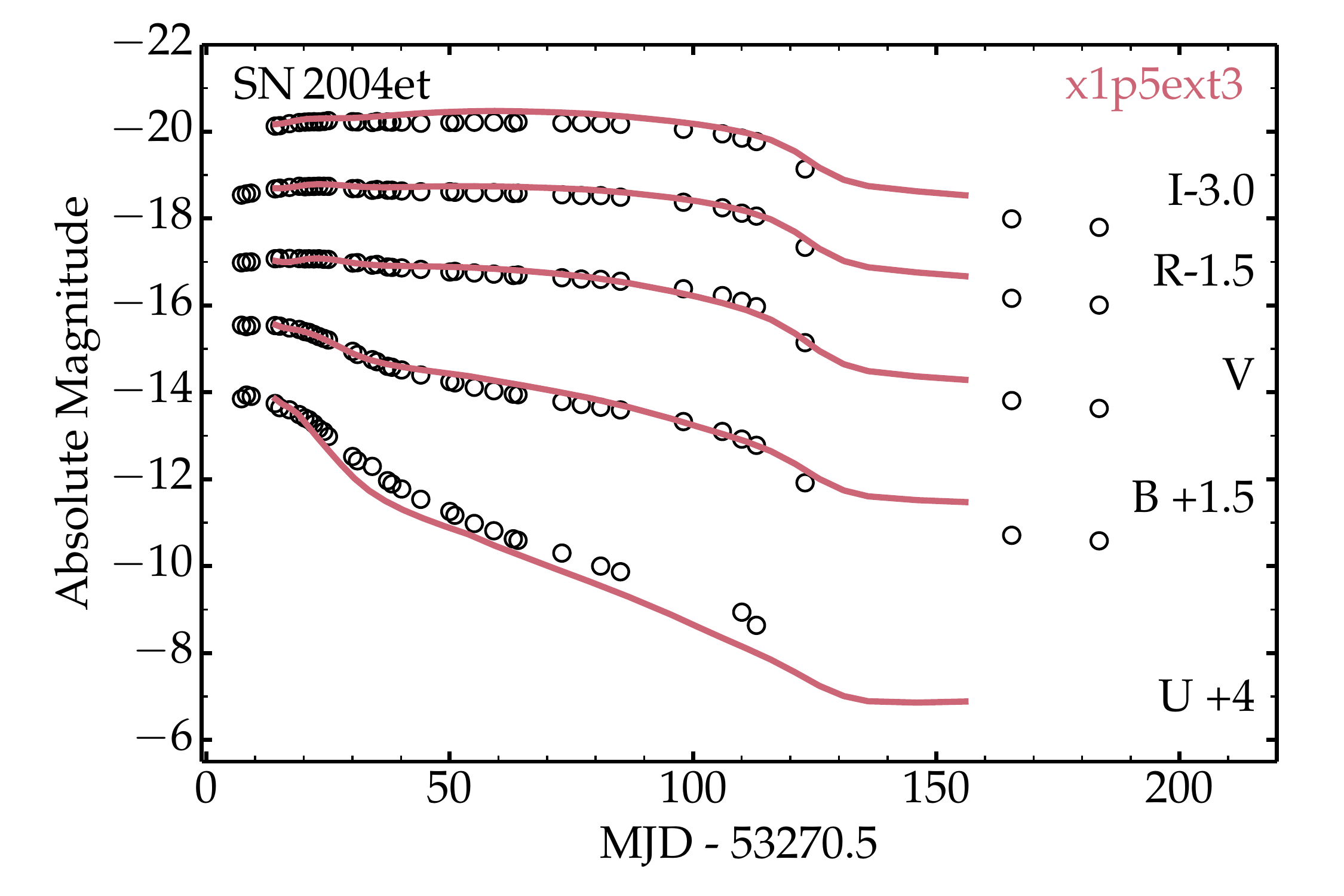, width=9.2cm}
\epsfig{file=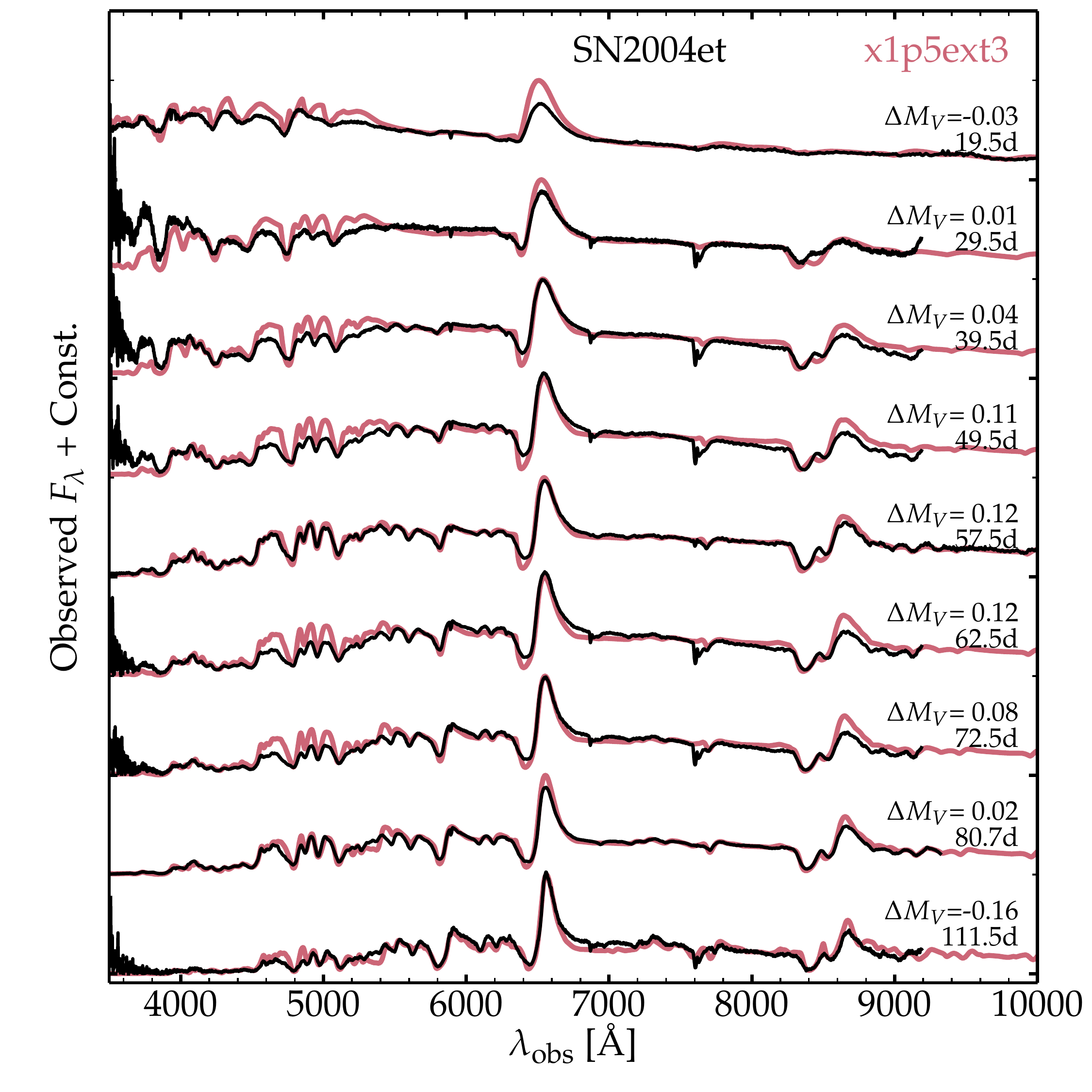, width=9.2cm}
\caption{Same as Fig.~\ref{fig_comp_99em}, but now showing a comparison of SN\,2004et with the model x1p5ext3 (with CSM). The adopted reddening is $E(B-V)=$\,0.3\,mag.
\label{fig_comp_04et}
}
\end{figure}

\subsubsection{SN\,2012\lowercase{aw}}

SN\,2012aw is very similar to SN\,1999em, although its expansion velocities are somewhat higher (by $\sim$600\,\kms) \citep{2013MNRAS.433.1871B}. Its progenitor was a RSG \citep{fraser_12aw_12, vandyk_12aw_12} which was later confirmed to have disappeared \citep{2016MNRAS.456L..16F}. \cite{jerkstrand_12aw_14} used the nebular spectrum and nucleosynthesis arguments to constrain the progenitor mass to the range 14 to 18\,\msun. Polarization observations indicate asymmetries typical of Type II-P SNe \citep{2012ATel.4033....1L,2013MNRAS.433.1871B}.

Figure~\ref{fig_comp_12aw}, similar to Fig.~\ref{fig_comp_99em} presented above for SN\,1999em, compares the observations of SN\,2012aw with a model without CSM (x1p5) and a model with CSM (x1p5ext3). As for SN\,1999em, the models reproduce satisfactorily the observed multi-band light curves and multi-epoch optical spectra but some differences are clearly visible at early times. Model x1p5 peaks at about 20\,d after explosion, which is later than observed. In the model x1p5ext3 (with 0.24\,\msun\ of CSM), all bands have reached their maximum at 15\,d (the simulation does not start earlier but the contrast between models x1p5 and x1p5ext3 is unambiguous). A concern though is that model x1p5ext3 is slightly bluer than observed for about a month. In practice, this color offset could be reduced if the CSM was more confined so that less mass is shocked at larger radii. Using a larger progenitor radius would exacerbate the color discrepancy. The model with CSM underestimates the depth of the H$\alpha$ absorption trough at 22.5\,d, which may indicate that the CSM is too massive. It is hard to conjecture here because there are numerous simplifications in the present exploration (1D; \cmfgen\ simulations started at 10-15\,d; simplistic CSM structure etc).

As SN\,2012aw is similar to SN\,1999em, \citet{morozova_sn2p_18} propose a similar model. While our model x1p5ext3 has a similar CSM mass, it has the same discrepancy with the results of \citet{morozova_sn2p_18}  -- our  progenitor radius is smaller and the explosion energy is larger. We surmise that the dynamical properties of their model would be in tension with the spectroscopic properties of SN\,2012aw.  SN\,2012aw has also been studied by \citet{dallora_12aw_14}, who propose an ejecta of 20\,\msun, a kinetic energy of $1.5 \times 10^{51}$\,erg, a \nifs\ mass of 0.06\,\msun. They do not discuss the adopted density structure of the progenitor. The spectral information used to constrain the ejecta expansion rate is primarily limited to a Sc\two\ line (a Fe\two\ line shows the
same velocity characteristics), with the first ``constraining" data point at 40\,d after the explosion.

Given the agreement obtained in Fig.~\ref{fig_comp_12aw}, a 15\,\msun\ progenitor model exploding with an H-rich envelope mass of about 9.5\,\msun, and producing and ejecta with about $1.2 \times 10^{51}$\,erg kinetic energy and 0.06\,\msun\ of \nifs\ seems compatible with observations. A small contribution from CSM improves the match to the brightness at $<15$\,d, but also causes a slight color discrepancy.

\begin{figure*}
\epsfig{file=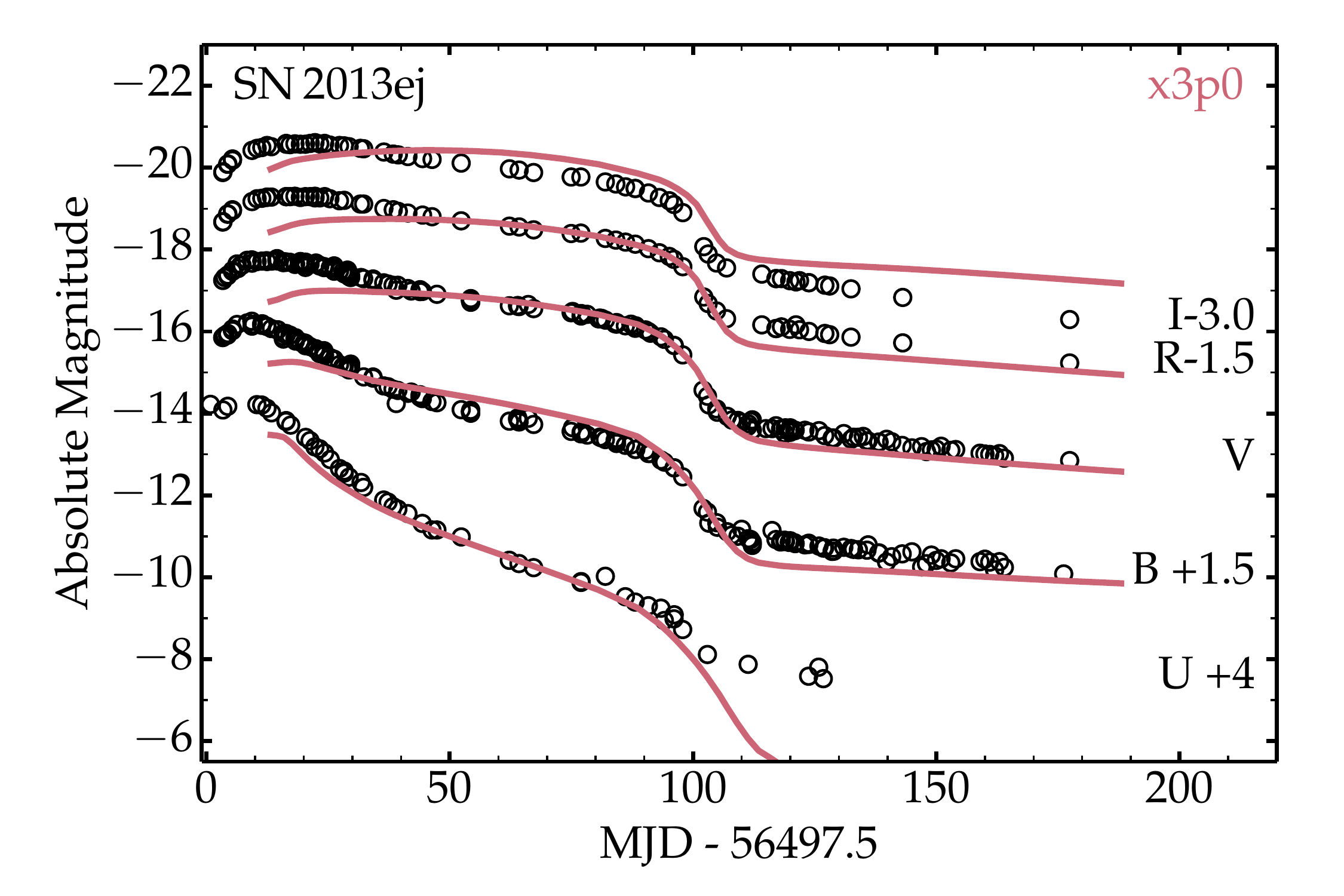, width=9.2cm}
\epsfig{file=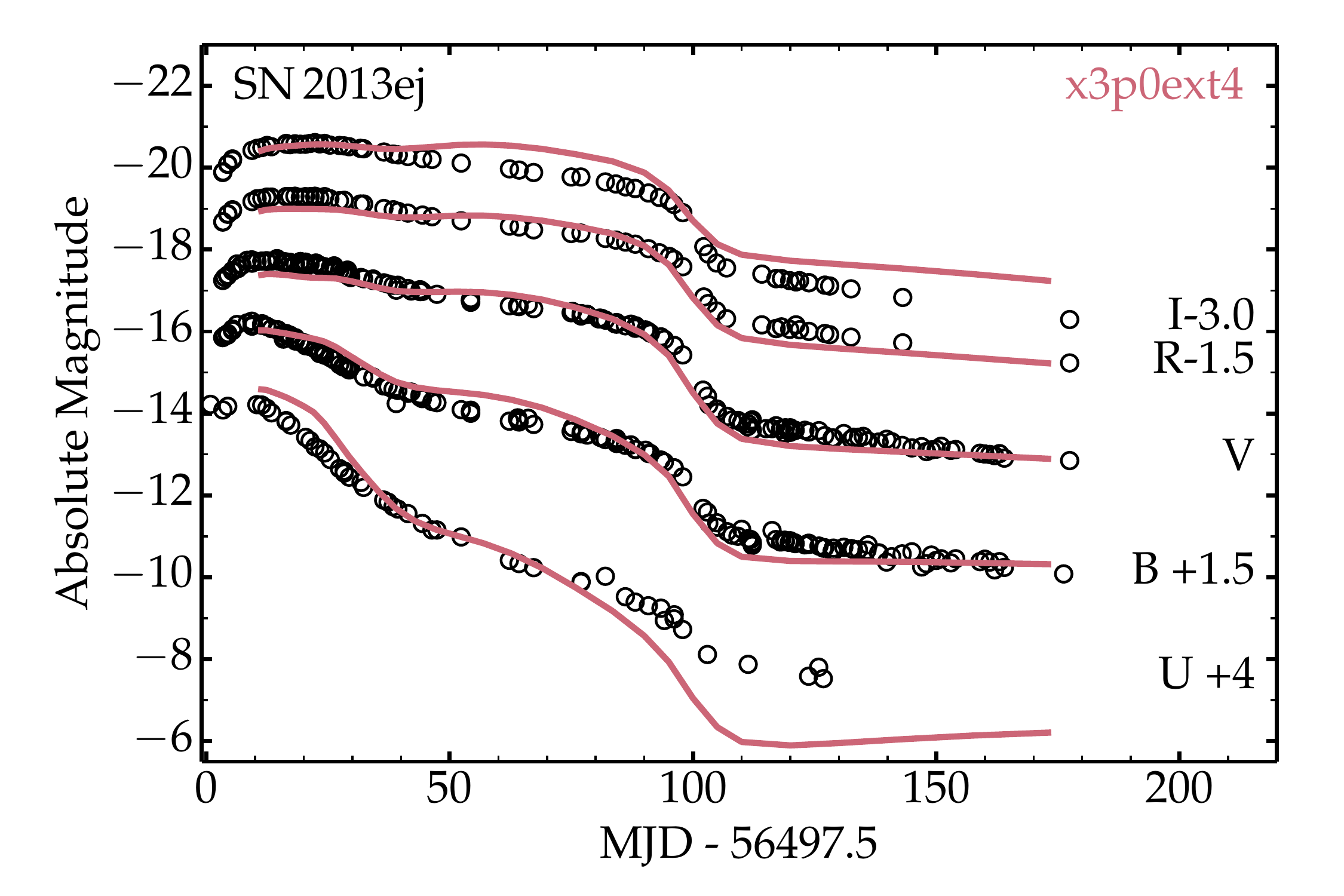, width=9.2cm}
\epsfig{file=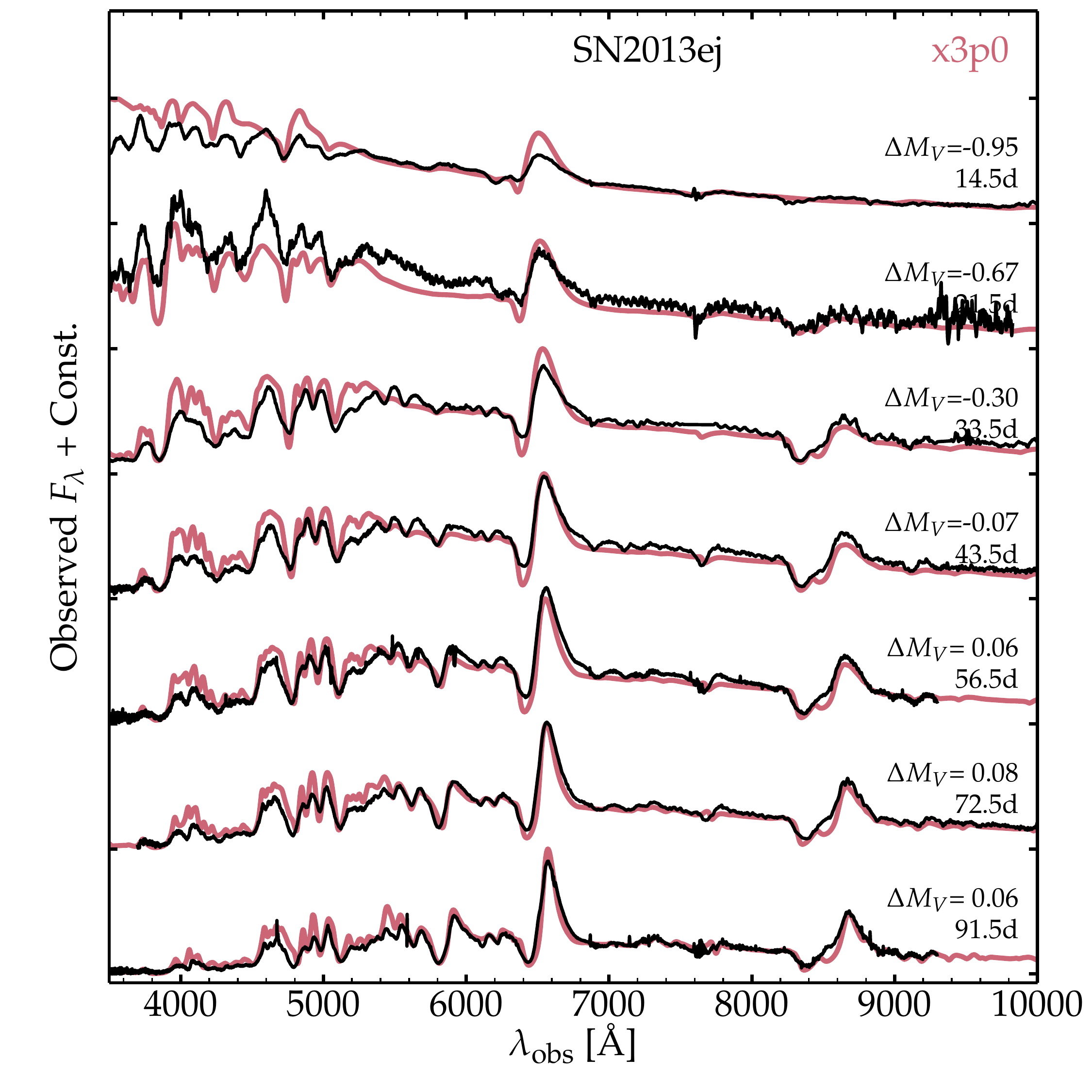, width=9.2cm}
\epsfig{file=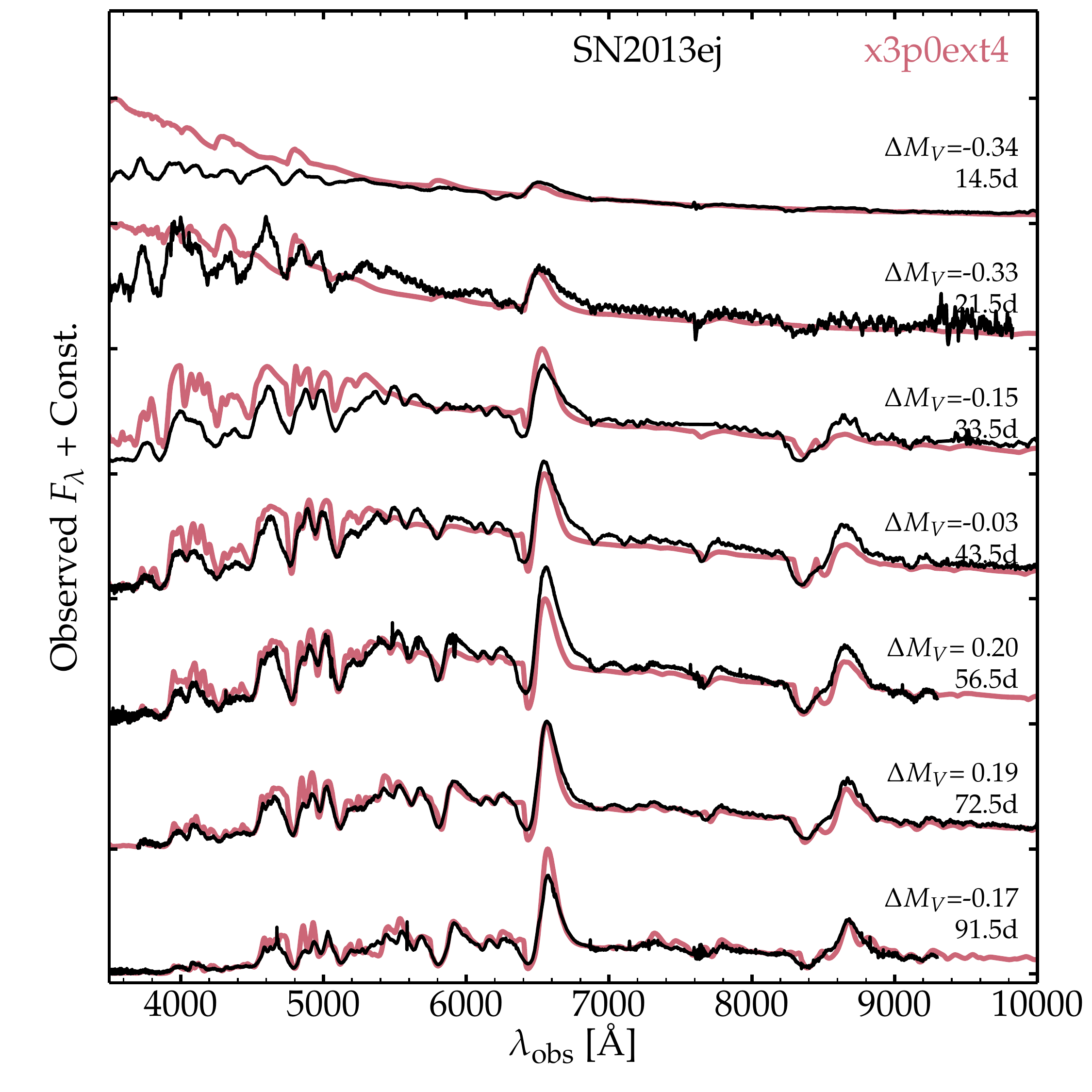, width=9.2cm}
\caption{Same as Fig.~\ref{fig_comp_99em}, but now showing a comparison of SN\,2013ej with the
model x3p0 (no CSM; left) and model x3p0ext4 (with CSM; right). The model with CSM provides a better match to the brightness at early times and the weakness of the emission features, but is too blue, especially at day 14.5. After 50 days the models are very similar.
\label{fig_comp_13ej}
}
\end{figure*}

\subsubsection{SN\,2004\lowercase{et}}

SN\,2004et is associated with NGC\,6946 and is one of 10 recent SNe known to have occurred in this galaxy \citep[e.g.,][]{2018MNRAS.481.2536K}. The estimate of the progenitor's mass of  $14^{+1}_{-2}$\,\msun\ by \cite{2019MNRAS.485L..58E} is higher than the earlier estimates of $10.7^{+0.9}_{-0.8}\,$\msun\ \citep{davies_beasor_18} and $12^{+3}_{-3}\,$\msun\  primarily due to an increase in the distance to NGC\,6946 \citep[from 5.5 to $\sim$7.8\,Mpc;][]{2018AJ....156..105A}. A detailed multi-wavelength study of SN\,2004et was done by  \cite{2007MNRAS.381..280M} who argue for a \isoni\ mass of $0.06 \pm 0.03$\,\msun\ (for $d=5.5$\,Mpc), an ejecta mass of 8 to 16\,\msun, and a progenitor mass of around 20\,\msun.

Figure~\ref{fig_comp_04et} compares the multi-band light curves and multi-epoch spectra of SN\,2004et with the model x1p5ext3 that was discussed in the previous section. Here, a reddening $E(B-V)=$\,0.3\,mag is used, and with this choice, SN\,2004et becomes very similar to SN\,2012aw. This lower reddening is more compatible with the color evolution of SN\,2004et. It also yields a reasonable match to the color evolution and to the multi-epoch spectra throughout the optical (some dates are less well fitted but it is also clear that some spectra have a problematic relative flux calibration). Other reddening values have been used in the literature --  \citet{morozova_sn2p_18} use $E(B-V)=$\,0.36\,mag while \citet{utrobin_04et_09} adopt $E(B-V)=$\,0.41\,mag.

The model parameters of \citet{morozova_sn2p_18} have the same offset as for the SNe discussed above, with a larger progenitor radius and a lower ejecta kinetic energy (the offset also partially arises from their adopted reddening).  The good match of model x1p5etx3 to the width of Doppler-broadened profiles in SN\,2004et does not seem compatible with the low kinetic energy proposed by \citet{morozova_sn2p_18}, who ignore spectral constraints.

Our model x1p5ext3 differs from that of \citet{utrobin_04et_09}, who use a non-evolutionary progenitor model. Their model parameters correspond to a 1500\,\rsun\ progenitor radius, an ejecta mass of 24.5\,\msun, an explosion energy of $2.3 \times 10^{51}$\,erg, and a \nifs\ mass of 0.068\,\msun. Only the \nifs\ mass is close to the 0.053\,\msun\ of model x1p5ext3, the offset resulting from the larger reddening used in \citet{utrobin_04et_09}. Our model suggests that an evolutionary model works well for SN\,2004et, and that there is no need to invoke a very large mass for the progenitor star. A 15\,\msun\ progenitor star (like our model x1p5ext3) is also proposed by \citet{jerkstrand_04et_12} based on nebular-phase spectral modeling.

\subsection{Comparison to fast decliners}
\label{sect_comp_2l}

\subsubsection{SN\,2013\lowercase{ej}}

SN\,2013ej is located in M74. \cite{fraser_13ej_14} identified a M supergiant as the possible progenitor. Archival studies of the SN field by \cite{2018MNRAS.480.1696J} indicate that the progenitor most likely did not have a major outburst in the decade prior to its death. Even at early times
SN\,2013ej was significantly polarized \citep[$\sim$1\%;][]{2013ATel.5275....1L,2017ApJ...834..118M}.
\cite{2017ApJ...834..118M} argue that the polarization data for SN\,2013ej are consistent with an oblate ellipsoidal photosphere viewed nearly edge on. They also find evidence in nebula spectra that interaction with a CSM is continuing. Evidence for asymmetries is also seen in the H$\alpha$ profile.  \cite{utrobin_chugai_13ej_17} argue that the H$\alpha$ asymmetry and the observed level of polarization arise from a strong asymmetry in the distribution of \isoni.

Figure~\ref{fig_comp_13ej} is analogous to Fig~\ref{fig_comp_12aw} but now compares a model without CSM (model x3p0) and a model with CSM (model x3p0ext4) with the multi-band light curves and multi-epoch spectra of SN\,2013ej. Both models do well after about 30\,d, but prior to that, only the model with CSM can capture approximately the bump in radiation (as compared to slow decliners; see Fig.~\ref{fig_obs_mv}).

Model x3p0 overestimates the line emission strengths early on. It is too faint in all bands (but not by much in $U$). However, after 50\,d, both multi-band light curves (and thus color curves) and optical spectra are well matched. Model x3p0ext4 resolves in part the brightness problem at $<30$\,d, but the color is too blue. Because of the CSM, there is less material at large velocities so the model under-predicts the width of some lines (note, however, that some line profiles have a complex morphology, such as the broad red shoulder in H$\alpha$). The early time spectra are in some ways better matched than with model  x3p0, in particular because the model captures the much reduced emission line strengths.

These trends suggest that CSM is indeed a necessary ingredient to reproduce the early time properties of SN\,2013ej but the exact properties of the CSM (mass, extend, or density structure), which may deviate from spherical symmetry, are likely an important component. A more confined CSM distribution would probably help resolving the color offset while preserving a fraction of the boost to the brightness.

\subsubsection{SN\,2014G}

The photometric and spectroscopic evolution of SN\,2014G has been extensively discussed by 
 \cite{terreran_14G_16}. Early spectra show high ionization features, such as He\,\two, C\,\four\ and a
 N\three/N\,\five\ blend indicating interaction of the ejecta with CSM, possibly a pre-existing wind.
 By comparing the strength of the [O\,\one] $\lambda\lambda 6300,6363$ doublet with synthetic spectra,
  \cite{terreran_14G_16} deduced a progenitor mass in the range 15 to 19\,\msun.
 
\begin{figure*}
\epsfig{file=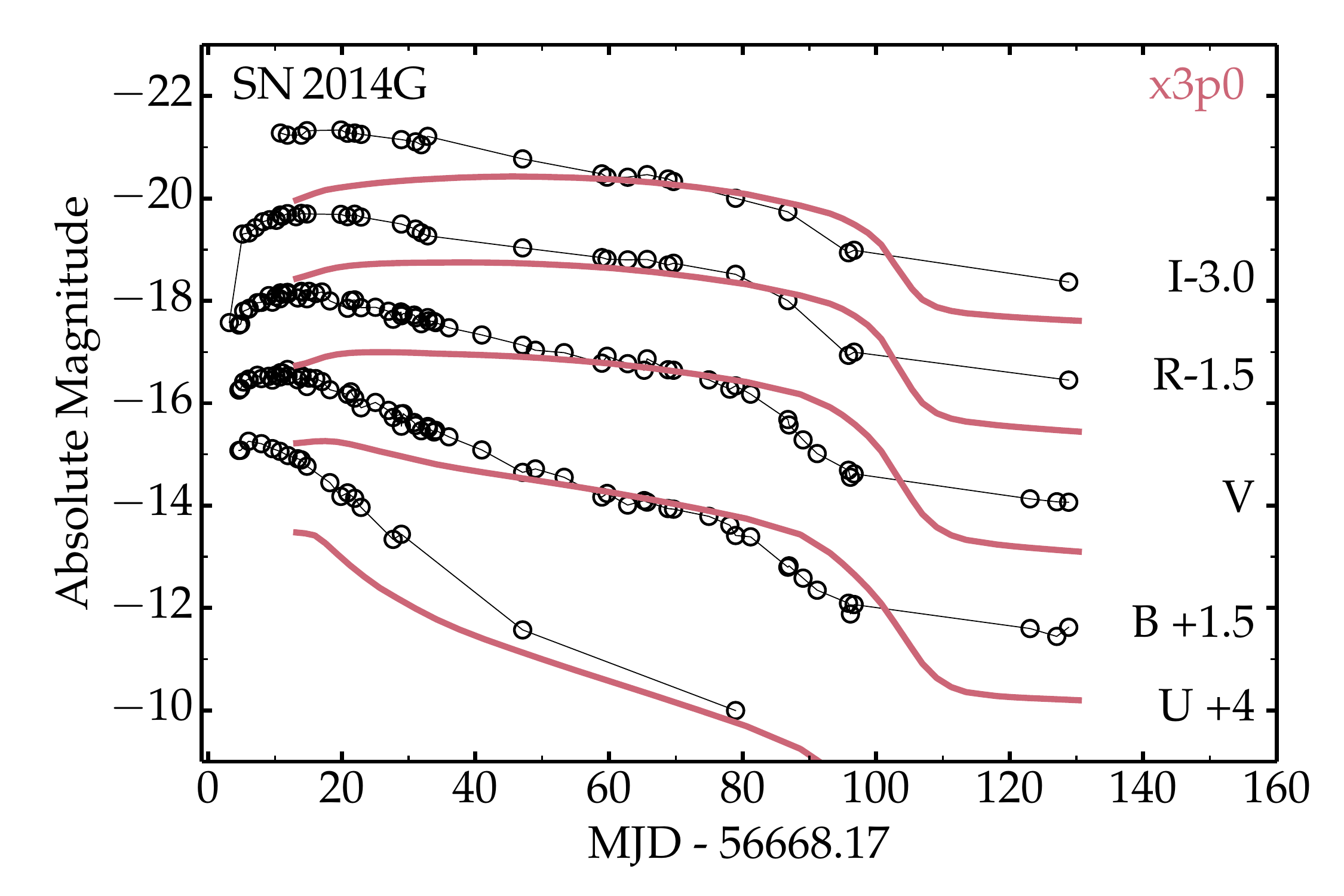, width=9.2cm}
\epsfig{file=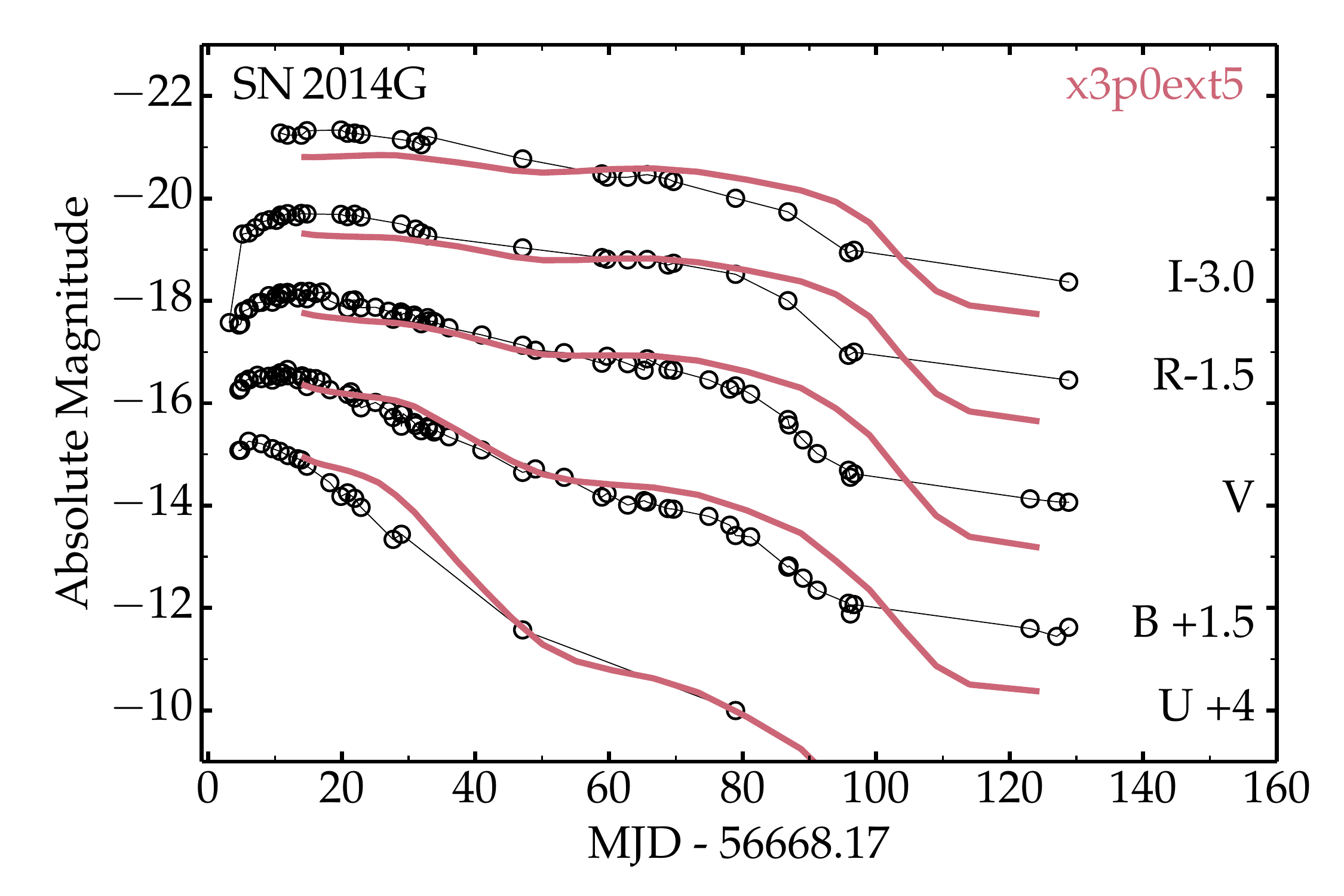, width=9.2cm}
\epsfig{file=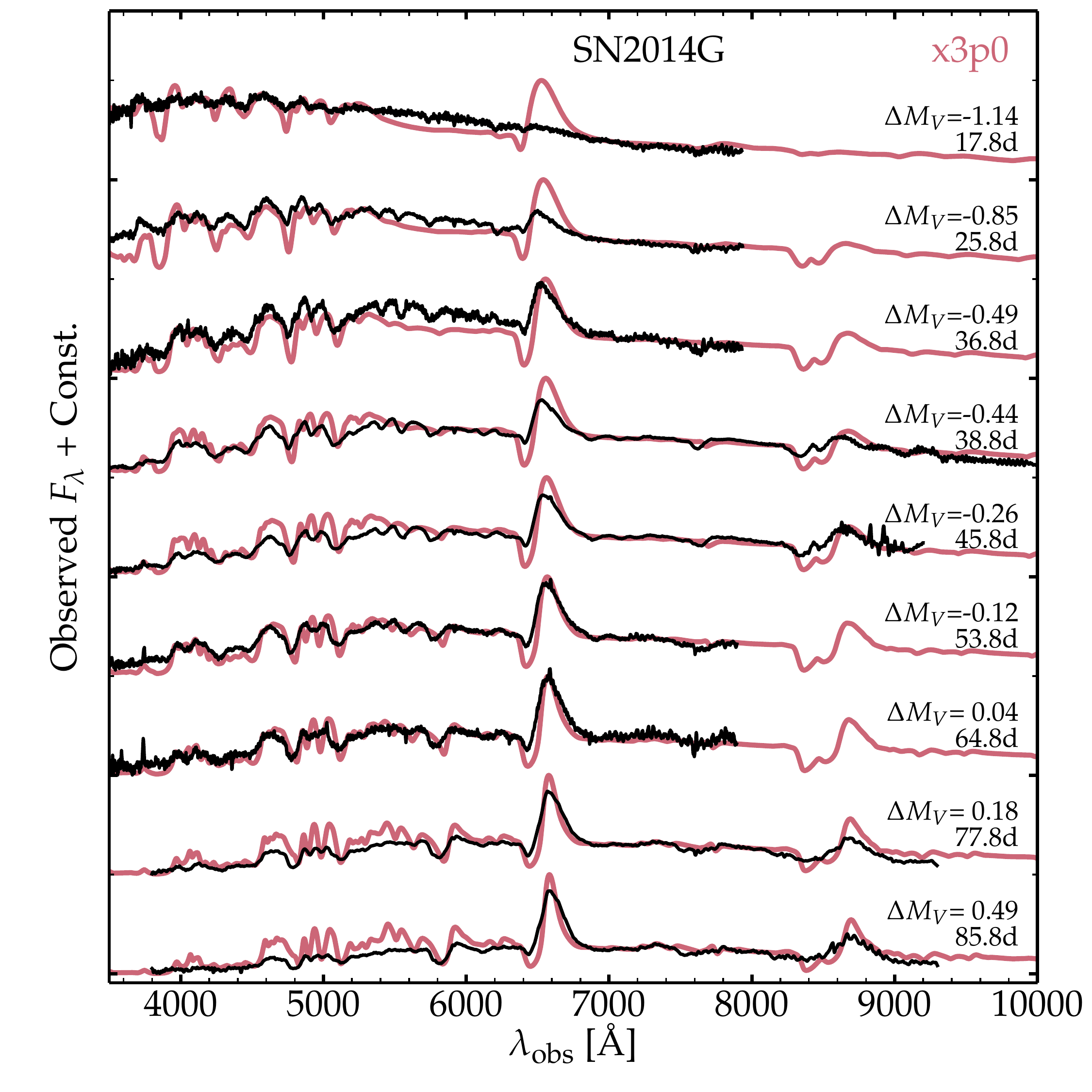, width=9.2cm}
\epsfig{file=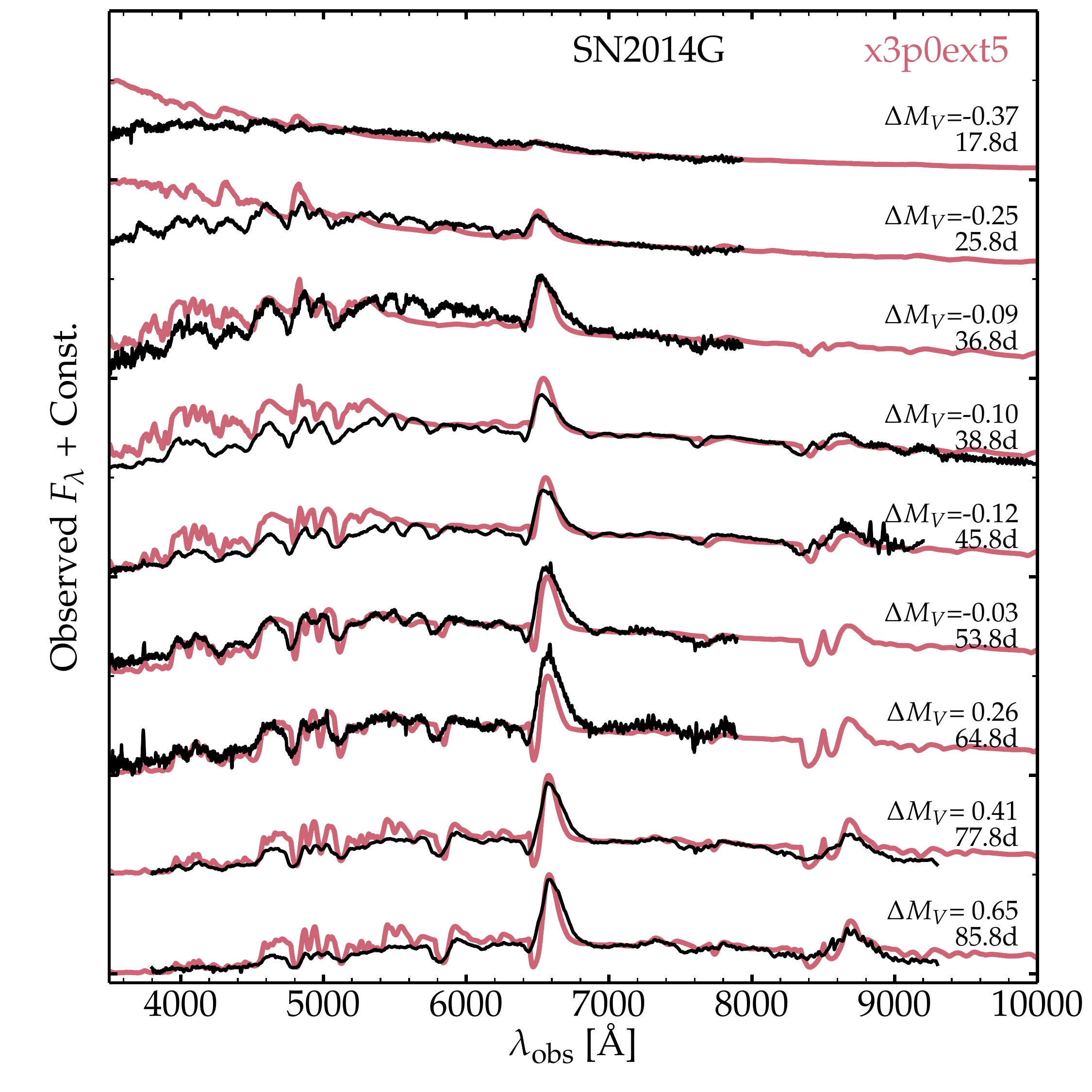, width=9.2cm}
\caption{Same as Fig.~\ref{fig_comp_99em}, but now showing a comparison of SN\,2014G with the
model x3p0 (no CSM; left) and model x3p0ext5 (with CSM; right). The model with the CSM shows much better agreement with the observations -- the light curve is better matched at earlier times, the weak H$\alpha$ P~Cygni profiles are in better agreement with observation, and the emission at early times is weak/absent.
\label{fig_comp_14G}
}
\end{figure*}

Figure~\ref{fig_comp_14G} compares the photometric and spectroscopic observations of SN\,2014G with the results from model x3p0 and x3p0ext5. Because the early-time brightness boost in SN\,2014G is greater than in SN\,2013ej, the brightness discrepancy with model x3p0 is larger. It is nearly resolved with model x3p0ext5 (CSM mass of 1.97\,\msun) but the model is now too blue up to about 30\,d. As before, using a more confined CSM would reduce the color offset. However, the model with CSM yields a better consistency with photometric observations, the featureless spectra at early times, and the weak absorption in H$\alpha$ at all times,

\section{Conclusions}
\label{sect_conc}

    We have presented  a set of simulations for Type II SNe arising from two different types of RSG star  progenitors. The {\it mdot} model set is characterized by progenitors having a range of H-rich envelope mass between 0.9 and 9.5\,\msun, but the same He core mass of about 4\,\msun. The {\it ext} set is characterized by a range of CSM mass between 0.02 and 1.97\,\msun\ enshrouding the same RSG progenitor star. In this study, we compare the SN ejecta and radiation properties for each set of models and confront these to multi-band light curves and multi-epoch spectra of Type II SNe characterized by a range of $V$-band decline rates (i.e. from slow to fast decliners).  All models have the same ejecta kinetic energy.

By reducing the H-rich envelope mass, the bolometric light curve transitions from slow to fast declining and from a long to a short high-brightness phase (i.e., photospheric phase). The $V$-band light curves are changed in a similar way to the luminosity. The boost at maximum is $<1$\,mag and the rise time is unchanged (here about $20-30$\,d). Optical colors (or the photospheric temperature) are also unaffected. The smaller the H-rich envelope mass, the larger the maximum ejecta velocity and the broader the P-Cygni profiles at early times.

By increasing the CSM mass located directly above $R_\star$, the luminosity increases at early times before eventually leveling off at the luminosity for the CSM-less counterpart. The photospheric phase duration is unaffected except for the model with the highest CSM mass. The interaction induced by the CSM reduces the outer ejecta kinetic energy and causes the formation of a dense shell. Hence, unlike for the {\it ext} model set, there is an anti-correlation between brightness boost and line width. For a large CSM mass, the early-time spectra are featureless, while at the recombination epoch, the line profiles show weaker absorptions. H$\alpha$ may show a pure emission profile.

Overall, the luminous fast decliners SNe 2013ej and 2014G are in better agreement with the {\it ext} model set, both concerning the multi-band light curves and the spectral evolution in the optical. These two SNe may require  0.5 to 1\,\msun\ of CSM, although the exact value depends on the CSM mass distribution. Interestingly, their H-rich envelope mass is about 1\,\msun\ lower than for the slow decliners, and thus still very massive. It may be that this envelope mass deficit corresponds to the mass excess residing directly above $R_\star$.

The slow decliners SNe 1999em, 2012aw, and 2004et probably require no more than 0.2\,\msun\ confined to the progenitor surface. Here, the CSM acts to reduce the rise time in the $V$ band.  We argue, however, that this rise time is nearly matched if one invokes no CSM but a relatively compact RSG star progenitor. Our models with zero up to 0.2\,\msun\ of CSM yield a compelling evidence that standard RSG star explosions as produced by stellar evolution models can reproduce with fidelity the observed properties of standard SNe II-P.

Because of the complexity inherent to the CSM, it is a challenge to obtain a good match to the early-time observations (light curves as well as spectra) of fast decliners. The influence of the CSM depends on its mass, its density structure, its maximum extent in radius beyond $R_\star$. For example, Type IIn spectral signatures can only occur if the photon mean free path in the CSM is large enough to allow for radiation escape through unshocked slow CSM (see e.g. \citealt{d18_13fs}).

The effects of the CSM when placed at $R_\star$ is very different from that at large distances. The main difference is that at $R_\star$, the CSM is shocked under optically-thick conditions and at a time when the shocked envelope has not yet reached its asymptotic kinetic energy. Interaction with such a CSM first transfers kinetic energy to radiative energy, but because of optical depth effects, this radiative energy is transferred back into kinetic energy.  By varying the extend and mass of the CSM, one can modulate the radiative losses at shock breakout and tune the luminosity boost. When interaction occurs at large distances and over large distances, the extracted kinetic energy is converted into radiation energy that escapes. Then, the boost to the luminosity can be very large. Hence, in events like SN\,1979C, or even more so for SN\,1998S, the much larger luminosity boost requires the CSM to be detached from $R_\star$, or extended far above $R_\star$, so that radiative losses can be much larger. As a result, the interaction model of \citet{D16_2n} proposed for SN\,1998S is much more luminous than the present model x3p0ext4 (which matches roughly the brightness of SN\,2013ej, which is much fainter than 98S) even though they have the same CSM mass. Hence, more work is needed to investigate the impact on SN observables of using different types of CSM, extent, density structure, extent etc. The CSM may also be clumpy and asymmetric, which could affect the predictions made so far assuming 1D.

The models presented here for slow decliners continue to support the notion that Type II SN progenitors may be more compact than typically obtained by stellar evolution models computed with a mixing length parameter of 1.5. There is evidence that inferred RSG radii are smaller if one models the full spectral energy distribution rather than the optical range alone \citep{davies_rsg_13}. The argument that stellar evolution models predict large RSG radii if one uses the default mixing length parameter is not convincing. This default is based on the Sun and may not apply for RSG stars. One can produce any RSG radius one wishes from a few hundreds to thousands of \rsun\ by tuning this parameter \citep{d13_sn2p}.  {\sc kepler} observations suggest that the mixing length differs between red giants and the Sun \citep{li_mlt_18}, although not by as much as adopted here. The situation in RSGs may also differ. That being said, we should reinvestigate to what extent CSM may alter this need for more compact RSG progenitors. It is clear that for events like SNe 1979C and 1998S, in which the radiation is strongly influenced by interaction, the progenitor radius has little impact on the observables.

This study is not exhaustive. Our small selection of slow decliners seems to be compatible with a 15\,\msun\ progenitor, 0.2\,\msun\ of CSM or less, and moderate variations in H-rich envelope mass around $9-10$\,\msun. In slow decliners, the inferred CSM, located directly above $R_\star$, seems to be very confined. It is also probably bound to the star and may be counted as part of the star mass. There is no evidence that it corresponds to a super-wind. The fast decliners (limited here to SN\,2013ej and 2014G) require slightly lower H-rich envelope masses to yield shorter photospheric phase durations. This may arise because they come from higher mass progenitors, which have a greater RSG wind mass loss rate, or from binaries, although binarity tends to produce pre-SN progenitors with a $< 1$\,\msun\ H-rich envelope mass \citep{yoon_ib_17}.

\begin{acknowledgements}
DJH acknowledges partial support from NASA theory grant NNX14AB41G and STScI theory grants HST- AR-12640.01 \& HST-AR-14568.001-A. LD thanks ESO-Vitacura for their hospitality.
\end{acknowledgements}

\bibliographystyle{aa}
\bibliography{new_sn_library_luc}

\appendix

\section{Energy conservation}
\label{App_en_con}

For a homologous flow we have the global comoving-frame energy constraint
\begin{eqnarray}
 r_{\rm max}^2 H(r_{\rm max}) &=& r^2 H(r) \nonumber \\
&+& \int_{\hbox{$r$}}^{\hbox{$r_{\rm max}$}}
{r^2 \over 4 \pi  } \left(\dot{e}_{\rm decay} - \rho {De \over Dt} + {P \over \rho} \frac{D\rho}{Dt} \right) \nonumber \\
 &-& {1 \over cr^2}  {D(r^4 J)  \over Dt}\,\,  dr
\label{eq_global_en}
\end{eqnarray}
\citep{HD12}.
Multiplying by $16 \pi^2 t$ and integrating from some initial time $t_0$ to $t$ yields
\begin{eqnarray}
 \int^t_{t_0} t L(t) \,dt = \int^t_{t_0} [t Q(t)  - t  I(t)] \,dt  + t_oE(t_o) -  t E(t) \,\,.
\label{eq_time_en}
\end{eqnarray}

\noindent
In the above
\begin{eqnarray}
I(t) = \int_{\hbox{$r$}}^{\hbox{$r_{\rm max}$}} \left({P \over \rho} \frac{D\rho}{Dt}-  \rho {De \over Dt}\right)\,dr \,\,,
\end{eqnarray}
\begin{eqnarray}
Q(t) =\int_{\hbox{$r$}}^{\hbox{$r_{\rm max}$}}  4\pi r^2 \dot{e}_{\rm decay} \,dr
\end{eqnarray}
and is the total energy emitted by radioactive decays\footnote{If we assume that $L(t)$ only represents the total IR/optical/UV luminosity then $Q(t)$ is the radioactive energy absorbed in the envelope.}, and
$E(t)$ is the total radiative energy of the envelope at time $t$. It arises because
\begin{eqnarray*}
& & \int^t_{t_0} 16 \pi^2 t  \int_{\hbox{$r_{\rm min}$}}^{\hbox{$r_{\rm max}$}} {1 \over cr^2}  {D(r^4 J)  \over Dt}\,\,  dr \, dt \\
=& &\int^t_{t_0} 16 \pi^2 t  \int_{\hbox{$V_{\rm min}$}}^{\hbox{$V_{\rm max}$}} {1 \over c V^2 t^2}  {D(r^4 J)  \over Dt}\,\,  t \,dV \, dt  \\
= & &\int_{\hbox{$r_{\rm min}$}}^{\hbox{$r_{\rm max}$}} 4\pi r^2 \left[ t 4\pi J/c \right]^t_{t_0} \,dr\\
=& & tE(t)-t_oE(t_o)
\end{eqnarray*}

\noindent
In equation \ref{eq_time_en} the luminosity at the inner boundary is taken as zero. The factor of $t$ in the equation allows  for the influence of adiabatic expansion. This equation would need to be modified in the presence of alternate energy sources such as a magnetar.

 An approximate form of equation \ref{eq_time_en} is
\begin{eqnarray}
 \int^t_{t_0} t L(t) \,dt = \int^t_{t_0} t Q(t) \,dt  + t_oE(t_o) -  t E(t)
\label{eq_time_approx}
\end{eqnarray}
since the gas-pressure terms are generally subservient to the radiation and decay terms, and
can be neglected.  Versions of the above equation (but in the observer's frame) have been provided by \cite{katz_13_56ni} who pointed out that one can derive the \isoni\ mass in Type Ia SNe by integrating the observed luminosity. \citet{nakar_56ni_16} later used a more general equation in a study of Type II-P SNe .

Valid for non-interacting SNe, Equation~(\ref{eq_time_approx}) highlights the two distinct mechanisms that produce the observed light curve. In a  generic Type Ia SN, the radiative energy in the ejecta is ``initially" small due to the rapid expansion from an Earth-size object ($r < 10^9$\,cm) to $10^{14}$\,cm on a time scale of one day. As a consequence the SN is initially faint. The luminosity of the SN increases as the energy deposited by radioactive decay in the interior diffuses to the surface, and the entire light curve is powered by nuclear decay -- primarily \isoni\ and its daughter isotope, \isoco.

\begin{figure}
\epsfig{file=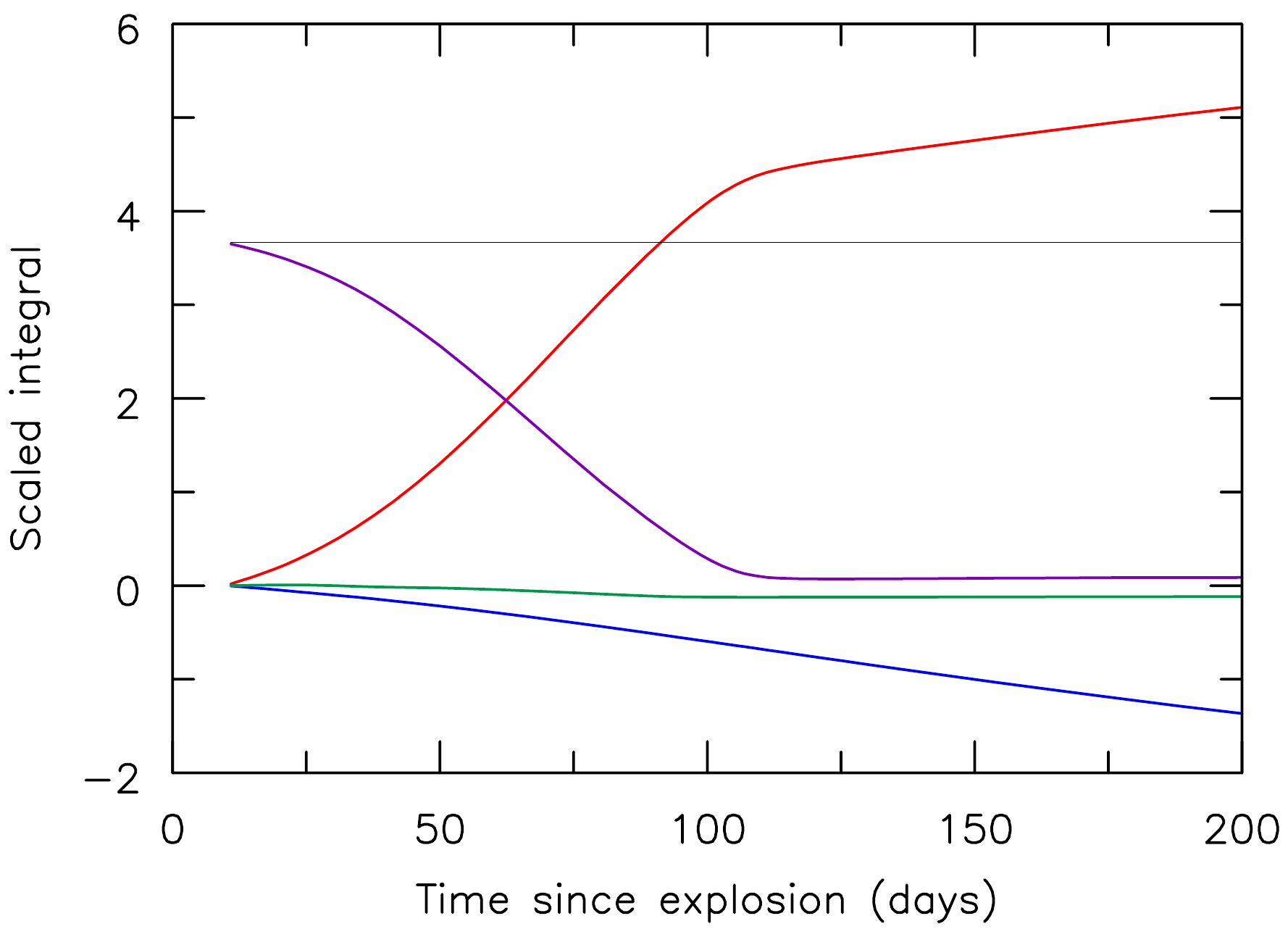,width=8cm}
\caption{ Comparison of various terms in Equation \ref{eq_time_en}. The red curve shows $\int^t_{t_0} t L(t) \,dt$,
the blue curve is $\int^t_{t_0} -t Q(t) \,dt$, the green curve is $\int^t_{t_0} t I(t) \,dt$, and the purple curve is $t E(t)$.
The sum of these 4 curves gives the initial energy $t_oE(t_o)$ (shown in black) to better than 1.5\% at all epochs.
}
\label{fig_en_con}
\end{figure}

Type II-P SNe represent the other extreme. The progenitor is a RSG with an initial radius of $\sim500$\,\rsun. With the large radius and \Teff\ $\sim 2\times 10^5$\,K at breakout, we see a rapid  brightening of the SN. To a large extent the light curve is determined by the initial temperature structure of the ejecta. A recombination wave moves into the ejecta allowing stored and trapped thermal energy to be released. This release controls the early part of the light curve. At later times energy released by radioactive decay becomes increasingly important, and it is the dominant power source in the nebular phase. In Fig~\ref{fig_en_con} we illustrate the various terms for model x3p0.

The practical importance of equation \ref{eq_time_en} is that it can be used to check the accuracy of the calculations. While equation \ref{eq_global_en} provide a check on the accuracy of the calculation at a single time step, it does not provide any indication of the accuracy over multiple time steps. However we can use equation~\ref{eq_time_en}, and such a check is now available with our \cmfgen\ calculations. Using this check we did discover a small, but systematic, energy loss in models (but which had very little influence on resultant spectra). In the zeroth-moment equation we have a term containing $Dr^4J/Dt$ which is equivalent to $rDr^3J/Dt + r^3 VJ$. For historical reasons we used the later form for differencing, however differencing the first form provides greater accuracy.

\section{Light curve comparison between \cmfgen\ and {\sc v1d} }
\label{sect_cmf_v1d_lc}

\begin{figure}
\epsfig{file=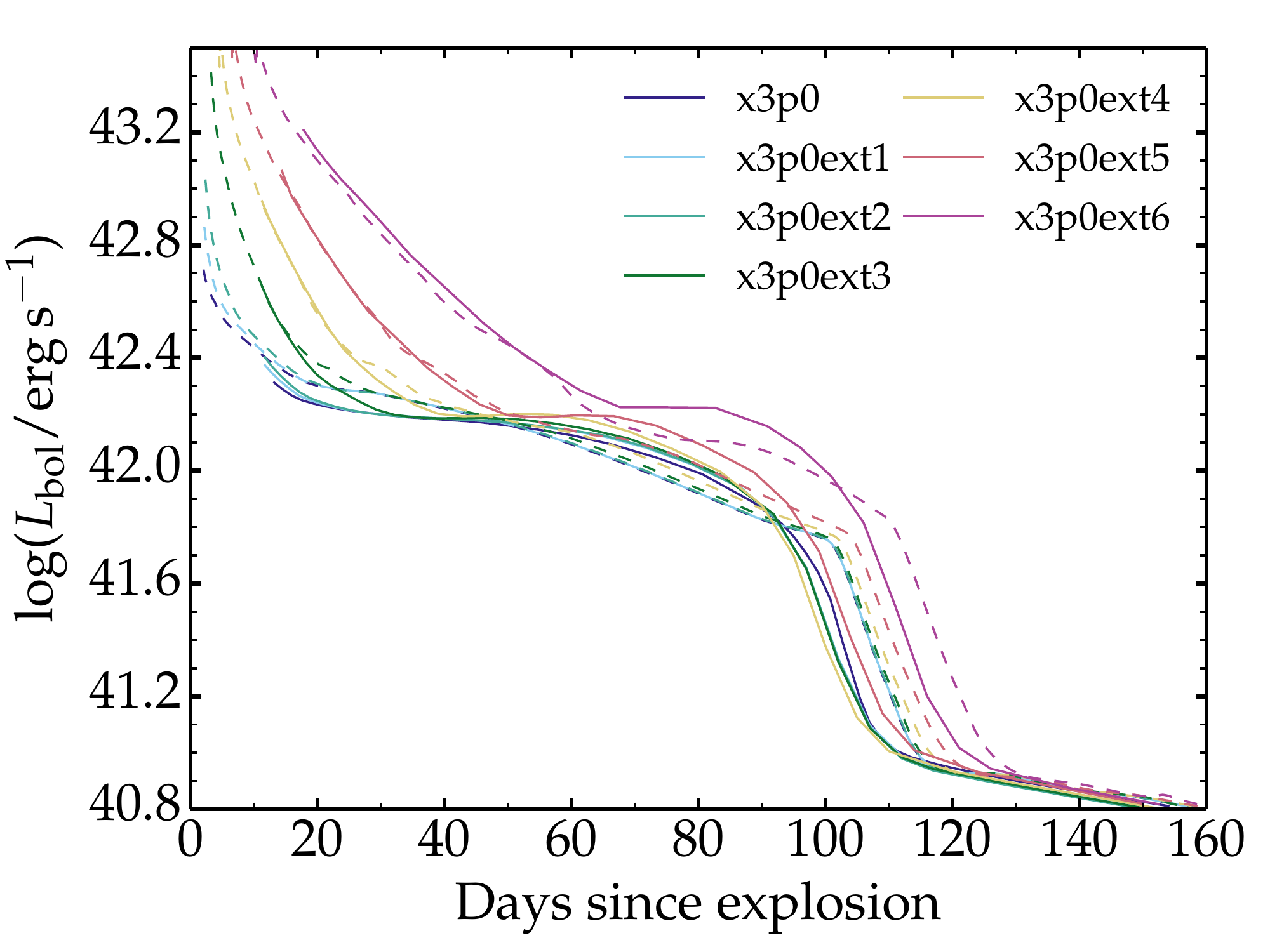,width=8cm}
\caption{ Bolometric light curve comparison between \cmfgen\ (solid) and {\sc v1d} (dashed) for model x3p0 and variants with a dense atmosphere.  We show the results for {\sc v1d} only past 1\,d after shock breakout.
}
\label{fig_lbol_v1d}
\end{figure}

   Figure~\ref{fig_lbol_v1d} compares the bolometric light curves obtained with \v1d\ and
   \cmfgen\ for the {\it ext} model set. \cmfgen\ simulations are started when the ejecta is close to homologous expansion, typically between 10 and 20\,d after explosion (it takes longer for models with a massive and extended CSM like model x3p0ext6). These two 1-D codes differ in many ways. The code \v1d\ solves the radiation hydrodynamics equations using gray flux-limited diffusion and assuming the gas is in LTE at each depth. \cmfgen\ ignores dynamical effects but solves the radiative transfer equation and the statistical equilibrium equations, accounting for the effects of line and continuum processes, as well as non-thermal and time-dependent effects. Despite these many differences, the bolometric light curves obtained with the two codes are in rough agreement. An offset is visible at the end of the photospheric phase, but it is small. In \cmfgen, the luminosity is higher at the end of the photospheric phase. This depletes the stored energy faster and causes an earlier transition to the nebular phase.

\section{Temperature evolution} 
\label{sect_temp_evol}

\begin{figure}[h]
\epsfig{file=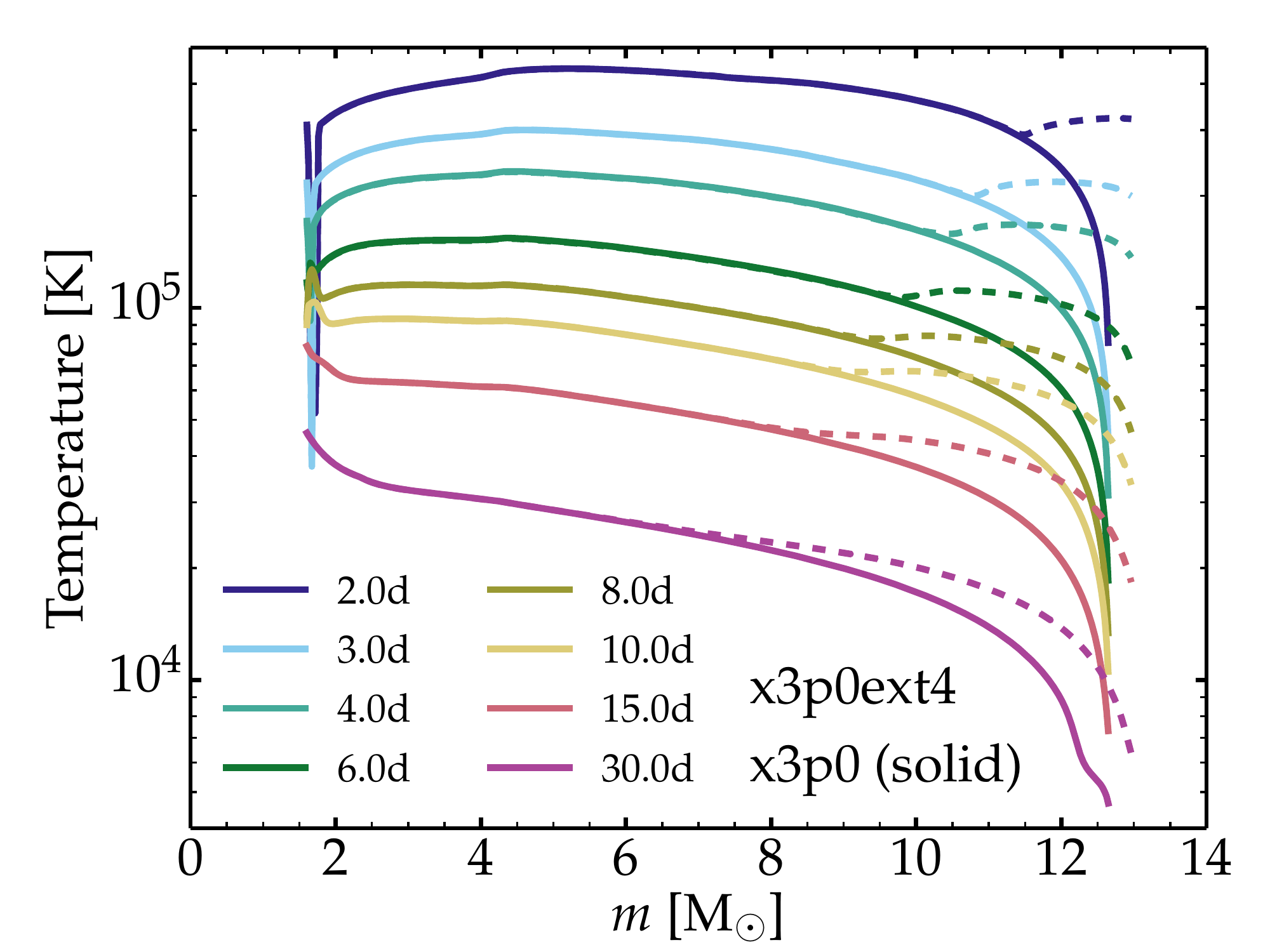,width=8cm}
\caption{Comparison of the temporal evolution of the temperature for model x3p0 and model x3p0ext4.
The influence of the CSM on the temperature structure is easily seen.}
\label{fig_temp_evol}
\end{figure}

In Figure~\ref{fig_temp_evol} we illustrate the temperature evolution for model x3p0 (solid), and model x3p0ext4 (dashed). As to be expected, the outer region of the x3p0ext4 model is hotter because of the interaction of the ejecta and radiation field with the CSM. At depth the temperature evolution is unchanged by the CSM.

\end{document}